\documentclass[12pt]{article}
\pdfoutput=1
\usepackage{subfigure}
\usepackage{graphicx,array}
\usepackage{color}
\usepackage{amsthm}
\usepackage{amsmath}
\usepackage{enumitem}
\usepackage{amssymb}
\usepackage{amsfonts}
\usepackage[hang,flushmargin]{footmisc} 
\usepackage{hyperref}
\usepackage{youngtab}
\usepackage{tikz}
\usepackage{graphicx}
\usepackage{ytableau}
\usepackage[T1]{fontenc} 
\usepackage{xcolor}
\usepackage{makecell}
\usepackage{longtable}
\usepackage{lipsum}
\def\simgt{\mathrel{\lower2.5pt\vbox{\lineskip=0pt\baselineskip=0pt
			\hbox{$>$}\hbox{$\sim$}}}}
\def\simlt{\mathrel{\lower2.5pt\vbox{\lineskip=0pt\baselineskip=0pt
			\hbox{$<$}\hbox{$\sim$}}}}

\newcommand{\beq}{\begin{equation}}
\newcommand{\eeq}{\end{equation}}
\newcommand{\bea}{\begin{eqnarray}}
\newcommand{\eea}{\end{eqnarray}}
\newcommand{\be}{\begin{eqnarray}}
\newcommand{\ee}{\end{eqnarray}}

\newcommand{\polp}{\epsilon^\parallel}
\newcommand{\polo}{\epsilon^\perp}

\definecolor{nicered}{rgb}{0.7,0.1,0.1}
\definecolor{nicegreen}{rgb}{0.1,0.5,0.1}

\definecolor{mGreen}{rgb}{0,0.6,0}
\definecolor{mgray}{rgb}{0.6,0.6,0.6}
\definecolor{mpurple}{rgb}{0.58,0,0.82}
\definecolor{backgroundColour}{rgb}{0.95,0.95,0.92}

\definecolor{mred}{rgb}{0.5,0.0,0.0}
\definecolor{mgreen}{rgb}{0.0,0.4,0.0}
\definecolor{mblue}{rgb}{0.0,0.0,0.6}
\definecolor{myellow}{rgb}{0.4,0.4,0.0}
\definecolor{mpink}{rgb}{0.4,0.0,0.4}
\definecolor{mcyan}{rgb}{0.0,0.4,0.4}
\definecolor{mblack}{rgb}{0.0,0.0,0.0}

\def\ve{{\varepsilon}}

\newcommand{\bd}[1]{\begin{fmffile}{#1}\begin{fmfgraph*}}
		\newcommand{\ed}{\end{fmfgraph*}\end{fmffile}}

\def\0{{(0)}}
\def\1{{(1)}}
\def\2{{(2)}}
\def\3{{(3)}}
\def\4{{(4)}}

\def\+{{(+)}}
\def\-{{(-)}}

\newcommand{\ba}{\begin{align}}
\newcommand{\ea}{\end{align}}
\def\be{\begin{equation}}
\def\ee{\end{equation}}
\def\beq{\be\begin{array}{c}}
	\def\eeq{\end{array}\ee}

\numberwithin{equation}{section}

\begin{document}
\thispagestyle{empty}
\hfill
TIFR/TH/20-1
\vspace{2cm}
\begin{center}
{\LARGE\bf
Classification of all 3 particle S-matrices quadratic in 
photons or gravitons}\\
\bigskip\vspace{1cm}{
{\large Soumangsu Chakraborty${}^{1,a}$, Subham Dutta Chowdhury${}^{2,a}$, Tushar Gopalka${}^{3,a,b}$, Suman Kundu$^{4,a}$,  Shiraz Minwalla${}^{5,a}$, Amiya Mishra${}^{6,a}$ }\let\thefootnote\relax\footnotetext{email:\\ ${}^{1}$soumangsu@theory.tifr.res.in,\,${}^{2}$subham@theory.tifr.res.in,\,${}^{3}$tushar.gopalka@students.iiserpune.ac.in,\\${}^{4}$suman@theory.tifr.res.in,\,${}^{5}$minwalla@theory.tifr.res.in,\,${}^{6}$amiya@theory.tifr.res.in.} 
} \\[7mm]
 {\it${}^{a}$ Department of Theoretical Physics, \\ 
 Tata Institute for Fundamental Research, Mumbai 400005}\\
[4mm]
{\it ${}^b$ Indian Institute of Science Education and Research, Pune - 411008}
 \end{center}
\bigskip
\centerline{\large\bf Abstract}
We explicitly construct every kinematically allowed three particle graviton-graviton-$P$  and photon-photon-$P$ S-matrix in every dimension and for every choice of the little group representation of the massive particle $P$.  We also explicitly construct the spacetime Lagrangian that generates each of these couplings. In the case of gravitons we demonstrate that this Lagrangian always involves (derivatives of) two factors of the Riemann tensor, and so is always of fourth or higher order in derivatives. This result verifies one of the assumptions made in the recent preprint \cite{Chowdhury:2019kaq} while attempting to  establish the rigidity of the Einstein tree level S-matrix within the space of local classical theories coupled to a collection of particles of bounded spin. 

\begin{quote} \small

\end{quote}

\newpage

\tableofcontents


\section{Introduction}

The authors of the recent paper \cite{Chowdhury:2019kaq} have conjectured that four particle scattering amplitudes in `consistent' classical theories can grow no faster that 
$s^2$, as $s$ is taken to infinity, at fixed $t$ ($s$ and $t$ are Mandlestam variables). They then went on to use this `Classical Regge Growth' (CRG) 
conjecture to classify consistent classical
\footnote{A classical S-matrix is one whose only non analyticities are simple poles corresponding to the exchange of a physical particle.} four graviton  S-matrices. In particular they demonstrated that  the classical Einstein S-matrix in $D \leq 6$ spacetime dimensions admits no polynomial deformations\footnote{i.e. a deformation 
	generated by a local addition to Einstein's action.} consistent with the CRG conjecture. Making some additional assumptions, they also argued that additions 
to the Einstein S-matrix with a finite number of additional poles \footnote{Corresponding to the exchange of a finite number of particles.}
also always violates the CRG scaling bound in $D \leq 6$. These results imply that, at least when $D \leq 6$,  the Einstein S-matrix is the unique classical four graviton S-matrix with a finite number of poles. 

We emphasize that this important sounding conclusion has so far only been established assuming both the validity of 
the CRG conjecture as well as an additional assumption. In this paper we will present no additional evidence either in favor of or against the CRG conjecture. We will
however prove the validity of the additional assumption (see below) that 
went into demonstrating that all finite pole exchange contributions 
to four particle gravitational scattering (other than the graviton pole in Einstein gravity) lead to S-matrices that grow faster than $s^2$ at fixed $t$.

We now describe the assumption that was made in \cite{Chowdhury:2019kaq} which 
we rigorously establish in this paper. In \cite{Chowdhury:2019kaq} the Regge growth of 
the pole exchange contribution of $P$ to the four graviton 
S-matrix was analyzed assuming that every kinematically allowed three particle graviton-graviton-$P$  S-matrix is generated by an interaction of the form 
\begin{equation}\label{tpf}
\int \sqrt{-g} \left( R R S \right) 
\end{equation}
(the extremely schematic notation used in \eqref{tpf} allows for a finite number of derivatives acting on the two Riemann tensors and on $S$, the field associated with the particle $P$). Note in particular that scattering amplitudes generated by this class of interactions are automatically of fourth or higher order in derivatives. 

While it is clear that Lagrangians of the form \eqref{tpf} 
generate graviton-graviton-$P$ S-matrices, it is less clear 
that all such S-matrices are generated by Lagrangians of 
the schematic form \eqref{tpf}. For example 
Lagrangians of the schematic form
\begin{equation}\label{scf}
\int \sqrt{-g} \left( R S \right), 
\end{equation} 
\footnote{Once again $R$ is a schematic for the Riemann tensor and $h$ is a schematic for the metric fluctuation.}
also contain cubic couplings of the form $ h h P$. The reader 
might wonder whether this simple observation is, in itself, 
already sufficient to falsify the assumption described above. 
This is not the case. In addition to 
the $hhP$ coupling described above, the Lagrangian  \eqref{scf} typically also generates a coupling of the form $hP$. Such a coupling results in 
an off diagonal term in $h$ and $P$ 
propagators. When this happens it follows that the field 
$h$ is a linear combination of the true graviton and (some 
derivatives on) the field associated with the particle $P$. In order to compute the 
true graviton-graviton-$P$ coupling one must first perform 
a field redefinition to the fields $h'$ and $P'$ 
that diagonalize the quadratic part of the action, and 
then compute the coefficient of $h' h' P'$. It is at least 
possible - and on physical grounds it seems plausible -  that 
this new coupling always vanishes
\footnote{The argument is the following. If we can perform 
	the required field redefinition not just linearly but also at the non-linear order, then the Lagrangian in the new variables 
	must presumably now have no couplings of the form \eqref{scf} (because there should be no quadratic cross terms in the Lagrangian involving the new graviton).}.

Considerations of the sort outlined in the previous paragraph motivated the authors of \cite{Chowdhury:2019kaq} to assume that all graviton-graviton-$P$ three particle scattering amplitudes 
are generated by Lagrangians of the form \eqref{tpf}. 
While plausible, these considerations do not seem completely infallible. \footnote{For example, similar considerations would lead one to 
	conjecture that all deformations of the Einstein graviton graviton graviton 3 particle S-matrix are generated by 
	Lagrangians of the form
	\begin{equation} \label{egm}
	\int \sqrt{-g} \left( R R R \right) 
	\end{equation} 
	(the analogue of \eqref{tpf}). 
	This result is famously untrue; in 5 and higher dimensions 
	the Gauss Bonnet term - which takes the schematic form 
	\begin{equation} \label{egm2}
	\int \sqrt{-g} \left( R R\right) 
	\end{equation}
	(the analogue of \eqref{scf}) also generates a new 3 graviton scattering amplitude. This action, which is quadratic in Riemann curvatures turns out not to modify the graviton propagator owing to a well known but remarkable algebraic fact; for the particular case of the Gauss-Bonnet action the term in \eqref{egm2} that is 
	quadratic in $h$ is a total derivative and 
	so does not modify the graviton propagator.}
In this note we provide a clear and rigorous proof that the final conclusion of these considerations is nevertheless true. All 
graviton-graviton-$P$ 3 point scattering amplitudes are 
indeed generated by Lagrangians of the form \eqref{tpf}, 
at least when the particle $P$ is massive. We establish this fact
in the most direct manner possible. We simply construct 
all kinematically allowed 3 particle graviton-graviton-$P$ 
S-matrices in every dimension and verify by inspection 
that these S-matrices are indeed generated by actions 
of the form \eqref{tpf}. 

The program of enumeration of all possible graviton
graviton $P$ three particle S-matrices is not difficult to carry through explicitly
 because it is an effectively 
finite problem. For any three particles $P_1$, $P_2$, $P_3$
it is well known that the vector space of kinematically 
allowed 3 particle S-matrices is finite dimensional. In the 
special case of graviton-graviton-$P$ scattering, it also turns out (see below) that the dimensionality of this space does not grow as a function of the complexity of the Lorentz 
representation of the particle $P$ or of the spacetime dimension in which we work, but is, instead, bounded to be less than or equal to eight. For this reason it is not too 
difficult to enumerate all possible couplings for particles 
$P$ transforming in any given representation of the 
massive little group $SO(D-1)$. 

Now the set of all possible representations of $SO(D-1)$ 
is not finite; representations of this form are specified by 
$\lfloor \frac{D-1}{2} \rfloor$, integers or half integers which we take to be 
the highest weights $(h_1, h_2, \ldots  h_{\lfloor \frac{D-1}{2} \rfloor})$ under 
rotations in orthogonal two planes 
(see Appendix \ref{DR} for notation and more details). It might thus seem that the 
task of scanning through all possible representations of 
$SO(D-1)$ is an onerous one, especially at large values 
of $D$. This is not the case for a simple reason. While the 
set of all possible representations of $SO(D-1)$ is 
indeed quite large, only a small subset of these representations can consistently scatter of two gravitons 
(see the end of subsection \ref{dgeeeg} for a listing). 
In particular this subset stabilizes for $D \geq 8$ (it does 
not further increase as $D$ is increased above $8$) and 
is manageable enough to allow us to systematically and explicitly  construct all non-zero  graviton-graviton-$P$ scattering in a case by case manner. 

Though we are primarily interested in graviton-graviton-$P$ couplings, through this paper 
we also explicitly construct all photon-photon-$P$ couplings, 
partly as a warm up to the problem of principal interest, 
and partly in the hope that the photon S-matrices that 
we construct will also turn out to find  their own applications (perhaps unrelated to the CRG conjecture) in future work. 

The method we use in our enumeration is explained in detail in section \ref{cptf} below.  We first explain how the number of distinct photon-photon-$P$ and graviton-graviton-$P$ three particle S-matrices can be counted using simple group theory considerations. We perform this counting using a method that is 
essentially identical to the 
procedure described in \cite{Kravchuk:2016qvl} for the counting of 
CFT correlators involving two stress tensor or two conserved currents in $d=D-1$. 
This counting problem is purely kinematical and makes no assumptions 
about the nature of the Lagrangian that generates the 
corresponding S-matrix . We then proceed to implement this counting procedure in every dimension. Once we know how many 
photon-photon-$P$ (or graviton-graviton-$P$) 3 particle 
S-matrices there are for particles $P$ transforming in 
any given representation of the massive little group $SO(D-1)$, we then proceed by trial and 
error to simply construct the same number of distinct 
interaction Lagrangians of the form \eqref{tpf} (in the 
case of gravitons) or a similar form with curvatures 
$R$ replaced by field strengths $F$ (in the case of photons). 
The key point is that in every example we are always able 
to saturate this counting. Restated, there are always as many 
independent Lagrangians of the form \eqref{tpf} (or its photon analogue) as the number of graviton-graviton-$P$ couplings (or photon-photon-$P$ couplings) enumerated through group theoretical means. 

In order to be sure that the Lagrangians we have computed 
are all actually independent, we have explicitly computed the
S-matrices generated by each of these Lagrangians (see Appendix \ref{pa} and \ref{D8Gr:Amp})  and verified their linear independence. As a further check of the completeness of our results, in some cases we have also algebraically (with no reference to a Lagrangian) constructed the most general Bose symmetric, Lorentz invariant and gauge invariant function of scattering data, and so have independently constructed the linear vector space of allowed 3 particle S-matrices. In each case for which we have performed this construction we have compared the resultant vector space 
of S-matrices with the space obtained from the Lagrangians 
described above (the ones we constructed by trial and error to match the counting). In each case we find perfect agreement. 

The set of Lagrangians that generate all 3 particle S-matrices stabilizes for $D \geq 8$; our final results 
for the corresponding  Lagrangians in these asymptotic 
dimensions are listed in Tables \ref{D8Ph} (in the case 
of photons) and \ref{D8gr} (in the case of gravitons). 
The results for each $D \leq 8$ are also presented in 
the text and tables of subsections \eqref{dese}, \eqref{desi}, \eqref{desfi}, \eqref{desfo} and \eqref{desth}
in the case of photons, and in subsections 
\ref{dgeeeg}, \ref{deseg}, \ref{desig} \ref{desfig}, \ref{desfog} for gravitons. 

To the best of our knowledge the systematic enumeration of all photon-photon-$P$ and graviton-graviton-$P$ three point functions (and corresponding Lagrangians) - for all possible particles $P$ and every 
dimension $D$ - has never before appeared in the literature. However 
special cases of this enumeration have appeared in previous work. 
In particular the complete enumeration of these three particle scattering amplitudes
is particularly simple in four dimensions for two separate reasons. First, the irreducible representations of the four dimensional massive little group  $SO(3)$ are particularly simple (the representations that can couple to two photons or two gravitons are all completely symmetric traceless tensors, labelled by a single integer $l$). Second four dimensions is particularly well suited to the spinor helicity formalism. Using both these 
advantages, the authors of \cite{Arkani-Hamed:2017jhn} have provided a very simple and completely general enumeration of all possible photon-photon-$P$ and 
graviton-graviton-$P$ amplitude in four dimensions in the language of the 
spinor helicity formalism. The paper \cite{Bonifacio:2018vzv} also studies photon-photon-$P$ and graviton-graviton-$P$ scattering in four dimensions, and generalizes this study to 5 dimensions for the special case that $P$ is a massless symmetric tensor. The spinor helicity approach has also been used to study photon-photon-$P$ and graviton-graviton-$P$ amplitudes in  the special case that $P$ is a symmetric tensor \cite{Jha:2018hag}. Though we have not carried out a completely systematic translation of the results of these earlier studies to the notation employed in this paper, as far as we can tell our general results agree with all earlier reported special cases that we are aware of atleast from the viewpoint of counting. \footnote{As we mention in the next paragraph, comparison with CFT correlators gives us an independent check of our results for the special 
	case $D=4$.}

The explicit construction of all graviton-graviton-(or photon-photon-) $P$ three point S matrices in $D$ dimensions is very closely tied to the construction 
of all $T_{\mu\nu}$ $T_{\mu\nu}$ (or $J_\mu J_\mu$) `corresponding operator' CFT three point functions in $d=D-1$ dimensions. This problem has been studied by several authors. In particular, the number of coefficients in the three point functions of two conserved symmetric currents of any rank with one non conserved tensor (again of any rank)
in $d=3$ was enumerated in \cite{Giombi:2011rz}.  In the special case that the conserved currents are of spin one and two, we have checked that the 
number of undetermined coefficients in the CFT ennumeration  match with the number of undetermined coefficients 
listed in this paper for  the scattering of two photons / two gravitons against a general massive particle in $D=4$, as expected from the AdS/CFT correspondence. 
The special case of parity even couplings 
in which the `corresponding operator' lies in the completely symmetric tensor representation was studied for $d \geq 4$  in \cite{Costa:2011mg}. The counting of three particle 
S-matrix structures presented in this paper agrees with 
the counting of the corresponding three point functions presented in Table 1 of \cite{Costa:2011mg}. While the 
three point functions of more  general representations of $SO(d)$ have been  studied in \cite{Costa:2014rya, Kravchuk:2016qvl} (see also \cite{Geyer:2000ig, Rejon-Barrera:2015bpa, Costa:2016hju, Lauria:2018klo}), we have not been able to locate a detailed CFT listing of the space of allowed three point functions of 
two stress tensors (or two conserved currents) and an operator transforming in general representations of $SO(d)$. 
It would be useful to perform 
such a study (if it indeed has not already been done) and to establish the detailed map - induced by the AdS/CFT correspondence -  from the S-matrix structures 
constructed in this paper to three point function CFT correlators.
\footnote{As we have mentioned above, the counting of the number of distinct 
S-matrix structures presented in this paper is identical to the counting 
of corresponding CFT correlators described in \cite{Kravchuk:2016qvl}, so the
number of independent structures on the two sides is guaranteed to match.}

To end this introduction we note that three particle S-matrices play roughly the same role for the Lorentz group that Clebsh-Gordon 
coefficients play in compact groups. In this paper we have provided a
detailed and completely explicit listing of this elementary group theoretic data 
for the special case of graviton-graviton-$P$ and photon-photon-$P$ scattering. Specifically  we have listed all kinematically allowed 
graviton-graviton-$P$ (and photon-photon-$P$) S-matrices. The single application we have made of this detailed listing so far - namely the tightening of the arguments of \cite{Chowdhury:2019kaq} - utilizes only a gross structural feature of our results. We hope that future investigations into diverse aspects of gravitational scattering will find several additional uses for the three point scattering amplitudes listed in this paper.

\section{Counting 3 point structures} \label{cptf}

\subsection{Labelling Massive Particles} \label{lmp}

Consider the three point scattering of two photons - or two gravitons - with 
a massive particle $P$ in $D$ spacetime dimensions. The particle $P$ transforms in an irreducible non-spinorial \footnote{The three point coupling between two vectors or two tensors  and an spinorial representation always vanishes.} representation of the massive little group $SO(D-1)$. As mentioned in the introduction (and explained in 
more detail in Appendix \ref{DR}), in this paper we label representations of $SO(2m)$
or $SO(2m+1)$ by their highest weights under rotations in the $m$ orthogonal two planes, $h_1, h_2 \cdots h_m$. For the non spinorial representations of interest 
to this paper, $h_i$ are positive integers for $i <m$. $h_m$ is also a positive  integer for the case $SO(2m+1)$, but can be either a positive or negative integer for the case $SO(2m)$. We work with the convention 
$$h_1 \geq h_2 \geq  \cdots \geq |h_{m}|,$$ and denote a representation labelled by 
these highest weights by $(h_1, h_2 \cdots h_m)$. 

We will sometimes find it convenient to associate Young Tableaux with non spinorial 
representations of $SO(2m+1)$ and $SO(2m)$. We use the symbol $Y_{(r_1, r_2 \cdots r_m)}$  to 
denote a Young Tableaux with $r_i$ boxes in the $i^{th}$ row; note, of course, that all $r_i$ are positive. In the case of $SO(2m+1)$ we associate the Young Tableaux $Y_{(r_1, r_2 \cdots r_m)}$ with the representation $(r_1, r_2, \cdots r_m)$; in this case the map between irreducible representations and Young Tableaux is one to one. In the case of $SO(2m)$ 
we associate the Young Tableaux  $Y_{(r_1, r_2 \cdots r_m)}$ with the direct sum 
of the representations  $(r_1, r_2 \cdots r_m)$ and $(r_1, r_2 \cdots -r_m)$ when 
$r_m>0$, but with the single representation $(r_1, r_2 \cdots 0)$ when $r_m=0$. In other words Young Tableaux with less than $m$ rows 
are associated with a unique $SO(2m)$ irreducible representation, while Young Tableaux 
with $m$ rows are associated with the direct sum of a pair of $SO(2m)$ irreducible representations.

Consider any particular Young Tableaux with a total of  $r$ boxes 
(i.e. $r_1+r_2 + \ldots r_m=r$). A set of spacetime fields that transform in the 
associated representations of $SO(D-1)$ may be constructed as follows. We first 
consider the set of all traceless $r$ index tensors of $SO(D-1)$. In general, 
this set of tensors transforms in a sum over several irreducible representations of 
$SO(D-1)$. It is, however, possible to project onto a subspace of the full space 
of $r$ index traceless symmetric tensors chosen so that this subspace transforms in the irreducible representation(s) associated with the given Young Tableaux. The choice of this subspace is not, in general, unique. In group theoretic terms the irreducible representation(s) associated with the Young Tableaux in question generally occur in the decomposition of traceless symmetric tensors with nontrivial multiplicity. \footnote{It is easy to make this statement quantitative in the case of the 
	group $SU(m)$. When we decompose the $r^{th}$ tensor product of $m$ dimensional 
	defining representations of the group $SU(m)$ - i.e $r$ index tensors of $SU(m)$
	- into irreducible representations of the group, we find that the set of distinct representation associated with a Tableaux $Y$ transform in the irreducible representation of $S_r$ labelled by the same Tableaux $Y$.}
When needed, in this paper we work with a particular choice obtained as follows.  
 We write down the Tableaux $Y$ 
and fill in each of its boxes with 
distinct integers ranging from $1$ to $r$. There are $r!$ distinct ways of assigning the integers to boxes; we simply pick any one. 
Any particular assignment of integers defines a projector  according to the following rules. Let $\Pi_1$ 
denote the operation of completely symmetrizing all entries in every particular row of a Young Tableaux (without any associated normalization). Let $\Pi_2$ denote the operation of  
completely anti symmetrizing all indices in each column of a Tableaux, again without 
any associated normalization. Let $h$ denote the product of hooks associated with the Young Tableaux. Then the projector ${\cal P}$, defined by 
\begin{equation}\label{projy}
 {\cal P}= \frac {\Pi_2 \Pi_1}{h} ,
 \end{equation} 
when acting on the  space of traceless rank $r$ tensors- is a projector onto a collection of tensors that transforms in 
the representation of $SO(D-1)$ associated with the Young 
Tableaux $Y$ \cite{Nutma:2013zea, 2009arXiv0901.0827E, Green:2005qr}. 
\footnote{ A second natural choice 
- one that we will not make in this paper - would be to 
associate a particular filling in of a Young Tableaux with 
the projector 
\begin{equation}\label{projalt}
{\cal P}'= \frac {\Pi_1 \Pi_2}{h}.
\end{equation} 
While ${\cal P}$ and ${\cal P}'$ are not the same, it can be shown that the 
set of tensors projected onto by ${\cal P}$ corresponding to all 
possible fillings of a Tableaux coincides with the set of 
tensors projected onto by ${\cal P}'$ corresponding to all possible choices of filling of the Tableaux. ${\cal P}$ and  ${\cal P'}$ respectively project onto the tensors denoted by $T^{{\bf YT'}}$ and $T^{{\bf YT}}$  in  \cite{Costa:2014rya}.
  } The subspace of tensors 
projected onto by ${\cal P}$ transform in the representation(s) 
associated with the Young Tableaux $Y$ (see Appendix \ref{rayt} for an 
intuitive explanation of this fact.). Similar projectors have been used to impose the symmetry of the Young's diagram on a tensor in CFT via ``Young symmetrizers'' in \cite{Costa:2014rya} following \cite{cvitanovi2008group}  (see Section 2.3 of \cite{Costa:2014rya} for more details).  

Different choices of filling the Tableaux $Y$ lead to 
different projectors ${\cal P}$. Every choice leads to a collection 
of tensors that transform in the same representation of $SO(D-1)$. 
As all results we present below care only about this final representation 
content, they are insensitive to the precise choice we make.

\subsection{The scattering plane and physical polarizations}  \label{sppp}

Consider the scattering of two massless particles - either photons or gravitons -  of momentum $k_1$ and $k_2$ with a single massive particle of momentum $k_3$ with 
$$k_1+k_2+k_3=0.$$  
Lorentzian $R^{D}$ can be viewed as a $D$ dimensional vector space. The three scattering momenta define a timelike two dimensional subspace of Lorentzian $R^{D}$. This subspace is spanned, for instance, by the two vectors $k_1$ and $k_2$. We refer to this two dimensional space as the `scattering plane'. 

\subsubsection{Polarizations of the massless particles}  \label{pmp} 

It is standard, and convenient, to label the polarizations of any given photon or graviton by a polarization vector $\epsilon$ (in the case of gravitons the 
polarization vector is null; see e.g. subsubsection 2.1.3 of \cite{Chowdhury:2019kaq} for details). Let the 
the polarization vector associated with the two photons (or gravitons) be 
denoted by $\epsilon_1$ and $\epsilon_2$ respectively. It follows from the 
Lorentz gauge condition (which we adopt in this paper in order to maintain manifest Lorentz invariance) that 
\begin{equation}\label{epp}
k_i . \epsilon_i=0         ~~~~~~~~(i=1 \cdots 2).
\end{equation} 
\footnote{Actually, the equation \eqref{epp} also holds 
for the massive particle, $i=3$.} 
As in subsubsection 2.1.3 of \cite{Chowdhury:2019kaq} we find it convenient to decompose $\epsilon_i$ 
into  its components orthogonal to the scattering plane and its components in the scattering plane
\begin{equation}\label{perpparde}
\epsilon_i = \polo_i + \polp_i .
\end{equation} 
By definition $\polp_i$ is the part of the polarization vector that lies 
in the scattering two plane. It follows that $\polp_i= a_i k_1 + b_i k_2$. 
The condition \eqref{epp} (and the fact that $k_1.k_2$ is generically non-zero 
while $k_1^2=k_2^2=0$) then implies that 
\begin{equation} \label{spolp}
\polp_i  \propto  k_i .
\end{equation} 
In other words $\polp_i$, the `in plane' part of the polarization vector $\epsilon_i$, is 
necessarily pure gauge, and can be set to zero by a gauge transformation. 
It follows that for the purpose of enumerating gauge invariant amplitudes, 
the photon (or graviton) polarization vectors can be taken to lie entirely 
orthogonal to the scattering plane, i.e. 
\begin{equation}\label{perpparden}
\epsilon_i = \polo_i, ~~~~~~k_1.\polo_i = k_2.\polo_i=0 .
\end{equation}

\subsubsection{States of the massive particle from $D$ dimensional fields} \label{pmap} 

In this paper we will construct the scattering amplitudes of particles $P$ 
that transform in $SO(D-1)$ irreducible representations(s) corresponding to any
given Young Tableaux. The scattering states of such particles are easily 
represented as the solutions to the linearized equations of motion of a 
(Lorentzian) $D$ dimensional spacetime field as follows.

Consider any Young Tableaux - like the one associated with the particle $P$ - 
that is linked to $SO(D-1)$ irreducible representation(s). Such a Young Tableaux - now thought of as a Tableaux associated with $SO(D)$ - 
can also be associated with a tensor field of $SO(D)$ \footnote{In the case when 
$D$ is even, a generic Young Tableaux associated with $SO(D)$ has $D/2$ rows. 
A Tableaux associated with $SO(D-1)$ representations has $D/2-1$ rows, and so 
are a subset of all possible $SO(D)$ Tableaux. In the case when $D$ is odd, on the 
other hand, the two sets of Tableaux simply coincide. The considerations of this 
subsection apply equally well to both cases.}   We impose two equations of motion on this tensor field. First, the mass shell condition $k_3^2=-m^2$.  Second, the condition that the contraction of $k_3$ with each of the tensorial indices vanishes. 

The space of solutions to the equations of motion of this $D$ dimensional tensor field at any 
given value of $k_3$ is then given by the space of $SO(D)$ tensors associated 
with $Y$ subject to the condition that all indices of these tensors are orthogonal 
to $k_3$. A moment's reflection will convince the reader that this space of 
solutions transforms under the $SO(D-1) \subset SO(D)$ that acts trivially on $k_3$ (i.e leaves it invariant). Under this $SO(D-1)$, moreover, the space of solutions 
transforms in the representation associated with the Young Tableaux $Y$ as we 
wanted.

\subsection{Counting three point structures}

In the previous subsection we have explained that the polarization tensors of photons (and gravitons) are effectively orthogonal to the scattering two plane and so each transform in a vector (traceless symmetric two tensor) of the $SO(D-2)$ that keeps the scattering two plane fixed. The polarizations of the massive particle $P$, on the other hand, transform in a single representation of the $SO(D-1)$ orthogonal to $k_1+k_2$, and so transform in the sum of a given collection of irreducible representations of $SO(D-2)$ (the precise list of representations is given by the branching rules of Appendix \ref{DR}). The most general kinematically allowed two photon (or two graviton) and one $P$ three point scattering amplitude is simply the most general Bose symmetric $SO(D-2)$ singlet that can be formed from the product of two $SO(D-2)$ 
vectors (or two $SO(D-2)$ symmetric tensors) and the $SO(D-2)$ representations 
obtained from the particle $P$. This is

  In the rest of this section, we will count the number of such invariants for every choice of the Young Tableaux $Y$ associated with $P$. Schematically one writes:

\section{Photons} 

Each photon polarization is a vector of $SO(D-2)$. The symmetric product (denoted by $S^2$) of two vectors transforms in the sum of a scalar and a traceless symmetric tensor. On the other hand, the antisymmetric product (denoted by $\Lambda^2$) of two vectors transforms in the antisymmetric two tensor representation. \footnote{This antisymmetric tensor vanishes when 
	$D=3$, is equivalent to a parity odd scalar when $D=4$ and is 
	equivalent to a parity odd vector when $D=5$. When $D=6$ the antisymmetric 
	tensor is a sum of two irreducible representations - self-dual and anti-self-dual. For $D > 6$ the antisymmetric tensor transforms in the single 
	$(1,1,0 \cdots 0 )$ representation. } Schematically one writes:
\begin{eqnarray}
 \Yvcentermath1 S^2\left(\yng(1)\otimes \yng(1)\right)&=&\Yvcentermath1 \yng(2) \oplus \bullet ~,\label{phdecompsym} \\
  \Yvcentermath1 \Lambda^2\left(\yng(1)\otimes \yng(1)\right)&=&\Yvcentermath1 \yng(1,1) ~ \label{phdecompasym}.
 \end{eqnarray}

The particle $P$ transforms in one (or two) particular representation(s) of 
$SO(D-1)$. It is useful to decompose these representation(s) of $SO(D-1)$ into the 
irreducible representations of the $SO(D-2)$ that leaves the scattering two plane 
invariant. This decomposition is given by the $SO(D-1) \rightarrow SO(D-2)$ 
branching rules reviewed in Appendix \ref{DR}. $SO(D-1)$ tensors are decomposed into  $SO(D-2)$ by either contracting every index with $k_1-k_2$ \footnote{The fact that every tensorial index is orthogonal to $k_1+k_2$ ensures that $k_1-k_2$ is the only vector available for branching purposes.} or projecting each index orthogonal to $k_1-k_2$ (or performing more complicated operations involving $\varepsilon$ (where $\varepsilon$ is the $SO(D)$ Levi-Civita tensor) tensors and possibly factors of $k_1-k_2$, see below for details). It is very important to keep in mind that  $k_1-k_2$ picks up a minus sign under the interchange $1 \leftrightarrow 2$. The $SO(D-2)$ representations that appear in the branching rules of Appendix \ref{DR} are of two sorts; those constructed only out of even powers of $k_1-k_2$ (such representations are invariant under the interchange of the indices $1$ and $2$) 
and those that are constructed out of a sum of terms, each of which is proportional 
to an odd power of $k_1-k_2$ (such representations pick up a minus sign under this interchange). We will call the first set of representations the `even' descendents of $Y$, and the second sort of representations the `odd' descendents of $Y$. 

It follows from the discussion of this subsection (and \eqref{phdecompsym}, \eqref{phdecompasym} in particular)  that the number of independent photon-photon-$P$ three point structures is the sum of the number of even scalar descendents of $P$, plus the number of even symmetric traceless two tensor descendents of $P$,  plus the number of odd antisymmetric two tensor descendents of $P$. 

Using the branching rules of Appendix \ref{DR}, it is now a simple matter to enumerate the number of independent photon-photon-$P$ structures for every choice of the Tableaux $Y$ in which $P$ transforms. To start with, it follows immediately from the rules of Appendix \ref{DR} that there are no three point structures  if $P$ transforms in a Tableaux with 4 or more rows, or if the number of boxes in the second plus third row exceed 2 \footnote{Intuitively this can be understood as follows. For two photons coupling to a mixed representation, we can form Lorentz invariants by contracting the polarizations of the mixed tensor with either one $\epsilon_1$, one $\epsilon_2$ or $k_1-k_2$. As we have only three independent vectors with which the $SO(D-1)$ tensor $Y$ can be dotted, it follows that no nonzero Lorentz scalar 
can be formed out of a representation with any columns of length 4 or greater. }. The most general representations that can couple with two photons have 
either 
\begin{itemize} 
	\item One box in the third row, one box in the second row, and one or more 
	boxes in the first row (see \hyperlink{D8Ph:s11:1}{Row 6} of Table \ref{D8Ph}) . They are denoted by $Y_{(r,1,1)}$\footnote{ We define the symbol $Y_{(a,b,c,d,\cdots)}$ to denote an Young Tableaux, which has $a$ boxes in the first row, $b$ boxes in the second row, $c$ boxes in the third row, $d$ boxes in the fourth row and so on.}.
	\item No boxes in the third row, either two (see \hyperlink{D8Ph:s2:1}{Row 5}  of Table \ref{D8Ph}), one (see \hyperlink{D8Ph:s1:1}{Row 3}, \hyperlink{D8Ph:s1:2}{Row 4} of Table \ref{D8Ph}) or zero boxes (see \hyperlink{D8Ph:s:1}{Row 1}, \hyperlink{D8Ph:s:2}{Row 2} of Table \ref{D8Ph}) in the 
	second row and correspondingly two or more, one or more and zero or more boxes 
	in the first row. They are denoted by $Y_{(r+2,2,0)}$, $Y_{(r+1,1,0)}$ and $Y_{(r,0,0)}$ respectively. 
	\end{itemize}  

The final result for the number of three point structures depends on the dimension in which we work. The dependence on dimension has its roots in the fact that a legal  Young Tableaux of $SO(2N+1)$ or $SO(2N)$ cannot have more than $N$ rows (and also the fact that $SO(2N)$ Tableaux with $N$ rows are special). Since we are only interested in  Tableaux of $SO(D-1)$ with a most 3 rows, the bounds on Young Tableaux (and the speciality of $SO(2N)$ Tableaux with $N$ rows) are unimportant when $D \geq 8$. As a consequence, we first study the counting problem for $D \geq 8$ and then individually consider every dimension $< 8$.  \footnote{The fact that the counting problem is different for $D \leq 7$ and for $D \geq 8$ may be understood more physically as follows. A photon-photon-$P$ scattering process has up to 7 independent vectors. These are the two polarizations of the two photons, up to 3 independent polarizations needed to characterize the state of the particle $P$ (we need up to three independent polarizations because the Young Tableaux for $P$ has upto 3 rows) and the two scattering momenta. It follows immediately that no scattering structure can involve an ${\varepsilon}$ tensor in $D \geq 8$; however scattering structures proportional to $\varepsilon$ do in general exist in $D \leq 7$.} 

\subsection{ $\mathbf{D \geq 8}$} \label{dgeee}

Let us first consider particles $P$ that transform under $SO(D-1)$ in the representation labelled by a completely symmetric Young Tableaux, $Y_{(r,0,0)}$ 
(LHS of \hyperlink{D8Ph:s:1}{Row 1}, \hyperlink{D8Ph:s:2}{Row 2} of Table \ref{D8Ph}). For every even $r \geq 0$, there exists one  photon-photon-$P$ 
structure corresponding to the fusion of the two photons 
into an $SO(D-2)$ scalar - this structure is obtained by contraction of all the indices 
of the traceless symmetric tensor with factors of $k_1-k_2$. This contraction 
structure is schematically depicted in the Young Tableaux on the LHS of 
 \hyperlink{D8Ph:s:1}{Row 1} (recall that yellow shading denotes contraction with $k_2-k_1$ in 
Fig \ref{D8Ph}). The 3 particle Lagrangian that generates this 3 point function 
is displayed in the right column of \hyperlink{D8Ph:s:1}{Row 1} of Table \ref{D8Ph}.

When $r$ is even and $r \geq 2$, there exists a second structure corresponding to the fusion of the two photons into the traceless symmetric tensor representation of $SO(D-2)$. 
This structure involves the contraction of  $r-2$ indices of the traceless 
symmetric tensor with $k_1-k_2$. The corresponding Lagrangian structure is given in RHS of  \hyperlink{D8Ph:s:2}{Row 2} of Table \ref{D8Ph}.

 In summary,
when $P \in Y_{(r,0,0)}$, 
  there are no photon-photon-$P$ structures when $r$ is odd
\footnote{The structures one can write down in this case are antisymmetric rather 
than symmetric under $1 \leftrightarrow 2$.}
one structure when $r=0$ and  $2$ such structures when $r$ is even and $r \geq 2$. 

Now consider particles $P$ that transform under $SO(D-1)$ in $Y_{(r+1,1,0)}$ 
 (LHS of \hyperlink{D8Ph:s1:1}{Row 3}, \hyperlink{D8Ph:s1:2}{Row 4} of Table \ref{D8Ph}). 
Contracting away $r$ boxes in the first row (see \hyperlink{D8Ph:s1:1}{Row 3} of Table \ref{D8Ph}) yields the antisymmetric tensor 
representation listed in RHS of \eqref{phdecompasym}; this contraction structure results in a Bose symmetric 
three point function only when $r$ is odd. The corresponding Lagrangian is listed 
in the right most column of \hyperlink{D8Ph:s1:1}{Row 3} of Table \ref{D8Ph}.
Contracting $r-1$ boxes in the first row and the single box 
in the second row  (\hyperlink{D8Ph:s1:2}{Row 4} of Table \ref{D8Ph}) yields the symmetric two tensor representation that appears on the RHS of \eqref{phdecompsym}. 
This contraction structure results in a Bose symmetric three point structure only when $r$ is even. The Lagrangian that gives rise to this 
S-matrix structure is listed in the rightmost column of \hyperlink{D8Ph:s1:2}{Row 4} of Table \ref{D8Ph} 
\footnote{Note that 
the rules of Appendix \ref{DR} tell us that we are not allowed to contract away 
all boxes in order to produce the singlet. Intuitively this is 
because two indices in the same column are mutually antisymmetrized, and so cannot 
be contracted with the same vector.}

Next consider the case of a Tableaux $Y_{(r+2,2,0)}$. Contracting both boxes in the 
second row and $r$ boxes in the first row (see LHS of \hyperlink{D8Ph:s2:1}{Row 5} of Table \ref{D8Ph}) produces the traceless symmetric 
two tensor representation (that appears in the RHS of \eqref{phdecompsym}) ; this manoeuvre yields a single 3 point structure when 
$r$ is even and no such structures when $r$ is odd. The explicit Lagrangian is listed in \hyperlink{D8Ph:s2:1}{Row 5} of Table \ref{D8Ph}. \footnote{In this case the 
	rules of Appendix \ref{DR} inform us that there is no legal way to obtain the two 
	index antisymmetric or the scalar representation. Again the intuitive reason 
	is that we are not allowed to contract two indices corresponding to the same 
	column of the Young Tableaux.} There are no other three particle S-matrices in this case.

Finally consider a Tableaux $Y_{(r,1,1)}$. Contracting the box in the third 
row and $r-1$ boxes in the first row (see LHS of \hyperlink{D8Ph:s11:1}{Row 6} of Table \ref{D8Ph}) yields the antisymmetric two tensor representation (that appears in the RHS of \eqref{phdecompasym}). Keeping track of Bose statistics, we find a single 3 point 
structure when $r$ is odd, and but no such structures when $r$ is even. Explicitly, the Lagrangian is given by RHS of  \hyperlink{D8Ph:s11:1}{Row 6} of Table \ref{D8Ph}. 
There are no other three particle S-matrices in this case.

The explicit scattering amplitudes generated by each of the Lagrangian structures listed in the 
last column of Table \ref{D8Ph} are explicitly presented in Appendix \ref{pa}.

\LTcapwidth=\textwidth
\begin{longtable}{|c|c|c|}
	\caption{{\bf Photons}, ${\bf D\geq 8}$:~~~{\small The yellow shaded part of the Young's Tableaux denotes contraction of the $SO(D-1)$ tensor with $k_1-k_2$. The Tableaux obtained 
    after omitting all shaded boxes gives the $SO(D-2)$ representation 
    in which the fusion with photons (in later tables gravitons) occurs. This convention is used in all the tables presented in this paper.}}\label{D8Ph}\\
	\hline
	\makecell{\bf Scattering \\ \bf Amplitude} & \textbf{Young Tableaux} & \textbf{Lagrangian Structure} \\
	\hline
	\endfirsthead
	\multicolumn{3}{c}%
	{\tablename\ \thetable\ :  $D\geq 8$ (\textit{Continued from previous page})} \\
	\hline
	\makecell{\bf Scattering \\ \bf Amplitude} & \textbf{Young Tableaux} & \textbf{Lagrangian Structure} \\
	\hline
	\endhead
	\hline \multicolumn{3}{r}{\textit{}} \\
	\endfoot
	\hline
	\endlastfoot
	& & \\
	\hypertarget{D8Ph:s:1}{1} & \ytableausetup{mathmode, boxsize=1.4em}
	\begin{ytableau}
		*(yellow) \mu_1 & *(yellow) \cdots & *(yellow) \mu_r \\
	\end{ytableau} & \makecell{$\nabla_{\mu_1} \cdots \nabla_{\mu_r} F_{\mu\nu} F_{\mu\nu} S_{\mu_1\cdots\mu_r}$ \\ $r$ is even \hspace{2mm} \& \hspace{2mm} $r\geq 0$}\\
	& & \\
	\hypertarget{D8Ph:s:2}{2} & \ytableausetup{mathmode, boxsize=1.4em}
	\begin{ytableau}
		\mu_1 & \mu_2 & *(yellow) \mu_3 & *(yellow) \cdots & *(yellow) \mu_r \\
	\end{ytableau} & \makecell{$\nabla_{\mu_3} \cdots \nabla_{\mu_r} F_{\alpha\mu_1} F_{\alpha\mu_2} S_{\mu_1\mu_2\mu_3\cdots\mu_r}$ \\ $r$ is even \hspace{2mm} \& \hspace{2mm} $r\geq 2$}\\
	& & \\
	\hypertarget{D8Ph:s1:1}{3} & \ytableausetup{mathmode, boxsize=1.4em}
	\begin{ytableau}
		\alpha & *(yellow) \mu_1 & *(yellow) \mu_2 & *(yellow) \cdots & *(yellow) \mu_r \\
		\beta \\
	\end{ytableau} & \makecell{$\nabla_{\mu_2} \cdots \nabla_{\mu_r} \nabla_{\mu_1} F_{\alpha\gamma} F_{\beta\gamma} S_{[\alpha\beta]\mu_1\mu_2\cdots\mu_r}$ \\ $r$ is odd \hspace{2mm} \& \hspace{2mm} $r\geq 1$}\\
	& & \\
	\hypertarget{D8Ph:s1:2}{4} & \ytableausetup{mathmode, boxsize=1.4em}
	\begin{ytableau}
		\alpha & \mu_1 & *(yellow) \mu_2 & *(yellow) \mu_3 & *(yellow) \cdots & *(yellow) \mu_r \\
		*(yellow) \beta \\
	\end{ytableau} & \makecell{$\nabla_{\mu_3} \cdots \nabla_{\mu_r} \nabla_{\mu_2} \nabla_{\beta} F_{\alpha\gamma} F_{\mu_1\gamma} S_{[\alpha\beta]\mu_1\mu_2\mu_3\cdots\mu_r}$ \\ $r$ is even \hspace{2mm} \& \hspace{2mm} $r\geq 2$}\\
	& & \\
	\hypertarget{D8Ph:s2:1}{5} & \ytableausetup{mathmode, boxsize=1.4em}
	\begin{ytableau}
		\alpha & \gamma & *(yellow) \mu_1 & *(yellow) \cdots & *(yellow) \mu_r \\
		*(yellow) \beta & *(yellow) \delta \\
	\end{ytableau} & \makecell{$\nabla_{\mu_1} \cdots \nabla_{\mu_r} F_{\alpha\beta} F_{\gamma\delta} S_{[\alpha\beta][\gamma\delta]\mu_1\cdots\mu_r}$ \\ $r$ is even \hspace{2mm} \& \hspace{2mm} $r\geq 0$}\\
	& & \\
	\hypertarget{D8Ph:s11:1}{6} & \ytableausetup{mathmode, boxsize=1.4em}
	\begin{ytableau}
		\mu_1 & *(yellow) \mu_2 & *(yellow) \cdots & *(yellow) \mu_r \\
		\alpha \\
		*(yellow) \beta \\
	\end{ytableau} & \makecell{$\nabla_{\mu_2} \cdots \nabla_{\mu_r} \nabla_{\beta} F_{\mu_1\delta} F_{\alpha\delta} S_{[\mu_1\alpha\beta]\mu_2\cdots\mu_r}$ \\ $r$ is odd \hspace{2mm} \& \hspace{2mm} $r\geq 1$}\\
	& & \\
\end{longtable}

\subsection{$\mathbf{D=7}$} \label{dese}

In this case $D-1=6$. The only difference between $D =7$ and $D \geq 8$ (see the previous subsection) comes from the fact that an $SO(6)$ Young Tableaux $Y_{(r_1,r_2,r_3)}$ with $r_3 >0$ corresponds to two irreducible representations of $SO(6)$. \footnote{Rather than 
one irreducible representation of $SO(D-1)$ as would have been the case for $D \geq 8$.  } As explained above, the highest weights of the two representations 
 are $(r_1, r_2, r_3)$ and $(r_1, r_2, -r_3)$ respectively.
 
 Now consider the scattering of two photons with the particle $P$ where $P$ transforms
 in the Tableaux $Y_{(r_1,r_2,r_3)}$. If $r_3=0$ then the 
$D  \geq 8$ discussion of the previous subsection carries over unchanged. The photon-photon-$P$ three point functions are once again generated by \hyperlink{D8Ph:s:1}{Row 1} to \hyperlink{D8Ph:s2:1}{Row 5} of Table \ref{D8Ph}). 

It only remains to study the case $r_2=r_3=1$, $r_1=r \geq 1$. In this case the Tableaux $Y_{(r,1,1)}$ corresponds to a sum of two irreducible 
representations of $SO(6)$. We can project onto these two representations in 
turn by demanding that the indices in the first row of the Tableaux are 
self or anti-self dual, i.e. obey the equation 
\begin{equation}\label{addeomfmm}
\varepsilon_{\alpha_1 \alpha_2 \alpha_3 \alpha_4 \alpha_5 
	\alpha_6} S_{[\alpha_4 \alpha_5 \alpha_6] \beta_2 \cdots \beta_{r}}=  \pm i S_{[\alpha_1 \alpha_2 \alpha_3] \beta_2 \cdots \beta_{r}}
\end{equation} 
($\alpha_1$ to $\alpha_3$ are the indices in the first column of the Tableaux, while 
$\beta_m$ is single index corresponding to the $m^{th}$ column of the Tableaux).
For either choice of sign in \eqref{addeomfmm}, the rules listed in Appendix \ref{DR} once again tell us that we have a single 
photon-photon-$P$ coupling when $r$ is odd and no such coupling 
when $r$ is even, corresponding to the contraction structure depicted in 
\hyperlink{D8Ph:s11:1}{Row 6} of Table \ref{D8Ph}.  The  Lagrangians  that describe the scattering of both  the `self dual' and the `anti-self dual' particles (the two 
signs in \eqref{addeomfmm}) both take the form listed in the third column of 
\hyperlink{D8Ph:s11:1}{Row 6} of Table \ref{D8Ph}. In the two cases above, however, the  field $S$ that participates in this coupling is constrained to obey the additional condition \begin{equation}\label{addeom}
\frac{ \partial_{\alpha_4}}{m}\varepsilon_{\alpha_1 \alpha_2 \alpha_3 \alpha_4 \alpha_5 \alpha_6 
	\alpha_7} S_{[\alpha_5 \alpha_6 \alpha_7] \beta_1 \cdots \beta_{r-1}}=  \pm S_{[\alpha_1 \alpha_2 \alpha_3] \beta_1 \cdots \beta_{r-1}}
\end{equation} 
(\eqref{addeom} is simply the spacetime equation that reduces to \eqref{addeomfmm}) 
in the particle's rest frame). 

From the Lagrangian point of view, the coupling between particles in the 
$Y_{(r,1,1)}$ and two photons can be understood from an alternate point of view. 
We could choose to work with a spacetime field $S_{[\alpha_5 \alpha_6 \alpha_7] \beta_1 \cdots \beta_{r-1}}$ that is not further constrained by either of 
the conditions \eqref{addeom}. The quantization of the field $S$ then produces both the self dual and the anti-self dual particle. The most general coupling of 
$S$ to two photons is given by \footnote{ The first term in \eqref{twocouplngs} is simply the Lagrangian listed in 
\hyperlink{D8Ph:s11:1}{Row 6} of Table \ref{D8Ph}. The second term in 
\eqref{twocouplngs} is the same Lagrangian with the replacement
$$S_{[\alpha_1 \alpha_2 \alpha_3] \beta_1 \cdots \beta_{r-1}} \rightarrow \frac{ \partial_{\alpha_4}}{m}\varepsilon_{\alpha_1 \alpha_2 \alpha_3 \alpha_4 \alpha_5 \alpha_6 
	\alpha_7} S_{[\alpha_5 \alpha_6 \alpha_7] \beta_1 \cdots \beta_{r-1}} : 
$$
compare with \eqref{addeom}.}

\begin{equation}\label{twocouplngs}\hypertarget{twocouplngs}{ } 
 A \left( \nabla_{\mu_2} \cdots \nabla_{\mu_r} \nabla_{\beta} F_{\mu_1\delta} F_{\alpha\delta} S_{[\mu_1\alpha\beta]\mu_2\cdots\mu_r}\right) + 
 B \left( \nabla_{\mu_2} \cdots \nabla_{\mu_r} \nabla_{\beta} F_{\mu_1\delta} F_{\alpha\delta} \left(\varepsilon_{\zeta \mu_1 \alpha \beta \theta_1 \theta_2 \theta_3 }
 	\frac{\partial_\zeta}{m} S_{[\theta_1\theta_2\theta_3]\mu_2\cdots\mu_r} \right)\right).
\end{equation} 

In the Lagrangian \eqref{twocouplngs} the coupling of the two photons to the 
self and anti-self dual particles, respectively are proportional to $A+B$ and 
$A-B$. Setting $A= \pm B$ we recover the coupling of the two photons 
to just the self dual or just the anti-self dual particle respectively.

\subsection{D=6} \label{desi}

In this case, the massive particle transforms in a representation of $SO(5)$. There are two differences 
between this case and the $D \geq 8$ case. The first 
difference is simply that Tableaux of $SO(5)$ never have
more than two rows, so there is no analogue of the three point function depicted in  \hyperlink{D8Ph:s11:1}{Row 6} of Table \ref{D8Ph}. 
The second difference arises from the fact that $D-2=4$, and $SO(4)$ 
Young Tableaux of the form $Y_{(r_1, r_2)}$ with $r_2 > 0$  correspond to the 
sum of two - rather than a single - irreducible 
representations of $SO(4)$. 
 It follows that the Tableaux depicted on the RHS of \eqref{phdecompasym}
 denotes two irreps. of $SO(4)$; these are self-dual and anti self-dual antisymmetric two tensor of $SO(4)$, i.e. the representations
with highest weights $(1,1)$ and $(1,-1)$ respectively.  
\footnote{In equations \begin{equation}\label{asymphd6}
	\Lambda^2 (1,0)= (1,1) \oplus (1,-1) .
	\end{equation}
The symmetric decomposition of the two photons, \eqref{phdecompasym} continues 
to work as in higher dimensions. In equations
\begin{equation} \label{symphd6}
S^2 (1,0)= (2,0) \oplus (0,0). 
\end{equation}}

Similar remarks hold for the branching rules from $SO(5)$ to $SO(4)$. 
$SO(4)$ representations with $|r_2| \neq 0$ appear on the RHS of the branching 
rules in pairs; every time the representation $(r_1, r_2)$ appears, the 
representation $(r_1, -r_2)$ also appears.  In other words the RHS of the $SO(5)$ 
branching rules produce $SO(4)$ representations that always group together into 
Tableaux of the form $Y_{(r_1, r_2)}$. 

When $P$ transforms in a Tableaux $Y_{(r,0)}$ or $Y_{(r+2,2)}$ - recall all Tableaux 
now have at most two rows - the $D \geq 8$ analysis carries through unchanged. 
The coupling of $P$ to two photons and corresponding Lagrangian structures 
continue to be listed in \hyperlink{D8Ph:s:1}{Row 1}, \hyperlink{D8Ph:s:2}{Row 2} and \hyperlink{D8Ph:s2:1}{Row 5} of Table \ref{D8Ph}.
When  $P$ transforms in the representation labelled by the Young Tableaux  $Y_{(r+1,1)}$, the three point coupling depicted in \hyperlink{D8Ph:s1:2}{Row 4} of Table \ref{D8Ph} also continues 
to work as for $D \geq 8$. However, the coupling in \hyperlink{D8Ph:s1:1}{Row 3} of Table \ref{D8Ph} splits up into two independent couplings, respectively corresponding to (linear combinations of) fusion of the anti-self-dual and the self 
dual parts of \eqref{phdecompasym} respectively. One of these two couplings 
is simply the one denoted in \hyperlink{D8Ph:s1:1}{Row 3} of Table \ref{D8Ph}. The second coupling uses the 
$SO(4)$ $\varepsilon$ tensor. Recall, however, that the covariant version of this 
structure is proportional to 
\begin{equation}\label{effep} 
k^a_1 k_2^b{\varepsilon}_{a b \mu_1 \mu_2 \mu_3 \mu_4 }
\end{equation} 
and so is odd under the interchange of $k_1$ and $k_2$. It follows that 
the second coupling exists only when $r$ is even (as opposed to the case 
recorded in \hyperlink{D8Ph:s1:1}{Row 3} of Table \ref{D8Ph} which exists only when $r$ is odd). The new contraction and 
Lagrangian structures for this parity odd coupling are presented in \hyperlink{Ph1D6}{Row 1} of Table \ref{D61} below.

\begin{longtable}{|c|c|c|}
	\caption{ {\bf Photons}, ${\bf D=6}$~~  {\small Recall that the Young Tableaux obtained after omitting the shaded boxes denotes the $SO(4)$ representation in which fusion with photons occurs. The representation in this case is the antisymmetric two tensor. The ${\Huge *}$ in the diagram indicates that this two tensor is Hodge dualized before fusing with the photons - or equivalently that this two tensor is fused 
    with the hodge dual of the two photon representation that appears on the RHS of 
\eqref{phdecompasym}. The symbol ${\Huge *}$ has an analogous meaning every time it appears in any table in this paper referring to an even spacetime dimension $D$. Note however 
that we assign the same symbol a slightly different meaning in tables referring to 
odd values of $D$.} }\label{D61}\\
	\hline
	\makecell{\bf Scattering \\ \bf Amplitude} & \textbf{Young Tableaux} & \textbf{Lagrangian Structure} \\
	\hline
	\endfirsthead
	\multicolumn{3}{c}%
	{\tablename\ \thetable\ :  $D=6$ (\textit{Continued from previous page})} \\
	\hline
	\makecell{\bf Scattering \\ \bf Amplitude} & \textbf{Young Tableaux} & \textbf{Lagrangian Structure} \\
	\hline
	\endhead
	\hline \multicolumn{3}{r}{\textit{}} \\
	\endfoot
	\hline
	\endlastfoot
	& & \\
	\hypertarget{Ph1D6}{1} & {\Huge *} \ytableausetup{mathmode, boxsize=1.4em}
	\begin{ytableau}
		e & *(yellow) \mu_1 & *(yellow) \dots & *(yellow) \mu_r\\
		f \\
	\end{ytableau}  & \makecell{$\varepsilon^{abcdef} \nabla_{\mu_1}\cdots\nabla_{\mu_r} F_{ab} F_{cd}  S_{[ef]\mu_1\cdots\mu_r}$  \\ $r$ is even \hspace{2mm} \& \hspace{2mm} $r\geq 0$}\\
	& & \\
\end{longtable}

All other couplings - and corresponding Lagrangians - are 
identical to those in subsection \ref{dgeee}.

\subsection{$\mathbf{D=5}$}\label{desfi}
In this case, the polarization vectors of the photons transform as vectors of $SO(D-2)=SO(3)$. The symmetric product of two 
photons continues to be given by \eqref{phdecompsym} (in the particular case of 
$SO(3)$ \eqref{phdecompsym} is simply the assertion that the symmetric product of 
two $j=1$ representations is the sum of a $j=0$ and $j=2$). The 
antisymmetric product of two photons also continues to be given by 
\eqref{phdecompasym}; in the case of $SO(3)$, however, the two box Tableaux on the RHS of \eqref{phdecompasym} can be dualized to a single box Tableaux, i.e. a $j=1$
representation. We thus 
recover the familiar statement that the antisymmetric product of two $j=1$ SO(3)
representations transforms in the $j=1$. 

The massive particle transforms in an irreducible representation of $SO(D-1)=SO(4)$. 
We will continue to label representations  of $SO(4)$ by the highest 
weights under the Cartans corresponding to rotations in 
orthogonal two planes, $(h_1, h_2)$. Note that $SO(4) \sim SU(2) \times SU(2)$, 
and so representations of $SO(4)$ can also be labelled by $(j_1, j_2)$, the two 
`$j$' values for the two $SU(2)$ factors. The reader who wishes to translate between 
our labelling of $SO(4)$ irreps. and the $SU(2)\times SU(2)$ notation can do so 
using the following dictionary\footnote{ Using the fact that 
$SO(3)$ is simply the diagonal combination of the two 
$SU(2)$ factors, it follows that the branching rule in $SU(2) \times SU(2)$ language is simply 
$$ (j_1, j_2) \rightarrow |j_1-j_2| \oplus \cdots \oplus (j_1+j_2).$$ }
\begin{equation}\label{jhh}
j_1= \frac{h_1+h_2}{2}, ~~~j_2=\frac{h_1-h_2}{2}.
\end{equation} 
The $SO(4)$ representations that can couple to two photons 
are those with $(h_1, h_2) =(r,0)$ or $(r+1, \pm 1)$ or 
$(r+2,\pm 2)$. When $r$ is even, the coupling of the first  of these representations to two photons 
is identical to the analogous coupling for $D \geq 8$ - The Lagrangians 
for even $r$ continue to be given by RHS of \hyperlink{D8Ph:s:1}{Row 1} and \hyperlink{D8Ph:s:2}{Row 2} of Table \ref{D8Ph}. In this special dimension, however, this particular representation has a nonzero coupling to two photons even when $r$ is odd. This parity odd coupling - listed in  \hyperlink{Ph1D5}{Row 1} of Table \ref{D51}; results from a fusion channel in which the  photons couple antisymmetrically  to $SO(3)$ spin 1.

\begin{longtable}{|c|c|c|}
	\caption{{\bf Photons}, ${\bf D=5}$,~~  {\small Here - and also for the analogous table for 
			gravitons in  $D=5$ later in this paper - the symbol ${\Huge *}$ has one of two meanings. Whenever the $SO(4)$ representation is symmetric traceless  - as in Row 1 in this table - this symbol indicates that fusion of the unshaded part of the Young Tableaux occurs 
	with the $SO(3)$ Hodge star of the representation on the RHS of \eqref{phdecompasym} (or analogous representation in the case of gravitons).
    In rows 2 and 3 of this table the same symbol indicates that the original 
$SO(4)$ Tableaux is Hodge stared before we shade in boxes and then fuse with photon 
representations.} }\label{D51}\\
	\hline
	\makecell{\bf Scattering \\ \bf Amplitude} & \textbf{Young Tableaux} & \textbf{Lagrangian Structure} \\
	\hline
	\endfirsthead
	\multicolumn{3}{c}%
	{\tablename\ \thetable\ :  $D=5$ (\textit{Continued from previous page})} \\
	\hline
	\makecell{\bf Scattering \\ \bf Amplitude} & \textbf{Young Tableaux} & \textbf{Lagrangian Structure} \\
	\hline
	\endhead
	\hline \multicolumn{3}{r}{\textit{}} \\
	\endfoot
	\hline
	\endlastfoot
	& & \\
	\hypertarget{Ph1D5}{1} & {\Huge *} \ytableausetup{mathmode, boxsize=1.4em}
	\begin{ytableau}
		\mu_1 & *(yellow) \mu_2 & *(yellow) \dots & *(yellow) \mu_r\\
	\end{ytableau}  & \makecell{$ \varepsilon^{abcd\mu_1} \nabla_{\mu_2}\cdots\nabla_{\mu_r} F_{ab} F_{cd} S_{\mu_1\mu_2\cdots\mu_r} $ \\ $r$ is odd \hspace{2mm} \& \hspace{2mm} $r\geq 1$}\\
	& & \\
	\hypertarget{D8Ph:s1:1*}{2} & {\Huge *} \ytableausetup{mathmode, boxsize=1.4em}
\begin{ytableau}
	\alpha & *(yellow) \mu_1 & *(yellow) \mu_2 & *(yellow) \cdots & *(yellow) \mu_r \\
	\beta \\
\end{ytableau} & \makecell{$\nabla_{\mu_2} \cdots \nabla_{\mu_r} \nabla_{\mu_1} F_{\alpha\gamma} F_{\beta\gamma} \left(\varepsilon^{\theta\alpha_1\beta_1\alpha\beta}\nabla_\theta S_{[\alpha_1\beta_1]\mu_1\mu_2\cdots\mu_r}\right)$ \\ $r$ is odd \hspace{2mm} \& \hspace{2mm} $r\geq 1$}\\
& & \\
\hypertarget{D8Ph:s1:2*}{3} & {\Huge *} \ytableausetup{mathmode, boxsize=1.4em}
\begin{ytableau}
	\alpha & \mu_1 & *(yellow) \mu_2 & *(yellow) \mu_3 & *(yellow) \cdots & *(yellow) \mu_r \\
	*(yellow) \beta \\
\end{ytableau} & \makecell{$\nabla_{\mu_3} \cdots \nabla_{\mu_r} \nabla_{\mu_2} \nabla_{\beta} F_{\alpha\gamma} F_{\mu_1\gamma} \left(\varepsilon^{\theta\alpha_1\beta_1\alpha\beta}\nabla_\theta S_{[\alpha_1\beta_1]\mu_1\mu_2\mu_3\cdots\mu_r}\right)$ \\ $r$ is even \hspace{2mm} \& \hspace{2mm} $r\geq 2$}\\
& & \\
	\hypertarget{D8Ph:s2:1*}{4} & {\Huge *} \ytableausetup{mathmode, boxsize=1.4em}
\begin{ytableau}
	\alpha & \gamma & *(yellow) \mu_1 & *(yellow) \cdots & *(yellow) \mu_r \\
	*(yellow) \beta & *(yellow) \delta \\
\end{ytableau} & \makecell{$\nabla_{\mu_1} \cdots \nabla_{\mu_r} F_{\alpha\beta} F_{\gamma\delta} \left(\varepsilon^{\theta\alpha_1\beta_1\alpha\beta}\nabla_\theta S_{[\alpha_1\beta_1][\gamma\delta]\mu_1\cdots\mu_r}\right)$ \\ $r$ is even \hspace{2mm} \& \hspace{2mm} $r\geq 0$}\\
& & \\
\end{longtable}
We now turn to the case in which $P$ transforms in the $(h_1, h_2)=(r+1,1)$ and $(r+1,-1)$ representations. Before studying the couplings of these fields to 
two photons, it is useful to first examine the free equations of motion of the field that create these particles. Let the field in question be denoted by 
$S_{[\alpha\beta]\mu_1\mu_2\cdots\mu_r}$. As usual $S$ is assumed to have 
the symmetry properties associated with the Young Tableaux in \hyperlink{D8Ph:s1:1}{Row 3} of Table \ref{D8Ph}. 
As usual, $S$ obeys the 5 dimensional spacetime equation of motion 
$$ \left( \nabla^2 -m^2 \right) S_{[\alpha\beta]\mu_1\mu_2\cdots\mu_r}=0.$$
The quantization of a tensor field with these symmetry properties, and subject this equation of motion, produces both the $(r+1,1)$ and the $(r+1,-1)$ representations. In order, to focus 
on just the $(r+1,1)$ representation (or just the $(r+1,-1)$ representation) we would need to impose an additional spacetime 
equation of motion on the field $S_{[\alpha\beta]\mu_1\mu_2\cdots\mu_r}$ that 
constrains it to be completely self-dual or completely anti-self-dual. The self duality condition in the little group $SO(4)$ is
\begin{equation}\label{selfdso4} 
\varepsilon_{ijkl} S_{[kl]m_1m_2\cdots m_r}= \pm 
 S_{[ij]m_1m_2\cdots m_r}.
\end{equation} 
The covariant 5 dimensional version of \eqref{selfdso4} is 
\begin{equation}\label{selfdso4n} 
i\varepsilon_{\alpha \beta \theta \gamma \delta }  \frac{\partial_\theta}{m} \left(  S_{[\gamma\delta]\mu_1\mu_2\cdots\mu_r} \right)  = \pm 
S_{[\alpha\beta]\mu_1\mu_2\cdots\mu_r}.
\end{equation} 
Note that in contrast with \eqref{addeom}, \eqref{selfdso4n} has a factor of $i$ on 
the LHS. \footnote{The presence or absence of this factor of $i$ is determined by 
	the following considerations. One can re-insert the 
	equation \eqref{selfdso4} into the LHS of the same equation. The resultant 
	equation takes the schematic form 
	$$ {\varepsilon} {\varepsilon} S= S.$$
	Taking care of the index contractions, it is easy to check that in this dimension 
	${\varepsilon} {\varepsilon}={\cal I}$.
	In other words the equation we get by iterating \eqref{selfdso4} twice is identically true; there are no obstructions to finding solutions for this equation. If we carry through the same procedure for $SO(2m)$ we find 
	 $${\varepsilon} {\varepsilon}=(-1)^{m}{\cal I}.$$
	 It follows that an equation of the form \eqref{selfdso4} has solutions 
	 only when $m$ is even. When $m$ is odd, on the other hand, the analogous 
	 equation must have an extra factor of $i$ on the RHS in order to admit 
	 solutions, explaining, for instance, the factor of $i$ in  \eqref{addeomfmm}. }

	 It follows that \eqref{selfdso4n} has no solutions if $S$ is a real 
field. In other words it is impossible to impose any equation on the real field 
$S$ that restricts its particle content to just the $(r+1, 1)$ excluding the 
$(r+1, -1)$ or vice versa. A real field $S$ always contains both of these 
representations together. For this reason the physically useful counting question 
is the following: how many 3 point couplings are there between two photons and 
either a $(r+1,1)$ or a $(r+1,-1)$?

At the level of counting, this question is easy to answer. It is useful to 
separately consider the case $r$ odd and $r$ even. Let us first focus on 
the case $r$ odd.  There is a single Bose symmetric $j=1$ $SO(3)$ representation in the branching rules of {\it each} of  the representations $(r+1,1)$ and $(r+1, -1)$. One of these three 
point structures, and its corresponding Lagrangian,  is listed in \hyperlink{D8Ph:s1:1}{Row 3} of Table \ref{D8Ph}  \footnote{ Recall that $j=1$ is the same as a Young Tableaux with a single column and two boxes in the case of $SO(3)$.}
 The second structure predicted by our counting is not difficult to work out. This structure - and the corresponding 
Lagrangian - are listed in \hyperlink{D8Ph:s1:1*}{Row 2} of Table \ref{D51}.  Now let us turn to the case $r$ even. 
From a counting point of view we once again have a single coupling corresponding 
to the branching of $(r+1,1)$ to $j=2$ and a second coupling corresponding to the 
descent of $(r+1,-1)$ to $j=2$.  One of these two structures was already listed 
in \hyperlink{D8Ph:s1:2}{Row 4} of Table \ref{D8Ph}. The second structure (and the Lagrangian that generates it) 
is new to this dimension. It is parity odd and is listed in \hyperlink{D8Ph:s1:2*}{Row 3} of Table \ref{D51}.

When $P$ transforms in the $(h_1, h_2)=(r+2,2)$ and $(r+2,-2)$ representations, 
the situation is very similar to the case dealt with in the previous paragraph 
(i.e. the $(r+1, 1)$ and $(r+1,-1)$ representations). As in the case above, the 
`equation of motion' that would restricts to one of the two representations 
$(r+2,2)$ or $(r+2,-2)$ is complex and has no real solutions. A real field 
$S$ produces particles transforming in both representations at the same time. As in the paragraph above 
we have an effective doubling of the $D \geq 8$ couplings corresponding to 
this symmetry. Concretely, we have two three point couplings for every even value 
of $r$ and no couplings when $r$ is odd. Each of these couplings correspond to branching of the representation to $j=2$. One of the three point couplings 
is that presented in \hyperlink{D8Ph:s2:1}{Row 5} of Table \ref{D8Ph} and the 
second (parity odd) three point coupling is listed in \hyperlink{D8Ph:s2:1*}{Row 4} of Table \ref{D51}

\subsection{$\mathbf{D=4}$}\label{desfo}

In this case the photon polarizations transform in the vector representation of $SO(D-2)=SO(2)$. This is a two dimensional (reducible) representation with 
$SO(2)$ charges $\pm 1$. The symmetric product of two photons transforms with $SO(2)$ charges $+2$, $0$, $-2$ and the antisymmetric product is a single 
spin zero state. This antisymmetric spin zero state is the Hodge dual of the RHS of \eqref{phdecompasym}; recall that the $SO(2)$ Hodge dual of an antisymmetric two 
tensor is a scalar. The covariant version of this dualization uses the two dimensional $\varepsilon$ tensor proportional to  $ k_1^\mu k_2^\nu \epsilon_{\mu\nu \alpha \beta}$ which itself picks up a minus sign under 
$ 1 \leftrightarrow 2$. It follows that this antisymmetric spin zero 
combination of photons is Bose symmetric rather than antisymmetric. 

 The most general non-spinorial field $S$, associated with the massive particle $P$, is labelled by its $SO(3) \sim SU(2)$ 
label $j$ (completely symmetric tensor with $j$ indices). To start with, for every even $j \geq 0$ there exists two photon-photon-$P$ coupling corresponding to the symmetric and anti-symmetric fusion of the photons to spin zero (all the free indices of the field associated with the 
particle $P$ are dotted with $k_1-k_2$.). The Lagrangian corresponding to the symmetric fusion continue to be given by \hyperlink{D8Ph:s:1}{Row 1} of Table \ref{D8Ph} while the anti symmetric fusion results in \hyperlink{Ph1D4}{Row 1} of Table \ref{PhD4}. 

In the case that $j \geq 2$ and $j$ continues to be even we have one additional three point structure corresponding to the branching to the parity even combination of the two spins with $\pm 2$ (in this case all but two of the 
indices of $P$ are contracted with $k_1-k_2$), as in $D \geq 8$. The Lagrangian structure continues to be given by \hyperlink{D8Ph:s:2}{Row 2} of Table \ref{D8Ph}. 

Finally, when $j$ is odd and $j \geq 3$ there exists a  single parity odd coupling 
corresponding to the branching of $P$ to the parity odd combination of 
spins $ \pm 2$. In equations this spin 2 combination of polarizations is given by 
\begin{equation}\label{struct} 
k_1^\mu k_2^\nu \epsilon_{\mu\nu \alpha \theta} 
\left( \epsilon_1^\theta \epsilon_2^\beta + \epsilon_2^\theta \epsilon_1^\beta
\right) .
\end{equation} 
Note that this combination is symmetric under the interchange $\epsilon_1 \leftrightarrow
\epsilon_2$ but antisymmetric under $k_1 \leftrightarrow k_2$, so overall picks up 
a minus sign under $ 1 \leftrightarrow 2$. 

In this case all but two indices of $P$ are contracted with $k_1-k_2$; Bose statistics is satisfied because the phase $-1$ obtained from the $k_1-k_2$ interchange is compensated for by the antisymmetry under interchange of the photon polarization structure \eqref{struct}. The corresponding Lagrangian is listed in \hyperlink{Ph2D4}{Row 2} of Table \ref{PhD4}


\begin{longtable}{|c|c|c|}
	\caption{{\bf Photons}, ${ \bf D=4}$,~~  {\small The meaning of the symbol ${\Huge *}$  was explained in the caption to Table \ref{D61}.}}\label{PhD4}\\
	\hline
	\makecell{\bf Scattering \\ \bf Amplitude} & \textbf{Young Tableaux} & \textbf{Lagrangian Structure} \\
	\hline
	\endfirsthead
	\multicolumn{3}{c}%
	{\tablename\ \thetable\ :  $D=4$ (\textit{Continued from previous page})} \\
	\hline
	\makecell{\bf Scattering \\ \bf Amplitude} & \textbf{Young Tableaux} & \textbf{Lagrangian Structure} \\
	\hline
	\endhead
	\hline \multicolumn{3}{r}{\textit{}} \\
	\endfoot
	\hline
	\endlastfoot
	& & \\
	\hypertarget{Ph1D4}{1} & {\Huge *} \ytableausetup{mathmode, boxsize=1.4em}
	\begin{ytableau}
		*(yellow) \mu_1 & *(yellow) \dots & *(yellow) \mu_r\\
	\end{ytableau}  & \makecell{$ \varepsilon^{abcd} \nabla_{\mu_1}\cdots\nabla_{\mu_r} F_{ab} F_{cd} S_{\mu_1\cdots\mu_r} $ \\ $r$ is even \hspace{2mm} \& \hspace{2mm} $r\geq 0$}\\
	& & \\
	\hypertarget{Ph2D4}{2} & {\Huge *} \ytableausetup{mathmode, boxsize=1.4em}
	\begin{ytableau}
		\mu_1 & \mu_2 & *(yellow) \mu_3 & *(yellow) \dots & *(yellow) \mu_r\\
	\end{ytableau}  & \makecell{ $ \varepsilon^{abc\mu_1} \nabla_{\mu_4}\cdots\nabla_{\mu_r}\nabla_\delta \nabla_{\mu_3} F_{ab} F_{\mu_2 \delta} \nabla_cS_{\mu_1\mu_2\mu_3\cdots\mu_r} $\\ $r$ is odd \hspace{2mm} \& \hspace{2mm} $r\geq 3$}\\
	& & \\
\end{longtable}

\subsection{$\mathbf{D=3}$}\label{desth}

In this case $SO(D-2)=SO(1)$ and so the $SO(D-2)$ singlet condition is empty. 
 Photon polarizations are just numbers; photons carry one scalar degree of freedom.  The antisymmetric product of two photon polarizations is empty, while the symmetric product of these polarizations is one dimensional. 
 
 The most general massive particle can be thought of as a two dimensional  traceless symmetric tensor with $r$ indices. All such two dimensional tensors have 
 only two nonzero components given by $C_{++ \cdots +}$ or $C_{--\cdots -}$  
 (tensor elements with some plus and some minus index components are not
 traceless - this follows from the fact that the metric in 2 dimensions has no 
 $++$ and no $--$ components but has $+-$ and $-+$ components). Self-dual (or anti-self dual) rank 
 $r$ traceless symmetric tensors - i.e. traceless symmetric tensors that obey 
 \begin{equation}\label{ppop}
 i \varepsilon_\mu^\nu C_{\mu \alpha_2 \cdots \alpha_r}=\pm  C_{\mu \alpha_2 \cdots \alpha_r}
 \end{equation} 
  have only $+$ (or only $-$) indices.
  The three dimensional 
 equation of motion that forces a symmetric traceless tensor to be self-dual or
 anti-self dual is given by  
\begin{equation}\label{3dew}
\varepsilon_{\theta_1 \theta_2 \theta_3} \partial_{\theta_2} C_{\theta_3 \alpha_1 
	\cdots \alpha_r} = \pm C_{\theta_1 \alpha_1 
	\cdots \alpha_r}.
\end{equation} 
Note that as in $D=7$ (but unlike $D=5$) this equation is real and so can be
meaningfully imposed on real fields $S$.

There is one nonzero three point photon photon self-dual tensor coupling, and also a corresponding photon photon anti self-dual tensor coupling for every even $r$. The corresponding contraction structure and Lagrangian is  given in
\hyperlink{D8Ph:s:1}{Row 1} of Table \ref{D8Ph} - once we remember that the field $S$ is constrained to obey 
the self duality (or anti self duality) condition \eqref{3dew}. The coupling between two photons and tensors with odd $r$ vanishes. 

As in $D=7$ we could choose to adopt an alternate point of view. We could work with a symmetric tensor field $S$ without imposing the self duality condition 
\eqref{3dew}. In this case the field $S$ corresponds to a particle of $SO(2)$ spin 
$r$ plus a particle of $SO(2)$ spin $-r$ (i.e. to both the self dual and the anti-self dual particle). In this case in addition to the coupling listed in \hyperlink{D8Ph:s:1}{Row 1} of Table \ref{D8Ph} 
there is an additional coupling listed in  \hyperlink{1D3}{Row 1} of Table \ref{PhD3}. The most general coupling of 
$S$ to two photons is given by $A$ times the coupling of \hyperlink{D8Ph:s:1}{Row 1} of Table \ref{D8Ph} plus 
$B$ times the coupling of \hyperlink{1D3}{Row 1} of Table \ref{PhD3}. Effectively, the two photons couple to the 
self dual particle (i.e. particle of spin $r$) with a coupling proportional to 
$A+B$ while they couple to the anti self dual particle  (i.e. the particle of spin 
-$r$) with a coupling proportional to $A-B$. We obtain the coupling only to 
the self or anti-self dual particle by setting $B= \pm A$.

\begin{longtable}{|c|c|c|}
	\caption{{\bf Photons}, ${\bf D=3}$,~~  {\small The symbol ${\Huge *}$ here signifies that one of the indices of the  $SO(2)$ symmetric tensor denoted in the second column of the table is Hodge dualized (i.e. RHS of \eqref{ppop} is replaced by LHS of \eqref{ppop})before we shade boxes and contract with the two photons.}}\label{PhD3}\\
	\hline
	\makecell{\bf Scattering \\ \bf Amplitude} & \textbf{Young Tableaux} & \textbf{Lagrangian Structure} \\
	\hline
	\endfirsthead
	\multicolumn{3}{c}%
	{\tablename\ \thetable\ :  $D=3$ (\textit{Continued from previous page})} \\
	\hline
	\makecell{\bf Scattering \\ \bf Amplitude} & \textbf{Young Tableaux} & \textbf{Lagrangian Structure} \\
	\hline
	\endhead
	\hline \multicolumn{3}{r}{\textit{}} \\
	\endfoot
	\hline
	\endlastfoot
	& & \\
\hypertarget{1D3}{1} & {\Huge *} \ytableausetup{mathmode, boxsize=1.4em}
\begin{ytableau}
	*(yellow) \mu_1 & *(yellow) \dots & *(yellow) \mu_r\\
\end{ytableau}  & \makecell{$  \nabla_{\mu_1}\cdots\nabla_{\mu_r} F_{ab} F_{ab} \left(\varepsilon_{\mu_1 \theta_2 \theta_3} \partial_{\theta_2} S_{\theta_3 \mu_2 
	\cdots \mu_r} \right) $ \\ $r$ is even \hspace{2mm} \& \hspace{2mm} $r\geq 2$}\\
& & \\
\end{longtable}

\section{Gravitons} 

Each graviton polarization is a symmetric traceless tensor of  $SO(D-2)$. For $D \geq 7$ the symmetric product of two gravitational polarizations transforms as 
\begin{eqnarray}\label{grdecompsym}
\Yvcentermath1 S^2\left(\yng(2)\otimes \yng(2)\right)&=&\Yvcentermath1 \yng(4) \oplus \yng(2) \oplus \yng(2,2) \oplus \bullet~.
\end{eqnarray}
The antisymmetric product of two graviton polarizations 
transforms as 

\begin{eqnarray}\label{grdecompasym}
\Yvcentermath1 \Lambda^2\left(\yng(2)\otimes \yng(2)\right)&=&\Yvcentermath1 \yng(3,1) \oplus \yng(1,1) ~.
\end{eqnarray}


As in the case of photons we count graviton-graviton-$P$ 
three point structures by using branching rules to decompose 
the $SO(D-1)$ representation $P$ into $SO(D-2)$ representations and enumerate those fusions with the two graviton representation content (\eqref{grdecompasym}
and \eqref{grdecompsym}) that are consistent with Bose symmetry. 

It follows immediately from the branching rules of Appendix \ref{DR} that there are no three point structures  if $P$ transforms in a Tableaux $Y_{(a,b,c,d,\cdots)}$ with $d>0$. Even when $d=0$,  no three point structures exist if $c>2 $ or if $b+c>4$. It follows that the  most general representations that can couple with two gravitons have Young Tableaux of one of the following structures :
\begin{itemize} 
	\item  $Y_{(r+m,m,0,\cdots,0,0)}$ with $0\leq m \leq 4$ and $r \geq 0$ (see LHS of \hyperlink{S00e1}{Row 1}-\hyperlink{S10e2}{Row 7},  \hyperlink{S20e1}{Row 10}-\hyperlink{S20o1}{Row 13}, \hyperlink{S30o1}{Row 17},  \hyperlink{S30e1}{Row 18}, and \hyperlink{S40e1}{Row 20} of Table \ref{D8gr} ).
	\item $Y_{(r+m,m+1,1,0,\cdots, 0)}$ where $0\leq m \leq 2$ and $r \geq 1$  (see LHS of \hyperlink{S11o1}{Row 8}, \hyperlink{S11o2}{Row 9}, \hyperlink{S21o1}{Row 14},  \hyperlink{S21e1}{Row 15} and \hyperlink{S31o1}{Row 19}   of Table \ref{D8gr})
	\item  $Y_{(r,2,2,0,\cdots, 0)}$ where $r \geq 2$ (see LHS of \hyperlink{S22e1}{Row 16} of Table \ref{D8gr}).
	\end{itemize} 
As in the case of photons we proceed to enumerate the 3 point structures dimension 
by dimension.

\subsection{ $\mathbf{D \geq 8}$} \label{dgeeeg}
\subsection*{$P$ in $Y_{(r, 0\cdots,0)}$ ~ : ~$r \geq 0$}
As in the case of photons, first consider $P$ that transforms under $SO(D-1)$ in the $Y_{(r,0\cdots,0)}$ Young Tableaux where $r\geq 0$. In this case the graviton-graviton-$P$ three point function is nonzero only if $r$ is even. If $r=0$ there exists a single coupling corresponding to the fusion of two gravitons into the $SO(D-2)$ singlet (i.e. the fusion to $Y_{(0,\cdots, 0)}$ which we denote by the symbol $\bullet$ on the RHS of \eqref{grdecompsym}). The corresponding Lagrangian is listed in  \hyperlink{S00e1}{Row 1} of Table \ref{D8gr}. 

When $r=2$ there exist two couplings corresponding to the fusion of two gravitons into the $SO(D-2)$ singlet and symmetric two tensor ($Y_{(0,\cdots, 0)}$ and $Y_{(2,0\cdots, 0)}$ on RHS of \eqref{grdecompsym}). The corresponding Lagrangians are listed in \hyperlink{S00e1}{Row 1} and \hyperlink{S00e2}{Row 2} of Table \ref{D8gr}. 

For every even $r \geq 4$, there are three couplings corresponding to the fusion of gravitons into the $SO(D-2)$ representations $Y_{(0,\cdots,0)}$, $Y_{(2,0,\cdots,0)}$ and $Y_{(4,0,\cdots,0)}$ (See RHS of \eqref{grdecompsym}). The corresponding Lagrangians are listed in \hyperlink{S00e1}{Row 1}, \hyperlink{S00e2}{Row 2} and \hyperlink{S00e3}{Row 3} of Table \ref{D8gr}. 

Bose symmetry ensures that there  are no graviton-graviton-$P$ couplings when $r$ is odd. An $SO(D-1)$ symmetric spin $r$ representation turns into an $SO(D-2)$ spin $s$ 
representation once $r-s$ of the tensor indices are dotted with $(k_1-k_2)$ (and the remaining $s$ indices are projected orthogonal to $k_1-k_2$). It follows that the corresponding couplings are Bose symmetric only when $r$ is even. 

\subsection*{$P$ in $Y_{(r+m, m, 0,\cdots,0)}$ ~: $1 \leq m \leq 4$, $r \geq 0$}

Now consider particles $P$ that transform under $SO(D-1)$ in a Tableaux $Y_{(r+m,m,0,\cdots, 0)}$ 
 with $1 \leq m \leq 4$. When $m=4$, contracting $r$ boxes in the first row and all 4 boxes in the second row with $k_1-k_2$ (see \hyperlink{S40e1}{Row 20} of Table \ref{D8gr})  yields the $SO(D-2)$ representation that transforms in  $Y_{(4,0, \cdots 0)}$. It follows there exists a single graviton-graviton-$P$ three point function corresponding to the fusion of the two gravitons 
 into the $Y_{(4,0, \cdots 0)}$ that appears on the RHS of \eqref{grdecompsym}. 
 This structure has $r+4$ factors of $k_1-k_2$ and so is Bose symmetric  when $r$ is any even 
 integer.  The $SO(D-1) \rightarrow SO(D-2)$ descent allow for no further three point S-matrices in this case. 
 In particular, Bose statistics forces the graviton-graviton-$P$ coupling to vanish 
 when $r$ is odd. The corresponding Lagrangian is listed in \hyperlink{S40e1}{Row 20} of Table \ref{D8gr}.
  
   When $m=3$, contracting $r-1$ boxes in the first row and all 3 boxes in the second row (\hyperlink{S30e1}{Row 18} of Table \ref{D8gr}) once again gives the $SO(D-2)$ representation that transforms in $Y_{(4,0, \cdots 0)}$ yielding one Bose symmetric graviton-graviton-$P$ three point function when $r$ is even and $r \geq 2$. On the other hand contracting $r$ boxes in the first row and 2 boxes in the second row yields the $SO(D-2)$ representation $Y_{(3,1,0 \cdots 0)}$ that 
   appears on the RHS of \eqref{grdecompasym} . This contraction structure, 
   depicted in \hyperlink{S30o1}{Row 17} of Table \ref{D8gr},  yields a single Bose symmetric graviton-graviton-$P$ three point function for every odd $r$. It follows in summary that when $P$ transforms in this representation we have a single 3 point structure for every $r$, even or odd, as long as $r \geq 1$.

      When $m=2$ we continue to have a single Bose symmetric structure for every odd $r$ (see \hyperlink{S20o1}{Row 13} of Table \ref{D8gr}), corresponding to contracting $r-1$ boxes in the first row and 1 box in the second row yielding the $SO(D-2)$ Tableaux $Y_{(3,1, 0, \cdots,0)}$ that appears on the RHS 
      of \eqref{grdecompasym}. For even $r$, on the other hand, we have two Bose symmetric structures when $r=0$ and three structures when $r \geq 2$. 
     The two structures that exist for all even $r \geq  0$  correspond to contracting only $r$ boxes in the top row yielding the $SO(D-2)$ Tableaux $Y_{(2,2,0,\cdots,0)}$ that appears on the RHS of \eqref{grdecompsym} (see \hyperlink{S20e1}{Row 10} of  Table \ref{D8gr})
      or contracting $r$ boxes in the top row and both boxes in the bottom row yielding the $SO(D-2)$ Tableaux $Y_{(2,0,0,\cdots,0)}$ (see \hyperlink{S20e2}{Row 11} of Table \ref{D8gr}). The structure that exists only for $r \geq 2$ corresponds to contracting $r-2$ boxes in the top row and both boxes in the second row yielding  the $SO(D-2)$ Tableaux $Y_{(4,0,0,\cdots,0)}$
      that appears on the RHS of \eqref{grdecompsym} (see \hyperlink{S20e3}{Row 12} of Table \ref{D8gr}).  In summary when $m=2$, we have two structures for $r=0$ (\hyperlink{S20e1}{Row 10}, \hyperlink{S20e2}{Row 11} of Table \ref{D8gr}), three structures for every even $r$ $ \geq 2$ (\hyperlink{S20e1}{Row 10}, \hyperlink{S20e2}{Row 11} and \hyperlink{S20e3}{Row 12} of Table \ref{D8gr}) and one structure for every odd $r \geq 1$ (\hyperlink{S20o1}{Row 13} of Table \ref{D8gr}).
    
     When $m=1$, we have one structure when $r=1$, listed in \hyperlink{S10o1}{Row 4} of Table \ref{D8gr} (corresponding to contracting one index in the top row to yield the $SO(D-2)$ Tableaux $Y_{(1,1,0, \cdots, 0)}$); and two structures when $r$ is odd and $r \geq 3$, listed in \hyperlink{S10o1}{Row 4} and \hyperlink{S10o2}{Row 5} of Table \ref{D8gr} (contracting $r$ boxes in the first row yields the $SO(D-2)$ Tableaux $Y_{(1,1,0, \cdots, 0)}$ while contracting $r-2$ boxes in the first row yields the $SO(D-2)$ Tableaux $Y_{(3,1,0, \cdots ,0)}$). Turning to even values of $r$, there are no Bose symmetric structures when $r=0$. There exists one structure when $r=2$, listed in \hyperlink{S10e1}{Row 6} of Table \ref{D8gr} (corresponding to contracting one index in the top row and one index in the second row yielding the $SO(D-2)$ Tableaux $Y_{(2,0, \cdots, 0)}$); and two such structures when $r \geq 4$, listed in \hyperlink{S10e1}{Row 6}  and \hyperlink{S10e2}{Row 7} of Table \ref{D8gr} (corresponding to contracting $r-1$ indices in the top row and the single box in the second row to get the $SO(D-2)$ Tableaux $Y_{(2,0,\cdots,0)}$ or contracting $r-3$ boxes in the first row and the single box in the second row to get the $SO(D-2)$ Tableaux $Y_{(4,0,\cdots,0)}$). In summary for $m=1$ there are no structures for $r=0$, one structure for $r=2$ (\hyperlink{S10e1}{Row 6} of Table \ref{D8gr})  and two structures for even $r \geq 4$ (\hyperlink{S10e1}{Row 6}  and \hyperlink{S10e2}{Row 7} of Table \ref{D8gr}). When $r$ is odd, we have one structure for $r=1$ ( \hyperlink{S10o1}{Row 4} of Table \ref{D8gr}) and two structures for $r \geq 3$ (\hyperlink{S10o1}{Row 4} and \hyperlink{S10o2}{Row 5} of Table \ref{D8gr}) . 

\subsection*{$P$ in $Y_{(r+m, m+1, 1, 0,\cdots,0)}$~: ~$0 \leq m \leq 2$, $r \geq 1$}

Now, consider the massive particle $P$ transforming in the $SO(D-1)$ representation labelled by the Young Tableaux $Y_{(r+m, m+1, 1,0, \cdots,0)}$, 
 where $0 \leq m \leq 2$  
(note $r \geq 1$). When $m=2$, no Bose symmetric structure exists when $r$ is even. When $r$ is odd on the other hand, there is always one Bose symmetric structure, obtained by contracting the single box in the third row, two boxes in the second row and all $r-1$ boxes in the top row to obtain the $SO(D-2)$ representation $Y_{(3,1,0,\cdots ,0)}$ (see \hyperlink{S31o1}{Row 19} of Table \ref{D8gr}).

 When $m=1$ and $r$ is even,  there is a single structure listed in \hyperlink{S21e1}{Row 15} of Table \ref{D8gr} (obtained by contracting the single box in the 3rd row and $r-1$ boxes in the first row to obtain the $SO(D-2)$ Tableaux $Y_{(2,2, 0,\cdots,0)}$). Turning to odd $r$, there is no structure when $r=1$ and one structure for $r \geq 3$,  listed in \hyperlink{S21o1}{Row 14} of Table \ref{D8gr} (corresponding to contracting the single box in the third row, one box in the second row and $r-2$ boxes in the top row to yield the $SO(D-2)$ Tableaux $Y_{(3,1,0,\cdots,0)}$). In summary, for $m=1$, we have one structure for every  $r \geq 2$.
 
  Finally, when $m=0$, there are no Bose symmetric structures when $r$ is even. Turning to odd $r$, for all  $r\geq 1$ there is a one such structure corresponding to contracting the single box in the third row  and $r-1$ boxes in the first row to yield the $SO(D-2)$ Tableaux $Y_{(1,1, 0,\cdots,0)}$ (see \hyperlink{S11o1}{Row 8} of Table \ref{D8gr}). For odd $r \geq 3$ there is one additional structure corresponding to contracting the single box in the third row and $r-3$ boxes in the first row to obtain the $SO(D-2)$ Tableaux $Y_{(3,1, 0, \cdots, 0)}$ (see \hyperlink{S11o2}{Row 9}  of Table \ref{D8gr}). In summary in this case there is one structure when $r =1$, two structures for odd $r \geq 3$ and no structures for even $r$.

\subsection*{$P$ in $Y_{(r, 2, 2, 0,\cdots,0)}$ ~:~ $r \geq 2$}
Finally consider Tableaux $Y_{(r, 2, 2, 0,\cdots,0)}$ 
 When $r$ is even there is a single Bose symmetric structure obtained by contracting $r-2$ boxes in the first row and both boxes in the third row to obtain $Y_{(2,2,0, \cdots, 0)}$ (see \hyperlink{S22e1}{Row 16} of Table \ref{D8gr}). When $r$ is odd on the other hand no three point structures exist. 

 As in the case of photons, it is also not difficult to find explicit expressions for the S-matrices generated by each of these Lagrangian structures. We present these explicit expressions in Appendix \ref{D8Gr:Amp}.


\begin{longtable}{|c|c|c|}
	\caption{{\bf Gravitons}, ${\bf D\geq 8}$~~ {\small The colouring principle of the Young's Tableaux, in the second column of this table, has been outlined in the caption to Table \ref{D8Ph}}}\label{D8gr}\\
	\hline
	\makecell{\bf Scattering \\ \bf Amplitude} & \textbf{Young Tableaux} & \textbf{Lagrangian Structure} \\
	\hline
	\endfirsthead
	\multicolumn{3}{c}%
	{\tablename\ \thetable\ :  $D\geq 8$ (\textit{Continued from previous page})} \\
	\hline
	\makecell{\bf Scattering \\ \bf Amplitude} & \textbf{Young Tableaux} & \textbf{Lagrangian Structure} \\
	\hline
	\endhead
	\hline \multicolumn{3}{r}{\textit{Continued on next page}} \\
	\endfoot
	\hline
	\endlastfoot
	& & \\
	\hypertarget{S00e1}{1} & \ytableausetup{mathmode, boxsize=1.4em}
	\begin{ytableau}
		*(yellow) \mu_1 & *(yellow) \cdots   & *(yellow) \mu_{r}
	\end{ytableau} & \makecell{$\nabla_{\mu_1}\cdots\nabla_{\mu_r} R_{\alpha\beta\gamma\delta} R_{\alpha\beta\gamma\delta} S_{\mu_1\cdots\mu_{r}}$  \\ $r$ is even \hspace{2mm} \& \hspace{2mm} $r\geq 0$}\\
	& & \\
	\hypertarget{S00e2}{2} & \ytableausetup{mathmode, boxsize=1.4em}
	\begin{ytableau}
		\mu_1  & \mu_2 & *(yellow) \mu_3 & *(yellow) \cdots   & *(yellow) \mu_{r}
	\end{ytableau} & \makecell{$\nabla_{\mu_3}\cdots\nabla_{\mu_r} R_{\mu_1\alpha\beta\gamma} R_{\mu_2\alpha\beta\gamma} S_{\mu_1\mu_2\mu_3\cdots\mu_{r}}$  \\ $r$ is even \hspace{2mm} \& \hspace{2mm} $r\geq 2$}\\
	& & \\
	\hypertarget{S00e3}{3} & \ytableausetup{mathmode, boxsize=1.4em}
	\begin{ytableau}
		\mu_1  & \mu_2 & \mu_3 &  \mu_4 & *(yellow)  \mu_5 & *(yellow) \cdots   & *(yellow) \mu_{r}
	\end{ytableau} & \makecell{$\nabla_{\mu_5}\cdots\nabla_{\mu_r} R_{\mu_1\alpha\mu_2\beta} R_{\mu_3\alpha\mu_4\beta} S_{\mu_1\mu_2\mu_3\mu_4\mu_5 \cdots \mu_{r}}$  \\ $r$ is even \hspace{2mm} \& \hspace{2mm} $r\geq 4$}\\
	& & \\
	\hypertarget{S10o1}{4} & \ytableausetup{mathmode, boxsize=1.4em}
	\begin{ytableau}
		a & *(yellow) \mu_1  & *(yellow)  \mu_2 & *(yellow) \cdots   & *(yellow) \mu_{r} \\
		c
	\end{ytableau} & \makecell{$\nabla_{\mu_2}\cdots\nabla_{\mu_r}\nabla_d R_{acef} R_{\mu_1 def} S_{[ac] \mu_1 \cdots \mu_{r}}$  \\ $r$ is odd \hspace{2mm} \& \hspace{2mm} $r\geq 1$}\\
	& & \\
	\hypertarget{S10o2}{5} & \ytableausetup{mathmode, boxsize=1.4em}
	\begin{ytableau}
		a &  \mu_1  & \mu_2 & *(yellow) \mu_3  & *(yellow)  \mu_4 & *(yellow) \cdots   & *(yellow) \mu_{r} \\
		e
	\end{ytableau} & \makecell{$\nabla_{\mu_4}\cdots\nabla_{\mu_r}\nabla_h R_{ae\mu_3 i} R_{h\mu_1 \mu_2 i} S_{[ae]\mu_1 \mu_2 \mu_3 \mu_4 \cdots \mu_r}$  \\ $r$ is odd \hspace{2mm} \& \hspace{2mm} $r\geq 3$} \\
	& & \\
	\hypertarget{S10e1}{6} & \ytableausetup{mathmode, boxsize=1.4em}
	\begin{ytableau}
		a & \mu_1 & *(yellow) \mu_2 & *(yellow) \mu_3 & *(yellow) \cdots & *(yellow) \mu_r \\
		*(yellow) d
	\end{ytableau} & \makecell{$\nabla_{\mu_3}\cdots\nabla_{\mu_r}R_{ef\mu_2h} \nabla _{\mu_1} \nabla _h R_{efad} S_{[ad]\mu_1\mu_2\mu_3\cdots\mu_r}$ \\ $r$ is even \hspace{2mm} \& \hspace{2mm} $r\geq 2$}\\
	& & \\
	\hypertarget{S10e2}{7} & \ytableausetup{mathmode, boxsize=1.4em}
	\begin{ytableau}
		a & \mu_1 & \mu_2 &  \mu_3 & *(yellow) \mu_4 & *(yellow) \mu_5 & *(yellow) \cdots & *(yellow) \mu_r \\
		*(yellow) f
	\end{ytableau} & \makecell{ $\nabla_{\mu_5}\cdots\nabla_{\mu_r} \nabla _{\beta}R_{\alpha \mu_1\mu_2h} \nabla _h\nabla _{\mu_4}\nabla _{\alpha }R_{\beta \mu_3af}  S_{[af] \mu_1\mu_2 \mu_3\mu_4\mu_5\cdots\mu_r} $ \\ $r$ is even \hspace{2mm} \& \hspace{2mm} $r\geq 4$} \\
	& & \\
	\hypertarget{S11o1}{8} & \ytableausetup{mathmode, boxsize=1.4em}
	\begin{ytableau}
		\mu_1 & *(yellow) \mu_2 & *(yellow) \cdots & *(yellow) \mu_r \\
		b \\
		*(yellow) c \\
	\end{ytableau} & \makecell{$\nabla_{\mu_2}\cdots\nabla_{\mu_r}\nabla _fR_{\mu_1bde}R_{cfde}S_{[\mu_1bc]\mu_2\cdots\mu_r}$ \\ $r$ is odd \hspace{2mm} \& \hspace{2mm} $r\geq 1$} \\
	& & \\
	\hypertarget{S11o2}{9} & \ytableausetup{mathmode, boxsize=1.4em}
	\begin{ytableau}
		\mu_1 & \mu_2  &  \mu_3  &  *(yellow) \mu_4 & *(yellow) \cdots & *(yellow) \mu_r\\
		b \\
		*(yellow) c \\
	\end{ytableau} & \makecell{$\nabla_{\mu_4}\cdots\nabla_{\mu_r}\nabla _h R_{\mu_1b\mu_2f} R_{ch\mu_3f} S_{[\mu_1bc]\mu_2\mu_3\cdots\mu_r}$ \\ $r$ is odd \hspace{2mm} \& \hspace{2mm} $r\geq 3$} \\
	& & \\
	\hypertarget{S20e1}{10} & \ytableausetup{mathmode, boxsize=1.4em}
	\begin{ytableau}
		r & t &  *(yellow) \mu_1 & *(yellow) \cdots & *(yellow) \mu_r \\
		s & u \\
	\end{ytableau} &
	\makecell{$\nabla_{\mu_1} \cdots\nabla_{\mu_r} R_{prqt} R_{psqu} S_{[rs][tu]\mu_1\cdots\mu_r}$ \\ $r$ is even \hspace{2mm} \& \hspace{2mm} $r\geq 0$} \\
	& & \\
	\hypertarget{S20e2}{11} & \ytableausetup{mathmode, boxsize=1.4em}
	\begin{ytableau}
		r & t &  *(yellow) \mu_1 & *(yellow) \cdots & *(yellow) \mu_r\\
		*(yellow) s & *(yellow) u \\
	\end{ytableau} &
	\makecell{ $\nabla_{\mu_1}\cdots\nabla_{\mu_r}R_{pqrs} R_{pqtu} S_{[rs][tu]\mu_1\cdots\mu_r}$ \\ $r$ is even \hspace{2mm} \& \hspace{2mm} $r\geq 0$} \\
	& & \\
	\hypertarget{S20e3}{12} & \ytableausetup{mathmode, boxsize=1.4em}
	\begin{ytableau}
		a & c & \mu_1 & \mu_2 &  *(yellow) \mu_3 & *(yellow) \cdots & *(yellow) \mu_r\\
		*(yellow) b & *(yellow) d \\
	\end{ytableau} & \makecell{$\nabla_{\mu_3}\cdots\nabla_{\mu_r}R_{ab\mu_1h}R_{cd\mu_2h}S_{[ab][cd]\mu_1\mu_2\mu_3\cdots\mu_r}$ \\ $r$ is even \hspace{2mm} \& \hspace{2mm} $r\geq 2$} \\
	& & \\
	\hypertarget{S20o1}{13} & \ytableausetup{mathmode, boxsize=1.4em}
	\begin{ytableau}
		c & i & \mu_1 & *(yellow) \mu_2 & *(yellow) \cdots & *(yellow) \mu_r \\
		a & *(yellow) j \\
	\end{ytableau} & \makecell{ $\nabla_{\mu_2}\cdots\nabla_{\mu_r}R_{ab\mu_1k}\nabla _k R_{bcij} S_{[ca][ij]\mu_1\mu_2\cdots\mu_r}$ \\ $r$ is odd \hspace{2mm} \& \hspace{2mm} $r\geq 1$} \\
	& & \\
	\hypertarget{S21o1}{14} & \ytableausetup{mathmode, boxsize=1.4em}
	\begin{ytableau}
		a & \mu_1 & \mu_2 & *(yellow) \mu_3 & *(yellow) \mu_4 & *(yellow) \cdots & *(yellow) \mu_r\\
		b & *(yellow) e \\
		*(yellow) c \\
	\end{ytableau} & \makecell{$\nabla_{\mu_4}\cdots\nabla_{\mu_r}\nabla_{\mu_3}R_{ab\mu_2j}R_{cj\mu_1e}S_{[abc][\mu_1e]\mu_2\mu_3\mu_4\cdots\mu_r}$ \\ $r$ is odd \hspace{2mm} \& \hspace{2mm} $r\geq 3$}\\
	& & \\
	\hypertarget{S21e1}{15} & \ytableausetup{mathmode, boxsize=1.4em}
	\begin{ytableau}
		a & \mu_1 & *(yellow) \mu_2 & *(yellow) \mu_3 & *(yellow) \cdots & *(yellow) \mu_r\\
		b & e \\
		*(yellow) c \\
	\end{ytableau} & \makecell{ $\nabla_{\mu_3}\cdots\nabla_{\mu_r}R_{ab\mu_2i}R_{ci\mu_1e}S_{[abc][\mu_1e]\mu_2\mu_3\cdots\mu_r}$ \\ $r$ is even \hspace{2mm} \& \hspace{2mm} $r\geq 2$}\\
	& & \\
	\hypertarget{S22e1}{16} & \ytableausetup{mathmode, boxsize=1.4em}
	\begin{ytableau}
		\mu_1 & \mu_2 & *(yellow) \mu_3 & *(yellow) \cdots & *(yellow) \mu_r\\
		b & e \\
		*(yellow) c & *(yellow) f \\
	\end{ytableau} &\makecell{ $\nabla_{\mu_3}\cdots\nabla_{\mu_r}R_{\mu_1b\mu_2h}R_{chef}S_{[\mu_1bc][\mu_2ef]\mu_3\cdots\mu_r}$ \\ $r$ is even \hspace{2mm} \& \hspace{2mm} $r\geq 2$}\\
	& & \\
	\hypertarget{S30o1}{17} & \ytableausetup{mathmode, boxsize=1.4em}
	\begin{ytableau}
		c & i & k & *(yellow) \mu_1 & *(yellow) \mu_2 & *(yellow) \cdots & *(yellow) \mu_r\\
		a & *(yellow) j & *(yellow) d \\
	\end{ytableau} & \makecell{ $ \nabla_{\mu_2}\cdots\nabla_{\mu_r}R_{abkd}\nabla _{\mu_1}R_{bcij}S_{[ca][ij][kd]\mu_1\mu_2\cdots\mu_r}$ \\ $r$ is odd \hspace{2mm} \& \hspace{2mm} $r\geq 1$}\\
	& & \\
	\hypertarget{S30e1}{18} & \ytableausetup{mathmode, boxsize=1.4em}
	\begin{ytableau}
		a & b & c & \mu_1 & *(yellow) \mu_2 & *(yellow) \mu_3 & *(yellow) \cdots & *(yellow) \mu_r \\
		*(yellow) f & *(yellow) i & *(yellow) j \\
	\end{ytableau} & \makecell{ $\nabla_{\mu_3}\cdots\nabla_{\mu_r}\nabla _kR_{afcj}\nabla _{\mu_2}R_{bi\mu_1k}S_{[af][bi][cj]\mu_1\mu_2\mu_3\cdots\mu_r} $ \\ $r$ is even \hspace{2mm} \& \hspace{2mm} $r\geq 2$}\\
	&& \\
	\hypertarget{S31o1}{19} & \ytableausetup{mathmode, boxsize=1.4em}
	\begin{ytableau}
		a & d & \mu_1 & *(yellow) \mu_2 & *(yellow) \cdots & *(yellow) \mu_r\\
		b & *(yellow) e & *(yellow) i \\
		*(yellow) c \\
	\end{ytableau} & \makecell{$\nabla_{\mu_2}\cdots\nabla_{\mu_r} \nabla _jR_{abde}R_{cj\mu_1i}S_{[abc][de][\mu_1i]\mu_2\cdots\mu_r}$ \\ $r$ is odd \hspace{2mm} \& \hspace{2mm} $r\geq 1$}\\
	& & \\
	\hypertarget{S40e1}{20} & \ytableausetup{mathmode, boxsize=1.4em}
	\begin{ytableau}
		a & c & e & i & *(yellow) \mu_1 & *(yellow) \cdots & *(yellow) \mu_r\\
		*(yellow) b & *(yellow) d & *(yellow) f & *(yellow) j \\
	\end{ytableau} & \makecell{ $\nabla_{\mu_1}\cdots\nabla_{\mu_r}R_{abcd}R_{efij}S_{[ab][cd][ef][ij]\mu_1\cdots\mu_r}$ \\ $r$ is even \hspace{2mm} \& \hspace{2mm} $r\geq 0$}\\
	&& \\
\end{longtable}


\subsection{$\mathbf{D=7}$} \label{deseg}

As for photons, the discussion of graviton scattering in $D=7$ is identical to 
the corresponding discussion for $D \geq 8$ for particles $P$ transforming in representations $(r_1, r_2, 0)$. For such representations the coupling of 
two gravitons to $P$ is listed in rows (\hyperlink{S00e1}{1}-\hyperlink{S10e2}{7}), (\hyperlink{S20e1}{10}-\hyperlink{S20o1}{13}) , \hyperlink{S30o1}{17}, \hyperlink{S30e1}{18} and \hyperlink{S40e1}{20} of Table \ref{D8gr}.

As in the case of photons, Tableaux $Y_{(r_1,r_2,r_3)}$ with $r_3 \neq 0$ correspond to the sum of two irreducible representations of $SO(6)$. We encounter this situation when $P$ transforms in the representations associated with the Tableaux $Y_{(r+1, 1,1)}$ (\hyperlink{S11o1}{row 8} and \hyperlink{S11o2}{9} of Table \ref{D8gr}), $Y_{(r+2, 3, 1)}$ (\hyperlink{S31o1}{row 19} of Table \ref{D8gr}) $Y_{(r+1, 2, 1)}$ with $r \geq 2$ (\hyperlink{S21o1}{row 14}, \hyperlink{S21e1}{15} of Table \ref{D8gr}) , $Y_{(r,2,2)}$ (\hyperlink{S22e1}{row 16} of Table \ref{D8gr}). All except the last of these Tableaux have only one column of length 3 - the last Tableaux, $Y_{(r,2,2)}$ has two columns of length 3. As we have mentioned above each of these Tableaux correspond to two rather than one irreducible representations of $SO(6)$  (see Appendix \ref{rayt}). In each case we can project onto the irreducible representations by demanding that the indices in the column of length 3 are self dual or anti self dual \footnote{In the special case of $Y_{(r,2,2)}$ which has two columns of length 3, we impose the same condition on any one of the columns - symmetry ensures that it does not matter which column we project. }. In equations we impose the condition 
\begin{equation}\label{addeomfmmg}
\varepsilon_{\alpha_1 \alpha_2 \alpha_3 \alpha_4 \alpha_5 
	\alpha_6} A_{[\alpha_4 \alpha_5 \alpha_6] \cdots }=  \pm i A_{[\alpha_1 \alpha_2 \alpha_3] \cdots }
\end{equation} 
where the first three indices of $A$ are the indices corresponding to the first column of the Young Tableaux, and the $\cdots$ refers to all the other indices 
of the tensor (corresponding to all the other columns of the Tableaux).
As in the case of photons, for  either choice of sign in \eqref{addeomfmmg} 
we have as many independent couplings to two gravitons as are listed in Table \ref{D8gr} for the corresponding Tableaux.  So each of the representations $(r+1, 1, \pm 1)$ can couple to two gravitons in the two ways tabulated in  (\hyperlink{S11o1}{row 8} and \hyperlink{S11o2}{9} of Table \ref{D8gr}). Each of the representations  $(r+2, 3, \pm 1)$ can couple to two gravitons in the unique way represented in (\hyperlink{S31o1}{row 19} of Table \ref{D8gr}), 
each of the representations $(r+1, 2, \pm 1)$ can couple to two gravitons in 
one of the two ways represented in \hyperlink{S21o1}{row 14}, \hyperlink{S21e1}{row 15} and each of the representations $(r,2,\pm 2)$ can couple in the unique manner represented in \hyperlink{S22e1}{row 16} of Table \ref{D8gr}. As in the case of photons, the  Lagrangians  for the particles corresponding to the two signs of \eqref{addeomfmmg} both take the form listed in the third column of the corresponding rows of Table \ref{D8gr}, with the one caveat that the field $S$ corresponding to the particle $P$ that participates in this coupling obeys the additional self / anti-self duality condition
 \begin{equation}\label{addeomg}
\frac{ \partial_{\alpha_4}}{m}\varepsilon_{\alpha_1 \alpha_2 \alpha_3 \alpha_4 \alpha_5 \alpha_6 
	\alpha_7} S_{[\alpha_5 \alpha_6 \alpha_7] \cdots }=  \pm S_{[\alpha_1 \alpha_2 \alpha_3] \cdots} ~.
\end{equation} 

As in the case of photons, from the Lagrangian point of view, the coupling between particles in Young Tableaux with atleast one column of length 3 and two gravitons can be alternately understood as follows. We start with $P_{[\alpha_5 \alpha_6 \alpha_7] \cdots}$ that is not further constrained by the condition \eqref{addeomg}. The quantization of $P$  produces both the self dual and the anti-self dual particle. The most general coupling of $P$ to two gravitons is is given either by the Lagrangian listed in the corresponding row of the third column of Table \ref{D8gr} or by the same Lagrangian structure with the replacement

\begin{equation}\label{reprule} 
S_{[\alpha_1 \alpha_2 \alpha_3 ] \cdots } 
\rightarrow \varepsilon_{\alpha_1 \alpha_2 \alpha_3 \alpha_4 \beta_1 \beta_2 \beta_3} 
\frac{\partial_{\alpha_4}}{m}  S_{ [\beta_1 \beta_2 \beta_3] \cdots} ~.
\end{equation} 
For concreteness we list all these `new' parity odd couplings in Table \ref{GrD7podd} below. 
As in the case of photons, we can, as a special case, obtain the coupling of 
two gravitons to the self dual or anti self dual particle by taking an the appropriate 
linear combinations of the new couplings of Table \ref{GrD7podd} and the old couplings of Table \ref{D8gr}.

\begin{longtable}{|c|c|c|}
	\caption{{\bf Gravitons}, ${\bf D=7}$,~~  {\small Here the symbol ${\Huge *}$ indicates that the first column of the original 
			$SO(6)$ Tableaux, in the second column of this table, is Hodge stared (i.e. RHS of \eqref{addeomg} is replaced by LHS of \eqref{addeomg}) before we shade in boxes and then fuse with graviton 
			representations.}}\label{GrD7podd}\\
	\hline
	\makecell{\bf Scattering \\ \bf Amplitude} & \textbf{Young Tableaux} & \textbf{Lagrangian Structure} \\
	\hline
	\endfirsthead
	\multicolumn{3}{c}%
	{\tablename\ \thetable\ : $D=7~~~~$  (\textit{Continued from previous page})} \\
	\hline
	\makecell{\bf Scattering \\ \bf Amplitude} & \textbf{Young Tableaux} & \textbf{Lagrangian Structure} \\
	\hline
	\endhead
	\hline \multicolumn{3}{r}{\textit{}} \\
	\endfoot
	\hline
	\endlastfoot
	& & \\
	\hypertarget{S11o1p}{1} & {\Huge *} \ytableausetup{mathmode, boxsize=1.4em}
	\begin{ytableau}
		\mu_1 & *(yellow) \mu_2 & *(yellow) \cdots & *(yellow) \mu_r \\
		b \\
		*(yellow) c \\
	\end{ytableau} & \makecell{$\nabla_{\mu_2}\cdots\nabla_{\mu_r}\nabla _fR_{\mu_1bde}R_{cfde} \left(\varepsilon_{\mu_1 b c \alpha_4 \beta_1 \beta_2 \beta_3} \frac{\nabla_{\alpha_4}}{m}  S_{ [\beta_1 \beta_2 \beta_3]\mu_2\cdots\mu_r}\right)$ \\  $r$ is odd \hspace{2mm} \& \hspace{2mm}  $r\geq 1$ } \\
	& & \\
	\hypertarget{S11o2p}{2} & {\Huge *} \ytableausetup{mathmode, boxsize=1.4em}
	\begin{ytableau}
		\mu_1 & \mu_2  &  \mu_3  &  *(yellow) \mu_4 & *(yellow) \cdots & *(yellow) \mu_r\\
		b \\
		*(yellow) c \\
	\end{ytableau} & \makecell{$\nabla_{\mu_4}\cdots\nabla_{\mu_r}\nabla _h R_{\mu_1b\mu_2f} R_{ch\mu_3f} \left(\varepsilon_{\mu_1 b c \alpha_4 \beta_1 \beta_2 \beta_3} \frac{\nabla_{\alpha_4}}{m}  S_{ [\beta_1 \beta_2 \beta_3]\mu_2\mu_3\cdots\mu_r}\right)$ \\ $r$ is odd \hspace{2mm} \& \hspace{2mm} $r\geq 3$} \\
	& & \\
	\hypertarget{S21o1p}{3} & {\Huge *}\ytableausetup{mathmode, boxsize=1.4em}
	\begin{ytableau}
		a & \mu_1 & \mu_2 & *(yellow) \mu_3 & *(yellow) \mu_4 & *(yellow) \cdots & *(yellow) \mu_r\\
		b & *(yellow) e \\
		*(yellow) c \\
	\end{ytableau} & \makecell{$\nabla_{\mu_4}\cdots\nabla_{\mu_r}\nabla_{\mu_3}R_{ab\mu_2j}R_{cj\mu_1e}$  \\
	\hspace{2mm} \hspace{2mm} $\left(\varepsilon_{a b c \alpha_4 \beta_1 \beta_2 \beta_3} \frac{\nabla_{\alpha_4}}{m}  S_{ [\beta_1 \beta_2 \beta_3][\mu_1e]\mu_2\mu_3\mu_4\cdots\mu_r}\right)$ \\ $r$ is odd \hspace{2mm} \& \hspace{2mm} $r\geq 3$}\\
	& & \\
	\hypertarget{S21e1p}{4} & {\Huge *} \ytableausetup{mathmode, boxsize=1.4em}
	\begin{ytableau}
		a & \mu_1 & *(yellow) \mu_2 & *(yellow) \mu_3 & *(yellow) \cdots & *(yellow) \mu_r\\
		b & e \\
		*(yellow) c \\
	\end{ytableau} & \makecell{ $\nabla_{\mu_3}\cdots\nabla_{\mu_r}R_{ab\mu_2i}R_{ci\mu_1e}  \left(\varepsilon_{a b c \alpha_4 \beta_1 \beta_2 \beta_3} \frac{\nabla_{\alpha_4}}{m}  S_{ [\beta_1 \beta_2 \beta_3][\mu_1e]\mu_2\mu_3\cdots\mu_r}\right)$ \\ $r$ is even \hspace{2mm} \& \hspace{2mm} $r\geq 2$}\\
	& & \\
	\hypertarget{S22e1p}{5} & {\Huge *} \ytableausetup{mathmode, boxsize=1.4em}
	\begin{ytableau}
		\mu_1 & \mu_2 & *(yellow) \mu_3 & *(yellow) \cdots & *(yellow) \mu_r\\
		b & e \\
		*(yellow) c & *(yellow) f \\
	\end{ytableau} &\makecell{ $\nabla_{\mu_3}\cdots\nabla_{\mu_r}R_{\mu_1b\mu_2h}R_{chef} \left(\varepsilon_{\mu_1 b c \alpha_4 \beta_1 \beta_2 \beta_3} \frac{\nabla_{\alpha_4}}{m}  S_{ [\beta_1 \beta_2 \beta_3][\mu_2ef]\mu_3\cdots\mu_r}\right)$ \\ $r$ is even \hspace{2mm} \& \hspace{2mm} $r\geq 2$}\\
	& & \\
	\hypertarget{S31o1p}{6} & {\huge *} \ytableausetup{mathmode, boxsize=1.4em}
	\begin{ytableau}
		a & d & \mu_1 & *(yellow) \mu_2 & *(yellow) \cdots & *(yellow) \mu_r\\
		b & *(yellow) e & *(yellow) i \\
		*(yellow) c \\
	\end{ytableau} & \makecell{$\nabla_{\mu_2}\cdots\nabla_{\mu_r} \nabla _jR_{abde}R_{cj\mu_1i}  \left(\varepsilon_{a b c \alpha_4 \beta_1 \beta_2 \beta_3} \frac{\nabla_{\alpha_4}}{m}  S_{ [\beta_1 \beta_2 \beta_3][de][\mu_1i]\mu_2\cdots\mu_r}\right)$ \\ $r$ is odd \hspace{2mm} \& \hspace{2mm} $r\geq 1$}\\
	& & \\
\end{longtable}

\subsection{D=6} \label{desig}

As the Young Tableaux of $SO(5)$ have no more than two rows, there is no analogue
in this dimension, of the couplings presented in \hyperlink{S11o1}{ Rows 8}, \hyperlink{S11o2}{9}, \hyperlink{S21o1}{14}, \hyperlink{S21e1}{15}, \hyperlink{S22e1}{16}, \hyperlink{S31o1}{19} of Table \ref{D8gr}. 

The analogues of the couplings in the remaining rows of of Table \ref{D8gr} (namely \hyperlink{S00e1}{Rows 1}-\hyperlink{S10e2}{7}, \hyperlink{S20e1}{10}-\hyperlink{S20o1}{13}, \hyperlink{S30o1}{17}, \hyperlink{S30e1}{18} and \hyperlink{S40e1}{20})
are slightly modified because Young Tableaux of $SO(4)$ with two rows correspond to 
two rather than one irreducible representations. In $D=6$ the formula for the symmetric  and antisymmetric tensor product for two gravitons (the analogue of \eqref{grdecompsym} and \eqref{grdecompasym}) 
become 
\begin{equation} \label{decompgravsi}\begin{split} 
 &S^2 (2,0)= (4,0) \oplus  (2,0) \oplus (0,0) \oplus (2,2)  \oplus (2, -2) ,\\
& \Lambda^2 (2,0)= (3,1) \oplus  (3, -1) \oplus (1,1 ) \oplus (1,-1) .\\
\end{split}\end{equation} 

When $P$ transforms in the representation $(r, 0)$ 
the graviton-graviton-$P$ three point functions are identical in number and structure to those for $D \geq 8$ and are listed in  \hyperlink{S00e1}{Row 1}-\hyperlink{S00e3}{3} of Table \ref{D8gr}.

Now let us consider the case that $P$ that transforms in the $(r+m,m)$ representation of $SO(5)$ with  $1 \leq m \leq 4$. When $m=4$ all the graviton-graviton-$P$ three point structures are once again identical to $D \geq 8$ and are listed in \hyperlink{S40e1}{Row 20} of Table \ref{D8gr}. We have one such structure for even $r$ and no such structures for odd $r$. When $m=3$ and $r$ is even, the $D \geq 8$ results once again apply unchanged 
and are listed in \hyperlink{S30e1}{Row 18} of Table \ref{D8gr}.  When $m=3$ and $r$ is odd, contracting away $r$ indices in the top row and $2$ indices in the bottom row leads to a Tableaux with 3 free indices in the first row and one free index in the second row, as for $D \geq 8$ (see \hyperlink{S30o1}{Row 17} of Table \ref{D8gr}). As for $D \geq 8$ this representation leads to one three point coupling for odd $r$. However as in the case of photons we can use the effective $SO(4)$ $\varepsilon$ tensor \eqref{effep} to dualize the column of the Tableaux to obtain another (parity odd) tensor that also transforms in a Tableaux with $3$ boxes in the first row and one box in the second row. This dualized tensor gives rise to another  - parity odd - three point structure, but this time only when $r$ is even ($r$ must be even because the effective $\varepsilon$  tensor \eqref{effep} picks up a minus sign under the $ 1 \leftrightarrow 2$ interchange). It follows that we now have two three point functions when $r$ is even - one of these is parity even while the other is parity odd. The Lagrangian for the additional (parity odd) structure is listed in \hyperlink{5D6}{Row 5} of Table \ref{D6}.  

When $m=2$ we continue to have a single three point structure corresponding to coupling to the antisymmetric product of two gravitons in a way that is very similar to the case $m=3$. Once again we contract $r-1$  boxes in the first row and 1 box in the second row to obtain  an $SO(4)$ Tableaux $Y_{(3,1)}$. As in the case $m=3$ this Tableaux leads to one three point  structure when $r$ is odd, and its dual leads to another (parity odd) three  point structure when $r$ is even. The doubling of structures is a reflection of the fact that the Tableaux $Y_{(3,1)}$ corresponds to 2 $SO(4)$ irreps, namely the $(3,1)$ and the $(3,-1)$ representations.The new parity odd structure is listed in \hyperlink{3D6}{Row 3} of Table \ref{D6}.

Still working with the case $m=2$, we now turn to the structures corresponding to the symmetric fusion of two gravitons.  The structures correspond to the coupling of the gravitons to the $(4,0)$ or $(2,0)$ representations of  $SO(4)$, and take the same structure as for $D \geq 8$. They are listed in  \hyperlink{S20e3}{Rows 12} and  \hyperlink{S20e2}{ Row 11} respectively of table \ref{D8gr}. The coupling of two gravitons to the $Y_{(2,2)}$ Tableaux has a new feature because this Tableaux corresponds to two irreps. - the $(2,2)$ and the $(2,-2)$ of $SO(4)$. This leads to a doubling of the $D\geq 8$ structures in this case. One of the two couplings is listed in  \hyperlink{S20e1}{Row 10} of Table \ref{D8gr}, for even $r$. The second coupling is obtained by using the effective $SO(4)$ Levi-Civita tensor to dualize any one of the columns of the  Tableaux $Y_{(2,2)}$; the corresponding new coupling exists only for odd values of $r$. The Lagrangian corresponding to this new structure is listed in \hyperlink{4D6}{Row 4} of Table \ref{D6}.

Finally when $m=1$ the structures corresponding to the symmetric fusion of two gravitons work exactly as for $D \geq 8$ and are listed in  \hyperlink{S10e1}{Rows 6} and \hyperlink{S10e2}{7} of Table \ref{D8gr}. We now examine the structures corresponding to the antisymmetric fusion of graviton polarizations. As in $D \geq 8$, the contraction structure depicted in \hyperlink{S10o2}{Row 5} of Table \ref{D8gr} leads to the $SO(4)$ Tableaux $Y_{(3,1)}$. As this Tableaux corresponds to two $SO(4)$ irreps., there is a doubling of three point structures. In addition to the structure tabulated in \hyperlink{S10o2}{Row 5} of Table \ref{D8gr} we have the parity odd three point couplings - for even $r$  - listed in \hyperlink{2D6}{Row 2} of Table \ref{D6}. Similarly the contraction structure depicted in \hyperlink{S10o1}{Row 4} of Table \ref{D8gr} is effectively doubled. Apart from the structure listed in Table \ref{D8gr} we have the new parity odd structure listed in row \hyperlink{1D6}{Row 1} of Table \ref{D6}, once again for even $r$.


\begin{longtable}{|c|c|c|}
	\caption{{\bf Gravitons}, ${\bf D=6}$,~~  {\small The meaning of the symbol ${\Huge *}$  was explained in the caption to Table \ref{D61}.}}\label{D6}\\
	\hline
	\makecell{\bf Scattering \\ \bf Amplitude} & \textbf{Young Tableaux} & \textbf{Lagrangian Structure} \\
	\hline
	\endfirsthead
	\multicolumn{3}{c}%
	{\tablename\ \thetable\ :  $D=6$ (\textit{Continued from previous page})} \\
	\hline
	\makecell{\bf Scattering \\ \bf Amplitude} & \textbf{Young Tableaux} & \textbf{Lagrangian Structure} \\
	\hline
	\endhead
	\hline \multicolumn{3}{r}{\textit{}} \\
	\endfoot
	\hline
	\endlastfoot
	& & \\
	\hypertarget{1D6}{1} & {\Huge *} \ytableausetup{mathmode, boxsize=1.4em}
	\begin{ytableau}
		e & *(yellow) \mu_1 & *(yellow) \dots & *(yellow) \mu_r\\
		f \\
	\end{ytableau}  & \makecell{$\varepsilon^{abcdef} \nabla_{\mu_1}\cdots\nabla_{\mu_r}R_{abgh}R_{cdgh} S_{[ef]\mu_1\cdots\mu_r}$  \\ $r$ is even \hspace{2mm} \& \hspace{2mm} $r\geq 0$}\\
	& & \\
	\hypertarget{2D6}{2} & {\Huge *}\ytableausetup{mathmode, boxsize=1.4em}
	\begin{ytableau}
		e & \mu_1  & \mu_2 & *(yellow) \mu_3 & *(yellow) \dots & *(yellow) \mu_r\\
		f \\
	\end{ytableau} & \makecell{$\varepsilon^{abcdef}\nabla_{\mu_3}\cdots\nabla_{\mu_r} R_{ab\mu_1h}R_{cd\mu_2h} S_{[ef]\mu_1\mu_2\mu_3\cdots\mu_r}$ \\ $r$ is even \hspace{2mm} \& \hspace{2mm} $r\geq 2$ }\\
	& & \\
	\hypertarget{3D6}{3} & {\Huge *}\ytableausetup{mathmode, boxsize=1.4em}
	\begin{ytableau}
		e & i & \mu_1 & *(yellow) \mu_2 & *(yellow) \mu_3 & *(yellow) \dots & *(yellow) \mu_r\\
		f & *(yellow) g \\
	\end{ytableau} & \makecell{$\varepsilon^{abcdef}\nabla_{\mu_3}\cdots\nabla_{\mu_r}\nabla_g\nabla_{\mu_2}R_{abik}R_{cd\mu_1k} S_{[ef][ig]\mu_1\mu_2\mu_3\cdots\mu_r}$ \\ $r$ is even \hspace{2mm} \& \hspace{2mm} $r\geq 2$ }\\
	& & \\
	\hypertarget{4D6}{4} & {\Huge *}\ytableausetup{mathmode, boxsize=1.4em}
	\begin{ytableau}
		e & g & *(yellow) \mu_1 & *(yellow) \mu_2 & *(yellow) \dots & *(yellow) \mu_r\\
		f & h \\
	\end{ytableau} & \makecell{$\varepsilon^{abcdef} \nabla_{\mu_2}\cdots\nabla_{\mu_r}  \nabla_{\mu_1} R_{abgj}R_{cdhj} S_{[ef][gh]\mu_1\mu_2\cdots\mu_r}$ \\ $r$ is odd \hspace{2mm} \& \hspace{2mm} $r\geq 1$ }\\
	& & \\
	\hypertarget{5D6}{5} & {\Huge *}\ytableausetup{mathmode, boxsize=1.4em}
	\begin{ytableau}
		a & g & h & *(yellow) \mu_1 & *(yellow) \cdots & *(yellow) \mu_r \\
		b & *(yellow) i & *(yellow) j \\
	\end{ytableau} & \makecell{$\varepsilon^{abcdef}\nabla_{\mu_1}\cdots\nabla_{\mu_r} R_{abgi} R_{cdhj} S_{[ef][[gi][hj]\mu_1\cdots\mu_r}$ \\ $r$ is even \hspace{2mm} \& \hspace{2mm} $r\geq 0$}\\
	& & \\
\end{longtable}


\subsection{$\mathbf{D=5}$}\label{desfig}
In this case $SO(D-2)=SO(3)$. Representations of $SO(3)$ are most conveniently labelled by their spin $j$. The symmetric product of two gravitons transforms in the sum of $j=0, 2, 4$. The antisymmetric product of two gravitons transforms in $j=1, 3$. These representations are obtained by dualizing the single column of length 
two in each of the Young Tableaux on the RHS of \eqref{grdecompasym} (this operation turns the first representation
on the RHS of \eqref{grdecompasym} into the $j=3$ while turning the second 
representation on the RHS of \eqref{grdecompasym} into the $j=1$ representation). 
Note that the effective $SO(3)$ $\varepsilon$ tensor is proportional to a single 
factor of $k_1-k_2$. It follows that the structures corresponding to the 
$j=3$ and $j=1$ fusion of two gravitons pick up a minus sign under the interchange 
$k_1 \leftrightarrow k_2$, a second minus sign under the interchange $\epsilon_1 
\leftrightarrow \epsilon_2$, and therefore are invariant under the full 
Bose interchange $ 1 \leftrightarrow 2$.

As in the case of photons, we choose to label allowed representations of $SO(D-1)=SO(4)$ by the highest weights for Cartans corresponding to rotations in orthogonal two planes, ($h_1$, $h_2$). The $SO(4)$ representations that can couple to two gravitons are those with $(h_1, h_2) =(r,0)$ or $(r+1, \pm 1)$, $(r+2,\pm 2)$, $(r+3,\pm 3)$ and  $(r+4,\pm 4)$.

When $r$ is even the coupling of first of these representations ($(r,0)$) to two gravitons is the same as for  $D \geq 8$ (see \hyperlink{S00e1}{Row 1} and \hyperlink{S00e2}{Row 2} and \hyperlink{S00e3}{Row 3} of table \ref{D8gr}) . The novelty in $D=5$ 
is that there also exist graviton-graviton-$P$ couplings in this case when $r$ 
is odd. More specifically there exists one such (parity odd) structure  for $r= 1$ (corresponding to the antisymmetric coupling of two gravitons to $SO(3)$ spin 1) and two structures for odd $r \geq 3$ (corresponding to the antisymmetric coupling of the two gravitons to $SO(3)$ spins 1 and 3 ). The corresponding new Lagrangians are listed in \hyperlink{6D5}{Row 2} and \hyperlink{5D5}{Row 1} of Table \ref{D5} respectively. 

As in the case of photons, the particles $(r+m, \pm m)$ (for any given $m \geq 1$) have to be considered together. The Lorentzian $SO(5)$ covariant equation of motion that projects onto either $+$ or $-$ is complex and has no real solutions.
As in $D=7$ we have an effective doubling of the couplings of this pair of particles 
to the product of two gravitons. The new couplings are obtained by making the 
replacement 
\begin{equation}\label{repruledf} 
S_{[\alpha_1 \alpha_2  ] \cdots } 
\rightarrow \varepsilon_{\alpha_1 \alpha_2 \alpha_3 \beta_1 \beta_2 } 
\frac{\partial_{\alpha_3}}{m}  S_{ [\beta_1 \beta_2 ] \cdots}~.
\end{equation} 
Explicitly, when $m=1$ the couplings listed  \hyperlink{S10o1}{Rows 4}-\hyperlink{S10e2}{7} of Table \ref{D8gr} are supplemented by their counterparts listed in rows \hyperlink{S10o1D5}{Row  3}-\hyperlink{S10e2D5}{6} of Table \ref{D5}. For the case $m=2$ the  couplings of \hyperlink{S20e1}{Rows 10}-\hyperlink{S20o1}{13}  of Table \ref{D8gr} are supplemented by the new couplings of \hyperlink{S20e1D5}{Rows 7}-\hyperlink{S20o1D5}{10}  of Table \ref{D5}. For the case $m=3$ the couplings of \hyperlink{S30o1}{Rows 17}-\hyperlink{S30e1}{18} of Table \ref{D8gr} are supplemented by the new couplings listed in \hyperlink{S30o1D5}{Rows 11}-\hyperlink{S30e1D5}{12} of Table \ref{D5}. Finally, when $m=4$ the coupling listed in \hyperlink{S40e1}{Row 20} of Table \ref{D8gr} is supplemented by its shadow coupling listed in \hyperlink{S40e1D5}{Row 13} of Table \ref{D5}

\begin{longtable}{|c|c|c|}
	\caption{{\bf Gravitons}, ${\bf D=5}$,~~  {\small The meaning of the symbol ${\Huge *}$  was explained in the caption to Table \ref{D51}.}}\label{D5}\\
	\hline
	\makecell{\bf Scattering \\ \bf Amplitude} & \textbf{Young Tableaux} & \textbf{Lagrangian Structure} \\
	\hline
	\endfirsthead
	\multicolumn{3}{c}%
	{\tablename\ \thetable\ :  $D=5$ (\textit{Continued from previous page})} \\
	\hline
	\makecell{\bf Scattering \\ \bf Amplitude} & \textbf{Young Tableaux} & \textbf{Lagrangian Structure} \\
	\hline
	\endhead
	\hline \multicolumn{3}{r}{\textit{}} \\
	\endfoot
	\hline
	\endlastfoot
		& & \\
	 \hypertarget{5D5}{1} & {\Huge *}\ytableausetup{mathmode, boxsize=1.4em}
		\begin{ytableau}
			\mu_1 & \mu_2 & \mu_3 & *(yellow) \mu_4 & *(yellow) \cdots & *(yellow) \mu_r \\
		\end{ytableau} & \makecell{$\varepsilon_{abcd\mu_3} \nabla_{\mu_4} \cdots  \nabla_{\mu_r}R_{ab\mu_1\gamma} R_{cd\mu_2\gamma} S_{\mu_1\mu_2\mu_3\mu_4\cdots\mu_r}$ \\ $r$ is odd \hspace{2mm} \& \hspace{2mm} $r\geq 3$}\\
		& & \\
		\hypertarget{6D5}{2} & {\Huge *}\ytableausetup{mathmode, boxsize=1.4em}
		\begin{ytableau}
			\mu_1 & *(yellow) \mu_2 & *(yellow) \cdots & *(yellow) \mu_r \\
		\end{ytableau} & \makecell{$\varepsilon_{abcd\mu_1} \nabla_{\mu_2} \cdots  \nabla_{\mu_r} R_{ab\mu \gamma } R_{cd\mu \gamma } S_{\mu_1\mu_2\cdots\mu_r}$ \\ $r$ is odd \hspace{2mm} \& \hspace{2mm} $r\geq 1$}\\
			& & \\
	\hypertarget{S10o1D5}{3} & {\Huge *} \ytableausetup{mathmode, boxsize=1.4em}
	\begin{ytableau}
		a & *(yellow) \mu_1  & *(yellow)  \mu_2 & *(yellow) \cdots   & *(yellow) \mu_{r} \\
		c
	\end{ytableau} & \makecell{$\nabla_{\mu_2}\cdots\nabla_{\mu_r}\nabla_d R_{acef} R_{\mu_1 def}   \left(\varepsilon_{a c \alpha_3 \beta_1 \beta_2 } \frac{\nabla_{\alpha_3}}{m}  S_{ [\beta_1 \beta_2 ] \mu_1 \cdots \mu_{r}}\right)$  \\ $r$ is odd \hspace{2mm} \& \hspace{2mm} $r\geq 1$}\\
	& & \\
	\hypertarget{S10o2D5}{4} & {\Huge *} \ytableausetup{mathmode, boxsize=1.4em}
	\begin{ytableau}
		a &  \mu_1  & \mu_2 & *(yellow) \mu_3  & *(yellow)  \mu_4 & *(yellow) \cdots   & *(yellow) \mu_{r} \\
		e
	\end{ytableau} & \makecell{$\nabla_{\mu_4}\cdots\nabla_{\mu_r}\nabla_h R_{ae\mu_3 i} R_{h\mu_1 \mu_2 i} $\\ $ \left(\varepsilon_{a e \alpha_3 \beta_1 \beta_2 } \frac{\nabla_{\alpha_3}}{m}  S_{ [\beta_1 \beta_2 ] \mu_1 \mu_2 \mu_3 \mu_4 \cdots \mu_r}\right)$  \\ $r$ is odd \hspace{2mm} \& \hspace{2mm} $r\geq 3$ } \\  
	& & \\
	\hypertarget{S10e1D5}{5} & {\Huge *} \ytableausetup{mathmode, boxsize=1.4em}
	\begin{ytableau}
		a & \mu_1 & *(yellow) \mu_2 & *(yellow) \mu_3 & *(yellow) \cdots & *(yellow) \mu_r \\
		*(yellow) d
	\end{ytableau} & \makecell{$\nabla_{\mu_3}\cdots\nabla_{\mu_r}R_{ef\mu_2h} \nabla _{\mu_1} \nabla _h R_{efad}$\\$\left(\varepsilon_{a d \alpha_3 \beta_1 \beta_2 } \frac{\nabla_{\alpha_3}}{m}  S_{ [\beta_1 \beta_2 ]\mu_1\mu_2\mu_3\cdots\mu_r}\right)$ \\ $r$ is even \hspace{2mm} \& \hspace{2mm} $r\geq 2$}\\
	& & \\
	\hypertarget{S10e2D5}{6} & {\Huge *} \ytableausetup{mathmode, boxsize=1.4em}
	\begin{ytableau}
		a & \mu_1 & \mu_2 &  \mu_3 & *(yellow) \mu_4 & *(yellow) \mu_5 & *(yellow) \cdots & *(yellow) \mu_r \\
		*(yellow) f
	\end{ytableau} & \makecell{ $\nabla_{\mu_4}\cdots\nabla_{\mu_r} \nabla _{\beta}R_{\alpha \mu_1\mu_2h} \nabla _h\nabla _{\alpha }R_{\beta \mu_3af}$ \\$ \left(\varepsilon_{a f \alpha_3 \beta_1 \beta_2 } \frac{\nabla_{\alpha_3}}{m}  S_{ [\beta_1 \beta_2 ] \mu_1\cdots\mu_r}\right)$ \\ $r$ is even \hspace{2mm} \& \hspace{2mm} $r\geq 4$} \\
	& & \\
		\hypertarget{S20e1D5}{7} & {\Huge *} \ytableausetup{mathmode, boxsize=1.4em}
	\begin{ytableau}
		r & t &  *(yellow) \mu_1 & *(yellow) \cdots & *(yellow) \mu_r \\
		s & u \\
	\end{ytableau} &
	\makecell{$\nabla_{\mu_1} \cdots\nabla_{\mu_r} R_{prqt} R_{psqu} \left(\varepsilon_{r s \alpha_3 \beta_1 \beta_2 } \frac{\nabla_{\alpha_3}}{m}  S_{ [\beta_1 \beta_2 ] [tu]\mu_1\cdots\mu_r}\right)$ \\ $r$ is even \hspace{2mm} \& \hspace{2mm} $r\geq 0$} \\
	& & \\
	\hypertarget{S20e2D5}{8} & {\Huge *} \ytableausetup{mathmode, boxsize=1.4em}
	\begin{ytableau}
		r & t &  *(yellow) \mu_1 & *(yellow) \cdots & *(yellow) \mu_r\\
		*(yellow) s & *(yellow) u \\
	\end{ytableau} &
	\makecell{ $\nabla_{\mu_1}\cdots\nabla_{\mu_r}R_{pqrs} R_{pqtu} \left(\varepsilon_{r s \alpha_3 \beta_1 \beta_2 } \frac{\nabla_{\alpha_3}}{m}  S_{ [\beta_1 \beta_2 ] [tu]\mu_1\cdots\mu_r}\right)$ \\ $r$ is even \hspace{2mm} \& \hspace{2mm} $r\geq 0$} \\
	& & \\
	\hypertarget{S20e3D5}{9} & {\Huge *}  \ytableausetup{mathmode, boxsize=1.4em}
	\begin{ytableau}
		a & c & \mu_1 & \mu_2 &  *(yellow) \mu_3 & *(yellow) \cdots & *(yellow) \mu_r\\
		*(yellow) b & *(yellow) d \\
	\end{ytableau} & \makecell{$\nabla_{\mu_3}\cdots\nabla_{\mu_r}R_{ab\mu_1h}R_{cd\mu_2h}  \left(\varepsilon\epsilon_{a b \alpha_3 \beta_1 \beta_2 } \frac{\nabla_{\alpha_3}}{m}  S_{ [\beta_1 \beta_2 ][cd]\mu_1\mu_2\mu_3\cdots\mu_r}\right)$ \\ $r$ is even \hspace{2mm} \& \hspace{2mm} $r\geq 2$} \\
	& & \\
	\hypertarget{S20o1D5}{10} & {\Huge *} \ytableausetup{mathmode, boxsize=1.4em}
	\begin{ytableau}
		c & i & \mu_1 & *(yellow) \mu_2 & *(yellow) \cdots & *(yellow) \mu_r \\
		a & *(yellow) j \\
	\end{ytableau} & \makecell{ $\nabla_{\mu_2}\cdots\nabla_{\mu_r}R_{ab\mu_1k}\nabla _k R_{bcij}\left(\varepsilon_{c a \alpha_3 \beta_1 \beta_2 } \frac{\nabla_{\alpha_3}}{m}  S_{ [\beta_1 \beta_2 ][ij]\mu_1\mu_2\cdots\mu_r}\right)$ \\ $r$ is odd \hspace{2mm} \& \hspace{2mm} $r\geq 1$} \\
		& & \\
	\hypertarget{S30o1D5}{11} & {\Huge *} \ytableausetup{mathmode, boxsize=1.4em}
	\begin{ytableau}
		c & i & k & *(yellow) \mu_1 & *(yellow) \mu_2 & *(yellow) \cdots & *(yellow) \mu_r\\
		a & *(yellow) j & *(yellow) d \\
	\end{ytableau} & \makecell{ $ \nabla_{\mu_2}\cdots\nabla_{\mu_r}R_{abkd}\nabla _{\mu_1}R_{bcij} \left(\varepsilon_{c a \alpha_3 \beta_1 \beta_2 } \frac{\nabla_{\alpha_3}}{m}  S_{ [\beta_1 \beta_2 ][ij][kd]\mu_1\mu_2\cdots\mu_r}\right)$ \\ $r$ is odd \hspace{2mm} \& \hspace{2mm} $r\geq 1$}\\
	& & \\
	\hypertarget{S30e1D5}{12} & {\Huge *} \ytableausetup{mathmode, boxsize=1.4em}
	\begin{ytableau}
		a & b & c & \mu_1 & *(yellow) \mu_2 & *(yellow) \mu_3 & *(yellow) \cdots & *(yellow) \mu_r \\
		*(yellow) f & *(yellow) i & *(yellow) j \\
	\end{ytableau} & \makecell{ $\nabla_{\mu_2}\cdots\nabla_{\mu_r}\nabla _kR_{afcj} R_{bi\mu_1k}\left(\varepsilon_{af \alpha_3 \beta_1 \beta_2 } \frac{\nabla_{\alpha_3}}{m}  S_{ [\beta_1 \beta_2 ][bi][cj]\mu_1\cdots\mu_r} \right)$ \\ $r$ is even \hspace{2mm} \& \hspace{2mm} $r\geq 2$}\\
	&& \\
		\hypertarget{S40e1D5}{13} & {\Huge *} \ytableausetup{mathmode, boxsize=1.4em}
	\begin{ytableau}
		a & c & e & i & *(yellow) \mu_1 & *(yellow) \cdots & *(yellow) \mu_r\\
		*(yellow) b & *(yellow) d & *(yellow) f & *(yellow) j \\
	\end{ytableau} & \makecell{ $\nabla_{\mu_1}\cdots\nabla_{\mu_r}R_{abcd}R_{efij} \left(\varepsilon_{ab \alpha_3 \beta_1 \beta_2 } \frac{\nabla_{\alpha_3}}{m}  S_{ [\beta_1 \beta_2 ][cd][ef][ij]\mu_1\cdots\mu_r}\right)$ \\ $r$ is even \hspace{2mm} \& \hspace{2mm} $r\geq 0$}\\
	&& \\
\end{longtable}

\subsection{$\mathbf{D=4}$}\label{desfog}

In this case $SO(D-2)$ is $SO(2)$. The symmetric product of two gravitons transforms with $SO(2)$ charges $+4$, $0$, $-4$ and the antisymmetric product is a single spin zero state. 

The antisymmetric spin zero state is constructed with the aid of an 
$SO(2)$ $\varepsilon$ tensor. As mentioned above, the covariant version of this 
tensor, $k_1 ^\mu k_2^\nu \varepsilon_{\mu\nu\alpha \beta}$,  is antisymmetric under
the interchange $ k_1 \leftrightarrow k_2$. As this antisymmetric structure also 
picks up a minus sign under the interchange $\epsilon_1 \leftrightarrow \epsilon_2$ 
it follows that this structure is invariant under the overall Bose exchange 
$1 \leftrightarrow 2$.  

The symmetric singlet of two gravitons is, of course, by itself Bose symmetric. 
The two symmetric product states with charges $4$ and $-4$ can be thought 
of in tensor language as a symmetric traceless four index tensor (recall that 
any such tensor has all $+$ or all $-$ indices (see the discussion in Section 
\ref{desfo}). This traceless symmetric tensor is given by 
\begin{equation}\label{aform} 
A_{\alpha \beta \gamma \delta} = \epsilon^1_{(\alpha} \epsilon^1_\beta 
\epsilon^2_\gamma \epsilon^2_{\delta)}  
\end{equation} 
where the brackets $($ and $)$in the subscript of \eqref{aform} denote complete symmetrization of indices together with a removal of all traces. Note that the tensor listed 
in \eqref{aform} is also manifestly Bose symmetric. 

The `dual' four index traceless symmetric tensor corresponding to $A_{\alpha \beta \gamma \delta}$ given by 
\begin{equation}\label{duagv} 
k_1^\mu k_2^\nu \varepsilon_{\mu \nu \alpha \theta} A_{\theta \beta \gamma \delta}.
\end{equation} 
Note that the dual tensor listed in \eqref{duagv} picks up a minus sign under the 
$1 \leftrightarrow 2$ interchange. 
\footnote{The fact that we can construct two independent four index traceless tensors 
- $A_{\alpha \beta \gamma \delta}$ and its dual - reflects the fact that 
the traceless symmetric tensor was made up of two irreducible representations 
of $SO(2)$, namely the spin $4$ and the spin $-4$ state. }

The most general non-spinorial field $S$, associated with the massive particle $P$, is labelled by its $SO(3) \sim SU(2)$ label $j$ (completely symmetric tensor with $j$ indices). For every even $j$ there exists one graviton-graviton-$P$ coupling corresponding to the symmetric fusing of the gravitons to spin zero (all the free indices of the field associated with the particle $P$ are dotted with $k_1-k_2$.) This is the coupling listed in \hyperlink{S00e1}{Row 1} of Table \ref{D8gr}. For every even $j$ there is also a second coupling corresponding to the antisymmetric fusing of the gravitons to spin zero; this structure is also 
Bose symmetric for even $j$ (see the discussion earlier in this subsection). This additional Lagrangian has no analogue an any higher dimension and is is listed in \hyperlink{2D4}{Row 2} of Table \ref{D4}.

For every even $j \geq 4$, there is another coupling corresponding to the fusion of the 4 index $SO(2)$ descendent of $P$ with $A_{\alpha \beta \gamma \delta} $ 
defined in \eqref{aform} (the remaining $j-4$ indices of $P$ are dotted with 
$k_1 -k_2$). This coupling is listed in \hyperlink{S00e3}{Row 3} of Table \ref{D8gr}. Similarly, for every odd $j \geq 4$ there is a yet another 
 coupling corresponding to the fusion of the 4 index $SO(2)$ descendent of $P$ with $A_{\alpha \beta \gamma \delta} $ defined in \eqref{aform}. Once again the remaining $j-4$ indices of $P$ are dotted with $k_1 -k_2$. This coupling is listed in \hyperlink{1D4}{Row 1} of Table \ref{D4}.  

This exhausts the list of graviton-graviton-$P$ couplings in $D=4$. 

\begin{longtable}{|c|c|c|}
	\caption{{\bf Gravitons}, ${\bf D=4}$,~~ {\small The meaning of the symbol ${\Huge *}$  was explained in the caption to Table \ref{D61}.}}\label{D4}\\
	\hline
	\makecell{\bf Scattering \\ \bf Amplitude} & \textbf{Young Tableaux} & \textbf{Lagrangian Structure} \\
	\hline
	\endfirsthead
	\multicolumn{3}{c}%
	{\tablename\ \thetable\ :  $D=5$ (\textit{Continued from previous page})} \\
	\hline
	\makecell{\bf Scattering \\ \bf Amplitude} & \textbf{Young Tableaux} & \textbf{Lagrangian Structure} \\
	\hline
	\endhead
	\hline \multicolumn{3}{r}{\textit{}} \\
	\endfoot
	\hline
	\endlastfoot
	& & \\
	\hypertarget{1D4}{1} & {\Huge *}\ytableausetup{mathmode, boxsize=1.4em}
	\begin{ytableau}
		\mu_1 & \mu_2 & \mu_3 & \mu_4 & *(yellow) \mu_5 & *(yellow) \mu_6 & *(yellow) \cdots &  *(yellow) \mu_r \\
	\end{ytableau} & \makecell{$\varepsilon_{abc\mu_1}\nabla_{\mu_6}\cdots\nabla_{\mu_r}\nabla_{\nu} R_{ab\mu_3\delta} \nabla_{\mu_5}\nabla_{c}R_{\mu_2\delta \nu\mu_4} S_{\mu_1\mu_2\mu_3\mu_4\mu_5\mu_6\cdots\mu_r}$ \\ $r$ is odd \hspace{2mm} \& \hspace{2mm} $r\geq 5$}\\
	& & \\
	\hypertarget{2D4}{2} & {\Huge *} \ytableausetup{mathmode, boxsize=1.4em}
	\begin{ytableau}
		*(yellow) \mu_1 & *(yellow) \cdots &  *(yellow) \mu_r \\
	\end{ytableau} & \makecell{$\varepsilon_{abcd} \nabla_{\mu_1}\cdots\nabla_{\mu_r} R_{abef} R_{cdef} S_{\mu_1\cdots\mu_r}$ \\ $r$ is even \hspace{2mm} \& \hspace{2mm} $r\geq 0$}\\
	& & \\
\end{longtable}

\section{Discussion}

As we have emphasized in the introduction, we have demonstrated in this paper that every graviton-graviton-$P$ 
3 particle S-matrix is generated by a Lagrangian of the form 
\eqref{tpf} and so is of fourth order in derivatives. 
It follows immediately from this observation that  every two derivative theory of gravity interacting
with other fields admits a consistent truncation to Einstein 
gravity at cubic order in amplitudes. It seems very likely that this result continues 
on to arbitrary order. A specification of all 3 point graviton-graviton-$P$
S-matrices completely specifies the Lagrangian \eqref{tpf}. Expanding  \eqref{tpf} in powers of 
the metric fluctuation $h$ then also specifies a class 
of $({\rm graviton})^n$ P couplings for $n \geq 3$ 
(these couplings are tied to the given graviton-graviton-$P$ 
couplings by diffeomorphism invariance). Of course 
$({\rm graviton})^n$ P couplings are not uniquely determined
by three particle S-matrix data. For instance in the case 
$n=4$ we could have additional couplings generated by 
Lagrangians of the schematic form 
\begin{equation}\label{tpfdisc}
\int \sqrt{-g} \left( R R R S \right). 
\end{equation}
However, every such Lagrangian is of 6 or higher order in 
derivatives. This discussion can be continued. Once we have fixed $({\rm graviton})^3$-P
scattering, the new data in $({\rm graviton})^4$-P scattering 
appears likely to lie at 8 and higher order in derivatives and so on. In particular it seems extremely likely to us that any two derivative theory of gravity coupling to any number of additional fields, just on kinematical grounds, always admits a consistent truncation to Einstein gravity at the full non-linear level. Note that 
a similar result does not hold for photons even at cubic order  - several of the 
photon-photon-$P$ couplings presented in this paper are nonzero at two derivative order.

In this paper we have classified all graviton-graviton-$P$ (and photon-photon-$P$)
S-matrices for all possible massive particles $P$. For completeness it would be useful to perform the same classification for massless particles. The study of massless particles introduces a few new complications. In particular the onshell 
decay of a massless particle $P$ to two photons or gravitons is very kinematically 
restricted. The two gravitons (or photons) can only be emitted collinearly with 
the initial particle $P$ (which, recall, cannot now be in its rest frame).
If we continue to work with real momenta as in this paper, the scattering two 
plane of our paper is replaced by a scattering line, modifying our analysis 
significantly. Another complication with massless particles is that they 
often enjoy additional gauge invariance. For all these reasons we postpone the (definitely doable) analysis of graviton graviton massless particle (or photon photon massless particle) coupling to future work. We note that the 
spinor helicity formalism is particularly well suited to this problem 
atleast in $D=4$ and $D=6$ \cite{Cheung:2009dc}.

Finally it would be useful to `sew' two identical copies of each of the 
graviton-graviton-$P$ three point functions, classified in this paper,  
through a $P$ propagator (see \cite{Costa:2016hju} for a discussion of the projectors that appear in the numerator of these projectors) in order to compute the explicit form for all kinematically allowed $P$ exchange contributions to four graviton 
scattering. Conceptually, these contributions are the classical scattering analogues of conformal blocks. Simple examples of these blocks were constructed in \cite{Chowdhury:2019kaq}. It would be useful to have explicit expressions for these blocks for the most general case. Though the process of obtaining these blocks 
may prove algebraically intensive, the procedure that needs to be followed in order 
to find them is completely straightforward. We leave this to future work.

\section*{Acknowledgements}

We would like to thank A. Gadde for initial collaboration and several very useful discussions, and A. Nair for invaluable mathematical 
advice. We would also like to thank I. Halder, L. Janagal and A. Zhiboedov for useful discussions. The work of all authors was supported by the Infosys Endowment for the study of the Quantum Structure of Spacetime.  The work of T.G. is  supported by a DST  Inspire fellowship. We would all also like to acknowledge our debt to the people of India for their steady support to the study of the basic sciences.

\appendix

\section{Branching Rules}\label{DR}

Irreducible representations of $SO(2m)$ and $SO(2m+1)$ are both labelled by the highest weights $(h_1, h_2 \cdots h_m)$. Here $h_i$ denotes 
the eigenvalue under rotations in the $i^{th}$ two plane. For any given irreducible 
representation either all $h_i$ are integers or all $h_i$ are half integers. 
In the case $SO(2m+1)$ (or more precisely $Spin(2m+1)$) 
\begin{equation}\label{allhs}
h_1 \geq h_2 \cdots \geq  h_m
\end{equation} 
the $h_i$ are all positive. In the  case of $SO(2m)$ (or more precisely $Spin(2m)$) a version of   
\eqref{allhs} still holds. In this case, however, while $h_1 \cdots h_{m-1}$ 
are all positive, $h_m$ can be either positive or negative,a and the condition \eqref{allhs} is replaced by \eqref{allhsn}
\begin{equation}\label{allhsn}
h_1 \geq h_2 \cdots \geq  |h_m|
\end{equation} 

For examples of use of our notation, the $n$ index traceless symmetric tensor 
representation of $SO(2m)$ has highest weights $ (n, 0 \cdots 0)$. The $n$ index 
antisymmetric tensor with $n  < m$ has highest weights $(1, 1, \cdots 1 , 0 \cdots 0)$
(i.e. $h_i=1$ for $ i = 1 \cdots n$ and $h_i=0$ for $i >n$). The unique $2^m$ dimensional spinor 
representation of $SO(2m+1)$ has highest weights $(\frac{1}{2}, \frac{1}{2}, \cdots, \frac{1}{2} )$. The two different $2^{m-1}$ dimensional spinor representations of $SO(2m)$ have highest weights 
$(\frac{1}{2}, \frac{1}{2}, \cdots, \pm \frac{1}{2} )$. The self dual and anti self dual $m$ index tensors of $SO(2m)$ respectively have highest weights 
$(1,1, \cdots , \pm 1)$.

The branching rules from $SO(2m)$ to $SO(2m-1)$ are given as follows (Theorem  8.1.3 of \cite{goodman2009symmetry}). The $SO(2m)$ 
representation $(h_1, h_2, \cdots h_m)$ decomposes into the sum of $SO(2m-1)$
representations labelled by highest weights $(h_1^a, h_2^a \cdots h_{m-1}^a)$ where 
\begin{itemize} 
	\item $h_i^a$ have the same integrality properties as $h_i$ (i.e. $h_i^a$ are all integers/ half integers if the same is true of $h_i$).
	\item The collection of numbers $h_i^a$ runs over all possibilities consistent 
	with the inequalities 
	\begin{equation} \label{inequa} 
	h_1 \geq h_1^a \geq  h_2 \geq h_2^a \cdots \geq  h_{m-1}^a \geq |h_{m}|.
	\end{equation} 
	\end{itemize} 

On the other hand branching rules from $SO(2m+1)$ to $SO(2m)$ are given as follows (Theorem 8.1.4 of \cite{goodman2009symmetry}). The $SO(2m+1)$ 
representation $(h_1, h_2, \cdots h_m)$ decomposes into the sum of $SO(2m)$
representations labelled by highest weights $(h_1^a, h_2^a \cdots h_{m}^a)$ where 
\begin{itemize} 
	\item $h_i^a$ have the same integrality properties as $h_i$ (i.e. $h_i^a$ are all integers/ half integers if the same is true of $h_i$).
	\item The collection of numbers $h_i^a$ runs over all possibilities consistent 
	with the inequalities 
	\begin{equation} \label{inequan} 
	h_1 \geq h_1^a \geq  h_2 \geq h_2^a \cdots \geq h_{m-1}^a \geq h_{m} \geq |h_{m}^a| .
	\end{equation} 
	\item Each of the representations described above occur 
	exactly once (i.e. with multiplicity unity).
\end{itemize} 

\section{Representations Associated with Young Tableaux} \label{rayt} 

Around \eqref{projy} we have defined a projection onto a subspace of traceless 
$SO(2m+1)$ or $SO(2m)$ tensors associated with a Young Tableaux  $Y_{(r_1, r_2 \cdots r_m)}$.We have also mentioned that this subspace of tensors transforms under $SO(2m+1)$ in the irreducible representation $(r_1, r_2 \ldots r_m)$ when $r_m=0$. 
or when the group is $SO(2m+1)$ (even when $r_m \neq 0$). In the case of the group 
$SO(2m)$ and when $r_m \neq 0$, however, the subspace of tensors described 
above transforms in the direct sum of two irreducible representations of $SO(2m)$, 
namely $(r_1, r_2 \cdots r_m) \oplus (r_1 , r_2 \ldots -r_{m})$. In this appendix 
we will give an intuitive explanation of these facts. 

Staying away from the case $SO(2m)$ with $r_m \neq  0$ for a moment, it is easy to 
convince oneself that the Tableaux $Y_{(r_1, r_2 \cdots r_m)}$ corresponds to a 
single representation with highest weights $(r_1 \cdots r_m)$. Tensors of $SO(N)$ 
differ from tensors of $SU(N)$ in two crucial ways. First $SO(N)$ indices can 
be contracted with each other - the same is not true of $SU(N)$. Second the 
$SO(N)$ $\varepsilon$ tensor can be used to impose group covariant equations 
(like self duality conditions) on $SO(N)$ tensors; the same is not true of $SU(N)$. 

All the tensors we deal with in this paper are traceless. It follows that the indices 
of these tensors cannot contract with each other. Consequently, the only significant 
difference between $SU(N)$ and $SO(N)$ traceless tensors lies in the possibility 
of new equations involving the ${\varepsilon}$ tensor. Now an $\varepsilon$ tensor 
has to act on antisymmetrized indices. In the case of $SO(2m+1)$ the highest number of mutually antisymmetrized indices is $\leq m$. The action of an $\varepsilon$ tensor 
on such a Tableaux always results in a tensor with more indices that the original 
tensor. It follows that $\varepsilon$ cannot be used to impose new equations 
on the tensors corresponding to any particular Young Tableaux. The same is true 
of the group $SO(2m)$ with $r_m=0$. It follows in these cases that the tensors 
associated with a given Tableaux transform in a single irreducible representation 
of $SO(N)$ (because this result is famously true of $SU(N)$ tensors and there is 
no significant difference between traceless $SO(N)$ and $SU(N)$ tensors in this case).

The highest weights of the corresponding representation are also easily deduced. We put the first $r_1$ vectors in the highest possible weight state, namely the state $(1,0, 0 \cdots 0)$. 
Putting the $(r+1)^{th}$ index into the same state gives zero (because of the symmetry 
properties of the Tableaux) so we do the next best thing but putting the next 
$r_2$ indices in the $(0, 1, 0 \cdots 0)$ state and so on. It follows that the
highest weight state of this representation is $(r_1, r_2 \cdots r_m)$.

We now turn to the case $SO(2m)$ with $r_m \neq 0$. 
Let us first consider the special case $Y_{(1,1,\cdots,1,1)}$. This Young Tableaux denotes  an antisymmetric tensor with $m$ indices. In precisely 
this case the $\varepsilon$ tensor can be used to impose a group covariant equation 
on the space of such tensors and so break it up into two irreducible representations
of $SO(2m)$. The first of these is the self-dual 
	tensor with highest weights given 
	by $(1,1,1, .. 1)$. The second is the anti self-dual tensor whose highest weights 
	are $(1,1, .. , 1, -1)$. 

Next let us consider the case of a Young Tableaux with two columns, each of length $m$, i.e. $Y_{(2,2,\cdots,2,2)}$. This representation can be thought of 
as the traceless part of the symmetric product of two copies of 
$Y_{(1,1,\cdots,1,1)}$. In this context traceless simply means the projection 
onto representations $(h_1, h_2 \cdots h_m)$ such that 
$$\sum_{i=1}^m |h_i|=2m.$$
Now the symmetric product of two self-dual tensors has a single traceless representation - its highest weights are  equal to $(2, 2 ... 2)$. 
Similarly the symmetric product of two anti self-dual tensors has a single 
traceless representation: its highest weights are $(2, 2 ... 2,-2)$. 
However there are no traceless representations in the product of the self dual and the anti self dual tensor. The `biggest' such representation has 
highest weights $(2, 2, 2, 0)$ and so has $\sum_i |h_i|=2m-2$ and so is not 
traceless (this representation is part of the Young Tableaux $Y_{(2,2,\cdots,2,2,0)}$). 
 
The generalization of these remarks to Young Tableaux $Y_{(p,p,\cdots,p,p)}$ is obvious. The $p^{th}$ symmetric product of $Y_{(1,1,\cdots,1,1)}$ has only two 
traceless representations (i.e. representations such that  
$\sum_{i=1}^m |h_i|=pm.$)
These are the representations $(p,p \cdots p)$ and $(p, p \cdots -p)$ which 
respectively occur in the $p^{th}$ product of self dual and the $p^{th}$ product of 
anti-self dual tensors respectively.  The generalization to more complicated representations of $SO(2m)$ can be worked out along similar lines.

\section{Photon Amplitudes} \label{pa}

In this Appendix we present explicit expressions for the photon-photon-$P$ S-matrix for a particular convenient choice of the polarization tensor of $P$.

Recall that in subsection \ref{lmp} we defined a projector onto a subspace of tensors that transforms under $SO(m)$ in the representation associated with the Young Tableaux $Y$. For the purpose of studying photon-photon-$P$ and graviton-graviton-$P$ scattering we focus on Young Tableaux with no more than three rows. Let the number of boxes in the $i^{th}$ row of the Tableaux be denoted by $r_i$. In subsection \ref{lmp} our algorithm for constructing a projector required us to fill in the boxes of the Tableaux with any choice of integers running from $1, \cdots ,r_1+r_2 +r_3$. We choose here to fill the first row of the Tableaux with the integers $ 1, \cdots, r_1$, the second row with $r_1+1, \cdots, r_1 + r_2$ and the third row with the integers $r_1+r_2+1, \cdots ,r_1 + r_2 + r_3$. This choice of filling of the Young Tableaux defines a projector ${\cal P}$ defined in \eqref{projy}. The (Lorentzian $SO(D)$) wave solution for the particle $P$ is chosen to be \footnote{In our notation for the field corresponding to the particle $P$ in 
	\eqref{Apol} we have not attempted to denote the symmetry properties of the various
	indices.}  
\begin{equation}\label{Apol}
S^{\mu_1 \mu_2 \cdots \mu_{r_1+r_2+r_3}}(x) 
= e^{-ik_3.x} {\cal P} \left( 
\epsilon_3^{\mu_1} \cdots \epsilon_3^{\mu_{r_1}}
{\epsilon}_{4}^{\mu_{r_1+1}} \cdots {\epsilon}_4^{\mu_{r_1 + r_2}} {\epsilon}_{5}^{\mu_{r_1+r_2+1}} \cdots {\epsilon}_5^{\mu_{r_1 + r_2+ r_3}} \right) 
\end{equation} 
where $\epsilon_m^\mu$ are Lorentzian $SO(D)$ vectors that obey the constraints 
\begin{equation}\label{condep} 
\epsilon_m \cdot \epsilon_n=0, ~~~~~k_3 \cdot \epsilon_m=0,  ~~~m, n= 3 \cdots 5,
\end{equation} 
In order to find an explicit expression for the scattering amplitude we simply plug \eqref{Apol} - along with a similar wave solution for the photon fields 
$A_\mu(x)$ into the explicit expression for the cubic interaction action- and then simply evaluate the onshell value of the cubic Lagrangian associated with each of the three point coupling. We list our final results below 
in terms of the variables
\begin{equation}\label{VarDef}
\begin{split}
A_1 = \epsilon_1.k_2, ~~~~~~~~~~~~~~ & b_{12} = \epsilon_1.\epsilon_2, ~~~~~~~~~~~~~~  A_2 = \epsilon_2.k_1,\\
A_3 = \epsilon_3.\left(k_1-k_2\right), ~~~~ & A'_3 = \epsilon_4.\left(k_1-k_2\right), ~~~~ A''_3 = \epsilon_5.\left(k_1-k_2\right),\\
b_{23} = \epsilon_2.\epsilon_3, ~~~~~~~~~~~~~ & b'_{23} = \epsilon_2.\epsilon_4, ~~~~~~~~~~~~~~ b''_{23} = \epsilon_2.\epsilon_5,\\
b_{13} = \epsilon_1.\epsilon_3, ~~~~~~~~~~~~~ & b'_{13} = \epsilon_1.\epsilon_4, ~~~~~~~~~~~~~~ b''_{13} = \epsilon_1.\epsilon_5.\\
\end{split}
\end{equation}
\subsection{$D\geq 8$} \label{bo}

In this subsection we present explicit results for the S-matrix corresponding to the 
parity even interactions listed in Table \ref{D8Ph}. The results below apply unrestrictedly in $D \geq 8$, and also apply for the allowed parity even amplitudes 
in $D \leq 8$. \footnote{The `disallowed' amplitudes are nonvanishing in $D \leq 8$ 
	but carry the same information as the collection of all allowed amplitudes. 
	For instance the Young Tableaux with a single column of three boxes is disallowed 
	in $D=5$. The amplitude for this Tableaux does not vanish - but 
	is simply equal to the amplitude for $P$ transforming in a Tableaux with a single 
	box (the vector) once we choose the effective value of $\epsilon_3^\mu$ for this 
	vector to equal 
	$$\epsilon_3^\mu  \rightarrow \varepsilon^{\mu \alpha \beta \gamma \delta } \epsilon_3^\alpha \epsilon_4^\beta \epsilon_5 ^\gamma (k_1+k_2)^\delta $$}

Each S-matrix is assigned a name of the form $\mathfrak{A}^{i,j}_{(a,b,c)}$. The subscript in this symbol gives the lengths of the first three rows of the $SO(D-1)$ Young Tableaux associated with $P$.  The first superscript is $e$ when the three point coupling is constructed via fusion to a representation in a Bose symmetric product of two photons (or gravitons below) 
and is $o$ when the  coupling is constructed via fusion to a representation in the 
Bose antisymmetric product of two photons (or gravitons). \footnote{In our notation
	we keep track of the symmetry or antisymmetry of structures under the 
	simultaneous interchange $\epsilon_1 \leftrightarrow \epsilon_2$ and 
	$k_1 \leftrightarrow k_2$.} 

The second superscript is a multiplicity 
label. The amplitudes we have listed are obtained from the interaction Lagrangian
terms recorded in the specified rows of Table \ref{D8Ph}.

In the rest of this appendix and the next appendix we present a systematic listing of 
the scattering amplitudes obtained from the Lagrangians listed in the various 
tables of the main text. The expressions for all amplitudes are presented
ignoring constant proportionality factors (i.e. factors of $i$, $2$ and 
various hook factors entering the projectors ${\cal P}$).

\begin{eqnarray}
\mbox{Row 1} &:& \hyperlink{D8Ph:s:1}{\mathfrak{A}^{e,1}_{(r,0,0)}} \propto - 2 A_3^r \left(2 A_1 A_2+b_{12} m^2\right)  \nonumber \\
\mbox{Row 2} &:& \hyperlink{D8Ph:s:2}{\mathfrak{A}^{e,2}_{(r,0,0)}} \propto A_3^{r-2} \left(A_3 \left(A_2 b_{13}-\frac{A_3 b_{12}}{2}\right)-b_{23} \left(A_1 A_3+b_{13} m^2\right)\right)\nonumber \\
\mbox{Row 3} &:& \hyperlink{D8Ph:s1:1}{\mathfrak{A}^{o,1}_{(r+1,1,0)}} \propto\frac{1}{4} A_3^r \left(b_{23} \left(A_1 A'_3+m^2 b'_{13}\right)+A_2 \left(b_{13} A'_3-A_3 b'_{13}\right)-b'_{23} \left(A_1 A_3+b_{13} m^2\right)\right) \nonumber \\
\mbox{Row 4} &:& \hyperlink{D8Ph:s1:2}{\mathfrak{A}^{e,1}_{(r+1,1,0)}} \propto\frac{1}{16} A_3^{r-1}\nonumber\\
&&~~~~~~~~~ \left(b_{23} \left(A_3 m^2 b'_{13}-A'_3 \left(A_1 A_3+2 b_{13} m^2\right)\right)+A_2 A_3 \left(b_{13} A'_3-A_3 b'_{13}\right)+A_3 b'_{23} \left(A_1 A_3+b_{13} m^2\right)\right)\nonumber\\
\mbox{Row 5} &:& \hyperlink{D8Ph:s2:1}{\mathfrak{A}^{e,1}_{(r+2,2,0)}} \propto-\frac{1}{2} A_3^r \left(b_{13} A'_3-A_3 b'_{13}\right) \left(b_{23} A'_3-A_3 b'_{23}\right)\nonumber\\
\mbox{Row 6} &:&\hyperlink{D8Ph:s11:1}{\mathfrak{A}^{o,1}_{(r,1,1)}} \propto\frac{1}{12} m^2 A_3^{r-1} \left(b'_{23} \left(A_3 b''_{13}-b_{13} A''_3\right)+b''_{23} \left(b_{13} A'_3-A_3 b'_{13}\right)+b_{23} \left(A''_3 b'_{13}-A'_3 b''_{13}\right)\right)\nonumber
\end{eqnarray}

\subsection{$D=7$}
Here we list the S-matrix corresponding to the single parity odd interaction 
term in  $D=7$ listed in \eqref{twocouplngs}.
\begin{eqnarray}
\hyperlink{twocouplngs}{\mathfrak{A}^{o,1}_{(r+1,1,0)}} &\propto& m A_3^{r-1} \varepsilon \left(k_1,\epsilon _1,\epsilon _2,k_2,\epsilon _3,\epsilon _4,\epsilon _5\right) \nonumber
\end{eqnarray}
where $\varepsilon \left( v_1, v_2, \cdots, v_n \right)$ is defined as 
\begin{eqnarray}
\varepsilon \left( v_1, v_2, \cdots, v_n \right) &\propto& \varepsilon_{\mu_1\mu_2 \cdots \mu_n} v_1^{\mu_1} v_2^{\mu_2} \cdots v_n^{\mu_n}.
\end{eqnarray}

\subsection{$D=6$}
Here we list the S-matrix corresponding to the single parity odd interaction 
term in  $D=6$ recorded in Table \ref{D61}.  
\begin{eqnarray}
\hyperlink{Ph1D6}{\mathfrak{A}^{e,1}_{(r+1,1,0)}} &\propto& 2 A_3^r ~\varepsilon \left(k_1,k_2,\epsilon_1,\epsilon_2,\epsilon_3,\epsilon_4\right) \nonumber
\end{eqnarray}

\subsection{$D=5$}
In this sub-section we list S-matrices corresponding to the $D=5$  parity odd interaction structures listed in Table \ref{D51}.  

\begin{eqnarray}
\mbox{Row 1} &:& \hyperlink{Ph1D5}{\mathfrak{A}^{o,1}_{(r,0,0)}} \propto 4 A_3^r ~\varepsilon \left(k_1,k_2,\epsilon_1,\epsilon_2,\epsilon_3\right) \nonumber\\
\mbox{Row 2} &:& \hyperlink{D8Ph:s1:1*}{\mathfrak{A}^{o,1}_{(r+1,1,0)}} \propto -\frac{1}{2} m A_3^r \varepsilon \left(\epsilon _1,\epsilon _2,k_3,\epsilon _3,\epsilon _4\right)\nonumber \\
\mbox{Row 3} &:& \hyperlink{D8Ph:s1:2*}{\mathfrak{A}^{o,2}_{(r+1,1,0)}} \propto -\frac{A_3^{r-2} \left(\varepsilon \left(k_2,\epsilon _3,\epsilon _4,\epsilon _1,k_1\right) \left(A_2 A_3-b_{23} m^2\right)-\varepsilon \left(k_1,\epsilon _3,\epsilon _4,\epsilon _2,k_2\right) \left(A_1 A_3+b_{13} m^2\right)\right)}{2 m}\nonumber \\
\mbox{Row 4} &:& \hyperlink{D8Ph:s2:1*}{\mathfrak{A}^{o,1}_{(r+2,2,0)}} \propto- \frac{A_3^r \left(\varepsilon \left(k_1,\epsilon _1,k_2,\epsilon _3,\epsilon _4\right) \left(b_{23} A'_3-A_3 b'_{23}\right)+\varepsilon \left(k_2,\epsilon _2,k_1,\epsilon _3,\epsilon _4\right) \left(A_3 b'_{13}-b_{13} A'_3\right)\right)}{m}\nonumber \\
\end{eqnarray}

\subsection{$D=4$}
In this subsection, we list the $D=4$ S-matrices corresponding to the parity odd interaction Lagrangians recorded in the corresponding rows of Table \ref{PhD4}.  
\begin{eqnarray}
\mbox{Row 1} &:& \hyperlink{Ph1D4}{\mathfrak{A}^{o,1}_{(r,0,0)}} \propto 4 A_3^r \varepsilon \left(k_1,k_2,\epsilon_1,\epsilon_2\right) \nonumber \\
\mbox{Row 2} &:& \hyperlink{Ph2D4}{\mathfrak{A}^{e,1}_{(r,0,0)}} \propto\frac{1}{2} A_3^{r-2} \left(\varepsilon \left(\epsilon_1,k_1,k_2,\epsilon_3\right) \left(A_2 A_3-b_{23} m^2\right)-\varepsilon \left(\epsilon_2,k_1,k_2,\epsilon_3\right) \left(A_1 A_3+b_{13} m^2\right)\right) \nonumber
\end{eqnarray}

\section{Graviton Amplitudes}\label{D8Gr:Amp}

\subsection{$D\geq 8$}\label{co}

In this subsection, we list the S-matrices corresponding to the parity even 
interaction Lagrangians (for $D\geq 8$ as well as for the allowed amplitudes in lower dimensions) listed
in the corresponding rows of Table \ref{D8gr}. 

\begingroup 
\allowdisplaybreaks
\begin{eqnarray}
\mbox{Row 1} &:&  \hyperlink{S00e1}{\mathfrak{A}^{e,1}_{(r,0,0)}} \propto \frac{1}{4} A_3^r \left(2 A_1 A_2+b_{12} m^2\right)^2  \nonumber \\
\mbox{Row 2} &:&  \hyperlink{S00e2}{\mathfrak{A}^{e,2}_{(r,0,0)}} \propto \frac{1}{16} A_3^{r-2} \left(2 A_1 A_2+b_{12} m^2\right) \left(A_3 \left(A_3 b_{12}-2 A_2 b_{13}+2 A_1 b_{23}\right)+2 b_{13} b_{23} m^2\right)\nonumber \\
\mbox{Row 3} &:&  \hyperlink{S00e3}{\mathfrak{A}^{e,3}_{(r,0,0)}} \propto\frac{1}{64} A_3^{r-4} \left(A_3 \left(A_3 b_{12}-2 A_2 b_{13}+2 A_1 b_{23}\right)+2 b_{13} b_{23} m^2\right){}^2 \nonumber \\
\mbox{Row 4} &:&  \hyperlink{S10o1}{\mathfrak{A}^{o,1}_{(r+1,1,0)}} \propto\frac{3}{32} A_3^r \left(2 A_1 A_2+b_{12} m^2\right) \nonumber\\
&&~~~~~~~~~~~\left(-A_2 b_{13} A'_3-A_1 b_{23} A'_3+A_2 A_3 b'_{13}+A_1 A_3 b'_{23}-b_{23} m^2 b'_{13}+b_{13} m^2 b'_{23}\right)\nonumber\\
\mbox{Row 5} &:&  \hyperlink{S10o2}{\mathfrak{A}^{o,2}_{(r+1,1,0)}} \propto -\frac{5}{128} A_3^{r-2} \left(A_3 \left(A_3 b_{12}-2 A_2 b_{13}+2 A_1 b_{23}\right)+2 b_{13} b_{23} m^2\right)\nonumber\\
&& ~~~~~~~~~~~\left(-A_2 b_{13} A'_3-A_1 b_{23} A'_3+A_2 A_3 b'_{13}+A_1 A_3 b'_{23}-b_{23} m^2 b'_{13}+b_{13} m^2 b'_{23}\right)\nonumber\\
\mbox{Row 6} &:&  \nonumber \hyperlink{S10e1}{\mathfrak{A}^{e,1}_{(r+1,1,0)}} \propto -\frac{1}{16} A_3^{r-1} \left(2 A_1 A_2+b_{12} m^2\right)\\
&&~~~~~~~~~\nonumber \left(b_{23} \left(A_3 m^2 b'_{13}-A'_3 \left(A_1 A_3+2 b_{13} m^2\right)\right)+A_2 A_3 \left(b_{13} A'_3-A_3 b'_{13}\right)+A_3 b'_{23} \left(A_1 A_3+b_{13} m^2\right)\right)\nonumber\\
\mbox{Row 7} &:&  \hyperlink{S10e2}{\mathfrak{A}^{e,2}_{(r+1,1,0)}} \propto -\frac{3}{128} A_3^{r-3} \left(A_1 A_3+b_{13} m^2\right) \left(A_2 A_3-b_{23} m^2\right) \nonumber\\
&&~~~~~~~~~\nonumber \left(b_{23} \left(A_3 m^2 b'_{13}-A'_3 \left(A_1 A_3+2 b_{13} m^2\right)\right)+A_2 A_3 \left(b_{13} A'_3-A_3 b'_{13}\right)+A_3 b'_{23} \left(A_1 A_3+b_{13} m^2\right)\right)\\
\mbox{Row 8} &:&  \hyperlink{S11o1}{\mathfrak{A}^{o,1}_{(r,1,1,0)}} \propto\frac{1}{12} m^2 A_3^{r-1}\left(2 A_1 A_2+b_{12} m^2\right) \nonumber \\
&& ~~~~~~~~~~~\left(b'_{23} \left(A_3 b''_{13}-b_{13} A''_3\right)+b''_{23} \left(b_{13} A'_3-A_3 b'_{13}\right)+b_{23} \left(A''_3 b'_{13}-A'_3 b''_{13}\right)\right) \nonumber\\
\mbox{Row 9} &:&  \hyperlink{S11o2}{\mathfrak{A}^{o,2}_{(r,1,1,0)}} \propto \nonumber \frac{1}{48} m^2 A_3^{r-3}\left(2 b_{23} \left(A_1 A_3+b_{13} m^2\right)+A_3 \left(A_3 b_{12}-2 A_2 b_{13}\right)\right)\\
&& ~~~~~~~~~~~\left(b'_{23} \left(A_3 b''_{13}-b_{13} A''_3\right)+b''_{23} \left(b_{13} A'_3-A_3 b'_{13}\right)+b_{23} \left(A''_3 b'_{13}-A'_3 b''_{13}\right)\right)\nonumber\\
\mbox{Row 10} &:&  \hyperlink{S20e1}{\mathfrak{A}^{e,1}_{(r+2,2,0)}} \propto \frac{1}{32} A_3^r \left(A_2 b_{13} A'_3+A_1 b_{23} A'_3-A_2 A_3 b'_{13}-A_1 A_3 b'_{23}+m^2 \left(b_{23} b'_{13}-b_{13} b'_{23}\right)\right){}^2\nonumber\\
\mbox{Row 11} &:&  \hyperlink{S20e2}{\mathfrak{A}^{e,2}_{(r+2,2,0)}}\propto  \frac{1}{8} A_3^r \left(2 A_1 A_2+b_{12} m^2\right) \left(b_{13} A'_3-A_3 b'_{13}\right) \left(b_{23} A'_3-A_3 b'_{23}\right)\nonumber \\
\mbox{Row 12} &:&  \hyperlink{S20e3}{\mathfrak{A}^{e,3}_{(r+2,2,0)}} \propto\frac{1}{32} A_3^{r-2} \left(A_3 \left(A_3 b_{12}-2 A_2 b_{13}+2 A_1 b_{23}\right)+2 b_{13} b_{23} m^2\right) \left(A_3 b'_{13}-b_{13} A'_3\right) \left(A_3 b'_{23}-b_{23} A'_3\right)\nonumber\\
\mbox{Row 13} &:&  \hyperlink{S20o1}{\mathfrak{A}^{o,1}_{(r+2,2,0)}}\propto-\frac{1}{16} A_3^{r-1} \left(b_{23} \left(A_1 A'_3+m^2 b'_{13}\right)+A_2 \left(b_{13} A'_3-A_3 b'_{13}\right)-b'_{23} \left(A_1 A_3+b_{13} m^2\right)\right)\nonumber\\
&& ~~~~~~~~~\left(b_{23} \left(A_3 m^2 b'_{13}-A'_3 \left(A_1 A_3+2 b_{13} m^2\right)\right)+A_2 A_3 \left(b_{13} A'_3-A_3 b'_{13}\right)+A_3 b'_{23} \left(A_1 A_3+b_{13} m^2\right)\right)\nonumber\\
\mbox{Row 14} &:&  \hyperlink{S21o1}{\mathfrak{A}^{o,1}_{(r+1,2,1)}} \propto-\frac{1}{192}A_3^{r-2} \left(b'_{23} \left(b_{13} A''_3-A_3 b''_{13}\right)+b''_{23} \left(A_3 b'_{13}-b_{13} A'_3\right)+b_{23} \left(A'_3 b''_{13}-A''_3 b'_{13}\right)\right)\nonumber\\
&&~~~~~~~~~\left(b_{23} \left(A'_3 \left(A_1 A_3+2 b_{13} m^2\right)-A_3 m^2 b'_{13}\right)+A_2 A_3 \left(A_3 b'_{13}-b_{13} A'_3\right)-A_3 b'_{23} \left(A_1 A_3+b_{13} m^2\right)\right) \nonumber \\
\mbox{Row 15} &:&  \hyperlink{S21e1}{\mathfrak{A}^{e,1}_{(r+1,2,1)}} \propto \frac{1}{96} A_3^{r-1} \left(b'_{23} \left(b_{13} A''_3-A_3 b''_{13}\right)+b''_{23} \left(A_3 b'_{13}-b_{13} A'_3\right)+b_{23} \left(A'_3 b''_{13}-A''_3 b'_{13}\right)\right)\nonumber \\
&&~~~~~~~~~~~\left(-b_{23} \left(A_1 A'_3+m^2 b'_{13}\right)+A_2 \left(A_3 b'_{13}-b_{13} A'_3\right)+b'_{23} \left(A_1 A_3+b_{13} m^2\right)\right) \nonumber \\
\mbox{Row 16} &:&  \hyperlink{S22e1}{\mathfrak{A}^{o,1}_{(r,2,2)}} \propto -\frac{1}{4} m^2 A_3^{r-2} \left(b'_{23} \left(A_3 b''_{13}-b_{13} A''_3\right)+b''_{23} \left(b_{13} A'_3-A_3 b'_{13}\right)+b_{23} \left(A''_3 b'_{13}-A'_3 b''_{13}\right)\right){}^2 \nonumber\\
\mbox{Row 17} &:&  \hyperlink{S30o1}{\mathfrak{A}^{o,1}_{(r+3,3,0)}} \propto \frac{1}{64} A_3^r \left(b_{13} A'_3-A_3 b'_{13}\right) \left(b_{23} A'_3-A_3 b'_{23}\right)\nonumber\\
&&~~~~~~~~~~~ \left(b_{23} \left(A_1 A'_3+m^2 b'_{13}\right)+A_2 \left(b_{13} A'_3-A_3 b'_{13}\right)-b'_{23} \left(A_1 A_3+b_{13} m^2\right)\right) \nonumber\\
\mbox{Row 18} &:&  \hyperlink{S30e1}{\mathfrak{A}^{e,1}_{(r+3,3,0)}} \propto \frac{1}{128} A_3^{r-1} \left(b_{13} A'_3-A_3 b'_{13}\right) \left(b_{23} A'_3-A_3 b'_{23}\right)\nonumber\\
&& ~~~~~~~~\left(b_{23} \left(A_3 m^2 b'_{13}-A'_3 \left(A_1 A_3+2 b_{13} m^2\right)\right)+A_2 A_3 \left(b_{13} A'_3-A_3 b'_{13}\right)+A_3 b'_{23} \left(A_1 A_3+b_{13} m^2\right)\right) \nonumber\\
\mbox{Row 19} &:&  \hyperlink{S31o1}{\mathfrak{A}^{o,1}_{(r+2,3,1)}} \propto \frac{1}{96} A_3^{r-1} m^2 \left(b_{13} A'_3-A_3 b'_{13}\right) \left(b_{23} A'_3-A_3 b'_{23}\right)\nonumber\\
&&~~~~~~~~~~~ \left(b'_{23} \left(A_3 b''_{13}-b_{13} A''_3\right)+b''_{23} \left(b_{13} A'_3-A_3 b'_{13}\right)+b_{23} \left(A''_3 b'_{13}-A'_3 b''_{13}\right)\right) \nonumber\\
\mbox{Row 20} &:&  \hyperlink{S40e1}{\mathfrak{A}^{e,1}_{(r+4,4,0)}} \propto \frac{1}{32} A_3^r \left(b_{13} A'_3-A_3 b'_{13}\right){}^2 \left(b_{23} A'_3-A_3 b'_{23}\right){}^2 \nonumber
\end{eqnarray}
\endgroup

\subsection{$D=7$}\label{D7Gr:Amp}

In this subsection, we list the $D=7$ S-matrices generated by parity odd Lagrangians listed in the corresponding rows of Table \ref{GrD7podd}.

\begingroup
\allowdisplaybreaks
\begin{eqnarray}
\mbox{Row 1} &:& \hyperlink{S11o1p}{\mathfrak{A}^{o,1}_{(r,1,1)}} \propto\frac{1}{2} m A_3^{r-1} \varepsilon \left(k_1,\epsilon _1,\epsilon _2,k_2,\epsilon _3,\epsilon _4,\epsilon _5\right) \left(2 A_1 A_2+b_{12} m^2\right)\nonumber \\
\mbox{Row 2} &:& \hyperlink{S11o2p}{\mathfrak{A}^{o,2}_{(r,1,1)}} \propto\frac{1}{8} m A_3^{r-3} \varepsilon \left(k_1,\epsilon _1,\epsilon _2,k_2,\epsilon _3,\epsilon _4,\epsilon _5\right) \left(2 b_{23} \left(A_1 A_3+b_{13} m^2\right)+A_3 \left(A_3 b_{12}-2 A_2 b_{13}\right)\right)\nonumber \\
\mbox{Row 3} &:& \hyperlink{S21o1p}{\mathfrak{A}^{o,1}_{(r+1,2,1)}} \propto-\frac{A_3^{r-2}}{4 m}\varepsilon \left(k_1,\epsilon _1,\epsilon _2,k_2,\epsilon _3,\epsilon _4,\epsilon _5\right)\nonumber\\
&&~~~~~~~~~  \left(\left(A_2 A_3-b_{23} m^2\right) \left(A_3 b'_{23}-b_{23} A'_3\right)+A_1 A_3 \left(b_{13} A'_3-A_3 b'_{13}\right)+b_{13}^2 m^2 A'_3-A_3 b_{13} m^2 b'_{13}\right)\nonumber \\
\mbox{Row 4} &:& \hyperlink{S21e1p}{\mathfrak{A}^{e,1}_{(r+1,2,1)}} \propto\frac{A_3^{r-2} \varepsilon \left(k_1,\epsilon _1,\epsilon _2,k_2,\epsilon _3,\epsilon _4,\epsilon _5\right)}{4 m} \nonumber\\
&&~~~~~~~~~ \left(\left(A_2 A_3-b_{23} m^2\right) \left(A_3 b'_{23}-b_{23} A'_3\right)+A_1 A_3 \left(A_3 b'_{13}-b_{13} A'_3\right)-b_{13}^2 m^2 A'_3+A_3 b_{13} m^2 b'_{13}\right)\nonumber \\
\mbox{Row 5} &:& \hyperlink{S22e1p}{\mathfrak{A}^{e,1}_{(r,2,2)}} \propto-\frac{1}{2} m A_3^{r-2} \varepsilon \left(k_1,\epsilon _1,\epsilon _2,k_2,\epsilon _3,\epsilon _4,\epsilon _5\right) \nonumber\\
&&~~~~~~~~~~~\left(b'_{23} \left(A_3 b''_{13}-b_{13} A''_3\right)+b''_{23} \left(b_{13} A'_3-A_3 b'_{13}\right)+b_{23} \left(A''_3 b'_{13}-A'_3 b''_{13}\right)\right)\nonumber \\
\mbox{Row 6} &:& \hyperlink{S31o1p}{\mathfrak{A}^{o,1}_{(r+2,3,1)}} \propto\frac{1}{2} m A_3^{r-1} \varepsilon \left(k_1,\epsilon _1,\epsilon _2,k_2,\epsilon _3,\epsilon _4,\epsilon _5\right) \left(A_3 b'_{13}-b_{13} A'_3\right) \left(A_3 b'_{23}-b_{23} A'_3\right)\nonumber \\
\end{eqnarray}
\endgroup

\subsection{$D=6$}\label{D6Gr:Amp}

In this subsection, we list the $D=6$ S-matrices generated by parity odd Lagrangians listed in the corresponding rows of Table \ref{D6}.

\begin{eqnarray}
\mbox{Row 1} &:& \hyperlink{1D6}{\mathfrak{A}^{e,1}_{(r+1,1,0)}} \propto-A_3^r \left(2 A_1 A_2+b_{12} m^2\right) \ve \left(k_1,k_2,\epsilon_1,\epsilon_2,\epsilon_3,\epsilon_4\right)  \nonumber \\
\mbox{Row 2} &:& \hyperlink{2D6}{\mathfrak{A}^{e,2}_{(r+1,1,0)}} \propto \frac{1}{2} A_3^{r-2} \left(A_3 \left(A_2 b_{13}-\frac{A_3 b_{12}}{2}\right)-b_{23} \left(A_1 A_3+b_{13} m^2\right)\right) \ve \left(k_1,k_2,\epsilon_1,\epsilon_2,\epsilon_3,\epsilon_4\right)   \nonumber\\
\mbox{Row 3} &:& \hyperlink{3D6}{\mathfrak{A}^{e,1}_{(r+2,2,0)}} \propto \frac{1}{32} A_3^{r-1} \ve \left(k_1,k_2,\epsilon_1,\epsilon_2,\epsilon_3,\epsilon_4\right)\nonumber\\
&&~~~~~~~~~ \left(b_{23} \left(A'_3 \left(-A_1 A_3-2 b_{13} m^2\right)+A_3 m^2 b'_{13}\right)+A_2 A_3 \left(b_{13} A'_3-A_3 b'_{13}\right)+A_3 b'_{23} \left(A_1 A_3+b_{13} m^2\right)\right) \nonumber\\
\mbox{Row 4} &:& \hyperlink{4D6}{\mathfrak{A}^{o,1}_{(r+2,2,0)}} \propto \frac{1}{8} A_3^r \ve \left(k_1,k_2,\epsilon_1,\epsilon_2,\epsilon_3,\epsilon_4\right)\nonumber\\
&& ~~~~~~~~~~~\left(b_{23} \left(A_1 A'_3+m^2 b'_{13}\right)+A_2 \left(b_{13} A'_3-A_3 b'_{13}\right)-b'_{23} \left(A_1 A_3+b_{13} m^2\right)\right) \nonumber\\
\mbox{Row 5} &:& \hyperlink{5D6}{\mathfrak{A}^{e,2}_{(r+3,3,0)}} \propto -\frac{1}{4} A_3^r \left(b_{13} A'_3-A_3 b'_{13}\right) \left(b_{23} A'_3-A_3 b'_{23}\right) \ve \left(k_1,k_2,\epsilon_1,\epsilon_2,\epsilon_3,\epsilon_4\right) \nonumber
\end{eqnarray}

\subsection{$D=5$}\label{D5Gr:Amp}

In this subsection, we list the $D=5$ S-matrices generated by parity odd Lagrangians listed in the corresponding rows of Table \ref{D5}.
\begingroup
\allowdisplaybreaks
\begin{eqnarray}
\mbox{Row 1} &:& \hyperlink{5D5}{\mathfrak{A}^{o,1}_{(r,0,0)}} \propto- A_3^{r-3} \ve \left(k_1,k_2,\epsilon_1,\epsilon_2,\epsilon_3\right) \left(A_3 \left(A_2 b_{13}-\frac{A_3 b_{12}}{2}\right)-b_{23} \left(A_1 A_3+b_{13} m^2\right)\right)\nonumber \\
\mbox{Row 2} &:& \hyperlink{6D5}{\mathfrak{A}^{o,2}_{(r,0,0)}} \propto 2 A_3^{r-1} \ve \left(k_1,k_2,\epsilon_1,\epsilon_2,\epsilon_3\right) \left(2 A_1 A_2+b_{12} m^2\right)  \nonumber\\
\mbox{Row 3} &:& \hyperlink{S10o1D5}{\mathfrak{A}^{o,1}_{(r+1,1,0)}} \propto\frac{A_3^{r-1}}{4 m}  \left(2 A_1 A_2+b_{12} m^2\right) \nonumber\\
&&~~~~~~~~~~~\left(\left(b_{23} m^2-A_2 A_3\right) \varepsilon \left(k_1,\epsilon _1,k_2,\epsilon _3,\epsilon _4\right)+\left(A_1 A_3+b_{13} m^2\right) \varepsilon \left(k_2,\epsilon _2,k_1,\epsilon _3,\epsilon _4\right)\right)\nonumber\\
\nonumber\\
\mbox{Row 4} &:& \hyperlink{S10o2D5}{\mathfrak{A}^{o,2}_{(r+1,1,0)}} \propto\frac{A_3^{r-3} \left(2 b_{23} \left(A_1 A_3+b_{13} m^2\right)+A_3 \left(A_3 b_{12}-2 A_2 b_{13}\right)\right)}{16 m}\nonumber\\
&& ~~~~~~~~~~~\left(\left(b_{23} m^2-A_2 A_3\right) \varepsilon \left(k_1,\epsilon _1,k_2,\epsilon _3,\epsilon _4\right)+\left(A_1 A_3+b_{13} m^2\right) \varepsilon \left(k_2,\epsilon _2,k_1,\epsilon _3,\epsilon _4\right)\right)\nonumber\\
\nonumber\\
\mbox{Row 5} &:& \hyperlink{S10e1D5}{\mathfrak{A}^{e,1}_{(r+1,1,0)}} \propto-\frac{A_3^{r-1} \left(2 A_1 A_2+b_{12} m^2\right)}{8 m}\nonumber\\
&& ~~~~~~~~~~~\left(\varepsilon \left(k_1,\epsilon _1,k_2,\epsilon _3,\epsilon _4\right) \left(A_1 A_3+b_{13} m^2\right)+\varepsilon \left(k_2,\epsilon _2,k_1,\epsilon _3,\epsilon _4\right) \left(A_2 A_3-b_{23} m^2\right)\right)\nonumber\\
\nonumber\\
\mbox{Row 6} &:& \hyperlink{S10e2D5}{\mathfrak{A}^{e,2}_{(r+1,1,0)}} \propto-\frac{A_3^{r-3} \left(A_1 A_3+b_{13} m^2\right) \left(A_2 A_3-b_{23} m^2\right)}{32 m}\nonumber\\
&&~~~~~~~~~~~ \left(\epsilon \left(k_1,\epsilon _1,k_2,\epsilon _3,\epsilon _4\right) \left(A_1 A_3+b_{13} m^2\right)+\epsilon \left(k_2,\epsilon _2,k_1,\epsilon _3,\epsilon _4\right) \left(A_2 A_3-b_{23} m^2\right)\right)\nonumber\\
\nonumber\\
\mbox{Row 7} &:& \hyperlink{S20e1D5}{\mathfrak{A}^{e,1}_{(r+2,2,0)}} \propto\frac{1}{8} m A_3^r \varepsilon \left(\epsilon _1,\epsilon _2,k_3,\epsilon _3,\epsilon _4\right)\nonumber\\
&& ~~~~~~~~~~~\left(-b_{23} \left(A_1 A'_3+m^2 b'_{13}\right)+A_2 \left(A_3 b'_{13}-b_{13} A'_3\right)+b'_{23} \left(A_1 A_3+b_{13} m^2\right)\right)\nonumber\\
\nonumber\\
\mbox{Row 8} &:& \hyperlink{S20e2D5}{\mathfrak{A}^{e,2}_{(r+2,2,0)}} \propto\frac{A_3^r \left(2 A_1 A_2+b_{12} m^2\right)}{2 m} \left(\varepsilon \left(k_1,\epsilon _1,k_2,\epsilon _3,\epsilon _4\right) \right.\nonumber\\
&&\left. ~~~~~~~~~~~\left(b_{23} A'_3-A_3 b'_{23}\right)+\varepsilon \left(k_2,\epsilon _2,k_1,\epsilon _3,\epsilon _4\right) \left(A_3 b'_{13}-b_{13} A'_3\right)\right)\nonumber\\
\nonumber\\
\mbox{Row 9} &:& \hyperlink{S20e3D5}{\mathfrak{A}^{e,3}_{(r+2,2,0)}} \propto-\frac{A_3^{r-2} \left(2 b_{23} \left(A_1 A_3+b_{13} m^2\right)+A_3 \left(A_3 b_{12}-2 A_2 b_{13}\right)\right)}{8 m}\nonumber\\
&& ~~~~~~~~~~~\left(\varepsilon \left(k_1,\epsilon _1,k_2,\epsilon _3,\epsilon _4\right) \left(A_3 b'_{23}-b_{23} A'_3\right)+\varepsilon \left(k_2,\epsilon _2,k_1,\epsilon _3,\epsilon _4\right) \left(b_{13} A'_3-A_3 b'_{13}\right)\right)\nonumber\\
\nonumber\\
\mbox{Row 10} &:& \hyperlink{S20o1D5}{\mathfrak{A}^{o,1}_{(r+2,2,0)}} \propto\frac{A_3^{r-1}}{8 m}\left(-A_3 \left(A_1 \varepsilon \left(k_1,\epsilon _2,k_2,\epsilon _3,\epsilon _4\right)+A_2 \varepsilon \left(k_2,\epsilon _1,k_1,\epsilon _3,\epsilon _4\right)\right)\right.\nonumber\\
&& \left. \left(b_{23} \left(A_1 A'_3+m^2 b'_{13}\right)+A_2 \left(b_{13} A'_3-A_3 b'_{13}\right)-b'_{23} \left(A_1 A_3+b_{13} m^2\right)\right) -\frac{1}{2} m^2 \varepsilon \left(\epsilon _2,\epsilon _1,k_3,\epsilon _3,\epsilon _4\right) \right.\nonumber\\
&&\left. \left(b_{23} \left(A_3 m^2 b'_{13}-A'_3 \left(A_1 A_3+2 b_{13} m^2\right)\right)+A_2 A_3 \left(b_{13} A'_3-A_3 b'_{13}\right)+A_3 b'_{23} \left(A_1 A_3+b_{13} m^2\right)\right)\right)\nonumber\\
\nonumber\\
\mbox{Row 11} &:& \hyperlink{S30o1D5}{\mathfrak{A}^{o,1}_{(r+3,3,0)}} \propto\frac{1}{8} m A_3^r \varepsilon \left(\epsilon _2,\epsilon _1,k_3,\epsilon _3,\epsilon _4\right) \left(b_{13} A'_3-A_3 b'_{13}\right) \left(b_{23} A'_3-A_3 b'_{23}\right)\nonumber\\
\nonumber\\
\mbox{Row 12} &:& \hyperlink{S30e1D5}{\mathfrak{A}^{e,1}_{(r+3,3,0)}} \propto-\frac{A_3^{r-1} \left(A_3 b'_{13}-b_{13} A'_3\right) \left(A_3 b'_{23}-b_{23} A'_3\right) }{8 m}\nonumber\\
&&~~~~~~~~~~~\left(\varepsilon \left(k_1,\epsilon _1,k_2,\epsilon _3,\epsilon _4\right) \left(A_2 A_3-b_{23} m^2\right)+\varepsilon \left(k_2,\epsilon _2,k_1,\epsilon _3,\epsilon _4\right) \left(A_1 A_3+b_{13} m^2\right)\right)\nonumber\\
\nonumber\\
\mbox{Row 13} &:& \hyperlink{S40e1D5}{\mathfrak{A}^{e,1}_{(r+4,4,0)}} \propto\frac{A_3^r \left(b_{13} A'_3-A_3 b'_{13}\right) \left(b_{23} A'_3-A_3 b'_{23}\right)}{2 m}\nonumber\\
&& ~~~~~~~~~~~\left(\varepsilon \left(k_1,\epsilon _1,k_2,\epsilon _3,\epsilon _4\right) \left(A_3 b'_{23}-b_{23} A'_3\right)+\varepsilon \left(k_2,\epsilon _2,k_1,\epsilon _3,\epsilon _4\right) \left(b_{13} A'_3-A_3 b'_{13}\right)\right)\nonumber\\
\nonumber\\
\end{eqnarray}
\endgroup

\subsection{$D=4$}\label{D4Gr:Amp}

In this subsection, we list the $D=4$ S-matrices corresponding to the parity odd  Lagrangians listed in the corresponding rows of Table \ref{D4}.

\begin{eqnarray}
\mbox{Row 1} &:&\hyperlink{1D4}{\mathfrak{A}^{o,1}_{(r,0,0)}} \propto \frac{1}{8} A_3^{r-4} \left(2 b_{23} \left(A_1 A_3+b_{13} m^2\right)+A_3 \left(A_3 b_{12}-2 A_2 b_{13}\right)\right) \nonumber\\
&&~~~~~~~~~~~ \left(\ve\left(\epsilon_1,k_1,k_2,\epsilon_3\right) \left(A_2 A_3-b_{23} m^2\right)+\ve\left(\epsilon_2,k_1,k_2,\epsilon_3\right) \left(A_1 A_3+b_{13} m^2\right)\right) \nonumber \\
\mbox{Row 2} &:&\hyperlink{2D4}{\mathfrak{A}^{e,1}_{(r,0,0)}} \propto -  A_3^r~ \ve \left(\epsilon_1,k_1,\epsilon_2,k_2\right) \left(2 A_1 A_2+b_{12} m^2\right)   \nonumber
\end{eqnarray}

\section{Bootstrap}

The set of all possible parity even,  gauge invariant and Lorentz invariant Bose symmetric photon-photon-$P$ and graviton-graviton-$P$ amplitudes- for the special choice of $P$ polarization listed at the beginning of Appendix \ref{pa} - can also be enumerated following the `bootstrap' method outlined (for special cases) in subsubsection 7.2.1 of \cite{Chowdhury:2019kaq}. Briefly, the most general amplitude of this form is 
a polynomial of first (for photons) or second (for gravitons) order individually in  $\epsilon_1$, $\epsilon_2$, and of  $r_1^{\mbox{th}}$ order in $\epsilon_3$,  $r_2^{\mbox{th}}$ order in $\epsilon_4$ and $r_3^{\mbox{th}}$ order in $\epsilon_5$, 
and can be of any order in the momenta. 

To begin with we ignore the requirement of gauge invariance. We impose the condition that our amplitude is consistent with the fact that $P$ is a tensor with the symmetry properties of $Y_{(r_1, r_2, r_3, 0 \cdots 0)}$ as follows. Following the discussion of subsection \ref{lmp} we imagine assigning
fictitious integer labels (running from $1 \cdots r_1$) to every occurrence of 
$\epsilon_3$, and similar labels for every occurrence of $\epsilon_4$ and $\epsilon_5$, and then 
completely antisymmetrizing the expressions along columns of the Tableaux. 
For every column of length 3 this procedure specifies that we have the 
unique factor given by
\begin{equation} \label{detofmat}
 {\rm Det}  \begin{pmatrix} A_3 & A'_3 & A''_3\\
                            b_{13}& b'_{13} & b''_{13}\\
                             b_{23}& b'_{23} & b''_{23}
                             \end{pmatrix}
\end{equation}
(see \eqref{VarDef} for notation).

Now turning to columns of length 2 - each such column can be associated with 
one of three factors depending on the way we have labelled the Young Tableaux. The 
allowed factors are the 3 minors of the elements of the last column  in the matrix \eqref{detofmat}. Which minor
we get depends on our particular labelling scheme. One way to proceed at this point 
could be to choose a `gauge'. We could, for instance, simply declare that every 
column of length 2 is associated with a factor of the minor of $b''_{23}$ - but this 
`gauge choice' will fail in an amplitude that contains no factor of - for instance 
- $b'_{13}$ (for such an amplitude we are forced to choose an alternate gauge). In 
order not to have to deal with these `gauge' subtleties in a case by case basis, 
we use an over complete basis in which we consider each of these possibilities 
as a separate basis vector (this is not much extra work as we do all our computations on Mathematica). 

Finally every column of length 1 in the Young Tableaux is then simply a factor of any of $A_3$, $b_{13}$ and $b_{23}$  (the number of occurrences of each of  the last two terms cannot be larger than one in the case of photons or two in the case of gravitons).  

The end result of this procedure is a set of polynomials that are 
not linearly independent (because we worked with an over complete basis). In the next 
step we construct a linearly independent basis for the vector space spanned by 
our collection of polynomials. This is not difficult to do and yields a complete, 
linearly independent basis for the most general Bose symmetric, Lorentz invariant (but not yet gauge invariant) graviton-graviton-$P$ or graviton-graviton-$P$ S-matrix. 

Finally, we demand that the variation under gauge transformations of the S-matrix vanish. This requirement yields a set of conditions that relates 
previously independent coefficients of our basis vectors. The most general solution 
to this constraint is the set of allowed gauge invariant S-matrices. 

We have explicitly implemented this program for parity even S-matrices in
$D \geq 8$, both for photons and for gravitons,  and have  verified that the results of this program agree exactly with 
the amplitudes in subsections \ref{bo} and \ref{co}. We view this agreement as 
a nontrivial algebraic check on the results presented in this paper.

\end{document}